\documentclass[journal]{IEEEtran}
\ifCLASSINFOpdf
 
\else

\fi

\usepackage[cmex10]{amsmath}

\usepackage{graphics,graphicx,epic,eepic,epsfig,times,units} % epstopdf
\usepackage{subcaption}

\usepackage{siunitx}
\DeclareSIUnit \VAr {VAr} %volt-ampere reactive  
\DeclareSIUnit \VA {VA} %volt-ampere 
\DeclareSIUnit \rad {Radians} % Radians

\usepackage{cite}

\usepackage[hidelinks]{hyperref} %should come before {cleverref}. Doesn't work otherwise.

\usepackage{cleveref}
\crefformat{equation}{(#2#1#3)}
\crefrangeformat{equation}{(#3#1#4)--(#5#2#6)}
\crefmultiformat{equation}{(#2#1#3)}%
{,~(#2#1#3)}{, (#2#1#3)}{,~(#2#1#3)}

\usepackage[ruled,vlined]{algorithm2e}

\usepackage{xcolor}

\usepackage{algpseudocode}
\usepackage{algcompatible}
\algrenewcommand\alglinenumber[1]{\scriptsize #1:}

\usepackage{tabularx}

\newcommand{\reals}{{\mbox{\bf R}}}

\DeclareMathOperator*{\argmin}{arg\,min}

\usepackage{psfrag}

\newtheorem{definition}{Definition}
\let\olddefinition\definition
\renewcommand{\definition}{\olddefinition\normalfont}

\allowdisplaybreaks

\begin{document}

\title{A Component-Based Dual Decomposition Method for the OPF Problem}

\author{Sleiman~Mhanna,~\IEEEmembership{~MIEEE}
	Gregor~Verbi\v{c},~\IEEEmembership{Senior~MIEEE}
	and~Archie~C.~Chapman,~\IEEEmembership{~MIEEE}
%\thanks{Sleiman~Mhanna, Gregor~Verbi\v{c}~and~Archie~C.~Chapman are with the School of Electrical and Information Engineering, The University of Sydney, Australia, (e-mail: sleiman.mhanna@sydney.edu.au; gregor.verbic@sydney.edu.au; archie.chapman@sydney.edu.au).}
%\thanks{Manuscript received December 19, 2014; revised December 27, 2015.}
}

%\markboth{IEEE TRANSACTIONS ON POWER SYSTEMS,~Vol.~, No.~, March~2017}
%{Mhanna \MakeLowercase{\textit{et al.}}: A Component-wise Dual Decomposition Method for the OPF Problem}

\maketitle

\begin{abstract}
	This paper proposes a component-based dual decomposition of the nonconvex AC optimal power flow (OPF) problem, where the modified dual function is solved in a distributed fashion. The main contribution of this work is that is demonstrates that a distributed method with carefully tuned parameters can converge to globally optimal solutions despite the inherent nonconvexity of the problem and the absence of theoretical guarantees of convergence. This paper is the first to conduct extensive numerical analysis resulting in the identification and tabulation of the algorithmic parameter settings that are crucial for the convergence of the method on 72 AC OPF test instances. Moreover, this work provides a deeper insight into the geometry of the modified Lagrange dual function of the OPF problem and highlights the conditions that make this function differentiable. This numerical demonstration of convergence coupled with the scalability and the privacy preserving nature of the proposed method makes it well suited for smart grid applications such as multi-period OPF with demand response (DR) and security constrained unit commitment (SCUC) with contingency constraints and multiple transmission system operators (TSOs).
\end{abstract}


\IEEEpeerreviewmaketitle

\setlength{\belowdisplayskip}{0pt} \setlength{\belowdisplayshortskip}{0pt}
\setlength{\abovedisplayskip}{0pt} \setlength{\abovedisplayshortskip}{0pt}
\section*{Notation}

\addcontentsline{toc}{section}{Notation}
\subsection{Input data and operators}
%\small
\begin{IEEEdescription}[\IEEEsetlabelwidth{$p_{i}^{g}/q_{i}^{g}$}\IEEEusemathlabelsep]
	\item[$\mathcal{B}$] Set of buses in the power network.
	\item[$\mathcal{B}_{i}$] Set of buses connected to bus $i$.
	\item[$b^\text{sh}_{i}$] Shunt susceptance (p.u.) at bus $i$.
	\item[$g^\text{sh}_{i}$] Shunt conductance (p.u.) at bus $i$.
	\item[$b^\text{ch}_{ij}$] Charging susceptance (p.u.) in the $\pi$-model of line $ij$.
	\item[$c0^{g}_{i}$] Constant coefficient ($\SI{}{\$}$) term of generator $g$'s cost function.
	\item[$c1^{g}_{i}$] Coefficient ($\SI{}{\$\per\mega\watt}$) of the linear term of generator $g$'s cost function.
	\item[$c2^{g}_{i}$] Coefficient ($\SI{}{\$\per\mega\watt\squared}$) of the quadratic term of generator $g$'s cost function.
	\item[$\mathcal{G}$] Set of all generators $(g,i)$ in the power network such that $g$ is the generator and $i$ is the bus connected to it.
	\item[$\mathcal{G}_{i}$] Set of all generators connected to bus $i$.
	\item[$\mathrm{i}$] Imaginary unit.
	\item[$\mathcal{L}$] Set of all transmission lines $ij$ where $i$ is the ``from'' bus.
	\item[$\mathcal{L}_{t}$] Set of all transmission lines $ij$ where $i$ is the ``to'' bus.
	\item[$p_{i}^{\text{d}}/q_{i}^{\text{d}}$] Active/reactive power demand ($\SI{}{\mega\watt}/\SI{}{\mega\VAr}$) at bus $i$.
	\item[$\overline{s}_{ij}$] Apparent power rating ($\SI{}{\mega\VA}$) of line $ij$.
	\item[$\underline{\theta}_{ij}^{\Delta}$] Lower limit of the difference of voltage angles of buses $i$ and $j$.
	\item[$\overline{\theta}_{ij}^{\Delta}$] Upper limit of the difference of voltage angles of buses $i$ and $j$.
	\item[$\theta_{i}^{\text{shift}}$] Phase shift ($\SI{}{\rad}$) of phase shifting transformer connected between buses $i$ and $j$ ($\theta_{i}^{\text{shift}}=0$ for a transmission line).
	\item[$\tau_{ij}$] Tap ratio magnitude of phase shifting transformer connected between buses $i$ and $j$ ($\tau_{ij}=1$ for a transmission line).
	\item[$T_{ij}$] Complex tap ratio of a phase shifting transformer ($T_{ij}=\tau_{ij}\mathrm{e}^{\mathrm{i} \theta_{i}^{\text{shift}}}$).
	\item[$Y_{ij}$] Series admittance (p.u.) in the $\pi$-model of line $ij$.
	\item[$\Im\left\{\bullet\right\}$] Imaginary value operator.
	\item[$\Re\left\{\bullet\right\}$] Real value operator.
	\item[$\underline{\bullet}/\overline{\bullet}$] Minimum/maximum magnitude operator.
	\item[$\left|\bullet\right|$] Magnitude operator/Cardinality of a set.
	\item[$\bullet^*$] Conjugate operator.
	\item[$k$] Iteration number.
	\item[$\rho_{v \theta}$] ADMM penalty parameter.
	\item[$\rho_{pq}$] Penalty parameter.
	\item[$\nu$] Proximal penalty parameter.
\end{IEEEdescription}

\subsection{Decision variables}
\begin{IEEEdescription}[\IEEEsetlabelwidth{$p_{i}^{g}/q_{i}^{g}$}\IEEEusemathlabelsep]
	\item[$p_{i}^{g}/q_{i}^{g}$] Active/reactive power ($\SI{}{\mega\watt}/\SI{}{\mega\VAr}$) generation of generator $g$ at bus $i$.
	\item[$p_{ij}/q_{ij}$] Active/reactive power ($\SI{}{\mega\watt}/\SI{}{\mega\VAr}$) flow along transmission line $ij$.
	\item[$V_{i}$] Complex phasor voltage (p.u.) at bus $i$ ($V_{i}=\left| V_{i} \right| \angle \theta_{i}=v_{i} \angle \theta_{i}$).
	\item[$v_{i_{(ij)}}$] Duplicate of $v_{i}$ at line $ij$ such that $j\in \mathcal{B}_{i}$.
	\item[$\theta_{i_{(ij)}}$] Duplicate of $\theta_{i}$ at line $ij$ such that $j\in \mathcal{B}_{i}$.		
	\item[$\boldsymbol{\lambda}$] Vector of Lagrange multipliers.
%	\item[$\text{G}_{1},\text{G}_{2},\text{G}_{3},\text{D}_{1},\text{D}_{2},\text{D}_{3},\text{L}_{1},\text{L}_{2},\text{L}_{3},\text{B}_{1},\text{B}_{2},\text{B}_{3}$] 
%	\item[$p_{1}^{g1},q_{1}^{g1},p_{2}^{g2},q_{2}^{g2},p_{3}^{g3},q_{3}^{g3},p_{1}^{d1},q_{1}^{d1},p_{2}^{d2},q_{2}^{d2},p_{3}^{d3},q_{3}^{d3}$]
%	\item[$v_{1},\theta_{1},v_{2},\theta_{2},v_{3},\theta_{3}$]
%	\item[$v_{2_{(23)}},\theta_{2_{(23)}},v_{3_{(32)}},\theta_{3_{(32)}}$]
%	\item[$v_{1_{(12)}},\theta_{1_{(12)}},v_{2_{(21)}},\theta_{2_{(21)}},v_{1_{(13)}},\theta_{1_{(13)}},v_{3_{(31)}},\theta_{3_{(31)}}$]
%	\item[$p_{12}+\mathrm{i} q_{12}$]
%	\item[$p_{21}+\mathrm{i} q_{21}$]
%	\item[$p_{13}+\mathrm{i} q_{13}$]
%	\item[$p_{31}+\mathrm{i} q_{31}$]
%	\item[$p_{23}+\mathrm{i} q_{23}$]
%	\item[$p_{32}+\mathrm{i} q_{32}$]
%	\item[$ $]
%	\item[$p_{1}^{g1}+\mathrm{i} q_{1}^{g1}$]
%	\item[$ $]
%	\item[$p_{2}^{g2}+\mathrm{i} q_{2}^{g2}$]
%	\item[$ $]
%	\item[$p_{3}^{g3}+\mathrm{i} q_{3}^{g3}$]
%	\item[$ $]
%	\item[$p_{1}^{d1}+\mathrm{i} q_{1}^{d1}$]
%	\item[$ $]
%	\item[$p_{2}^{d2}+\mathrm{i} q_{2}^{d2}$]
%	\item[$ $]
%	\item[$p_{3}^{d3}+\mathrm{i} q_{3}^{d3}$]
%	\item[$ $]
%	\item[$=$]
\end{IEEEdescription}

\subsection{Acronyms}
\begin{IEEEdescription}[\IEEEsetlabelwidth{$p_{i}^{g}/q_{i}^{g}$}\IEEEusemathlabelsep]
	\item[AC] Alternating current.
	\item[ADMM] Alternating direction method of multipliers.
	\item[DR] Demand response.
	\item[IPM] Interior-point optimization methods.
	\item[GNLP] Global nonlinear programming.
	\item[KKT] \emph{Karush-Kuhn-Tucker}.
	\item[NLP] Nonlinear programming.
	\item[OCD] Optimality conditions decomposition.
	\item[OPF] Optimal power flow.
	\item[SCUC] Security-constrained unit commitment.
	\item[SDP] Semidefinite programming.
	\item[SOCP] Second-order cone programming.
	\item[TSO] Transmission system operator.
\end{IEEEdescription}

\section{Introduction}

The alternating current (AC) power flow equations, which model the steady-state physics of power flows, are the linchpins of a broad spectrum of optimization problems in electrical power systems. Unfortunately, these nonlinear equations are the main sources of nonconvexity, which makes these problems notorious for being extremely challenging to solve using global nonlinear programming (GNLP) solvers. Therefore, the research community has focused on improving interior-point nonlinear optimization methods (IPM) to compute feasible solutions efficiently \cite{IPMforOPF,MATPOWER}. Although these methods only (theoretically) guarantee local optimality, they have been shown, thanks to tight convex relaxations \cite{Kocuk_strongSOCP,Coffrin_Strengtheningwithboundtightening,Hijazi_PolynomialSDPcuts,Coffrin_StrengtheningSDP,Kocuk_Matrixminorreformulations}, to reach near-optimal (if not globally optimal) solutions on all the known test cases in the literature. This paper capitalizes on this to numerically show that the proposed distributed method solves the modified dual problem of the nonconvex AC OPF problem to near optimality, if not to global optimality.

In particular, the second-order cone programming (SOCP) and the semidefinite programming (SDP) relaxations have garnered considerable attention. The increased interest in this line of research stems from the fact that the SDP relaxation is shown to be exact, \textit{i.e.}, yields a zero optimality gap, on a variety of case studies \cite{Lavai_0dualitygapinOPF}. However, in many practical OPF instances, the SDP relaxation yields inexact solutions \cite{Molzahn_investigationofnon0gap,Kocuk_InexactnessofSDP}. In these scenarios, an AC feasible solution cannot be recovered from the SDP relaxed solution. Nonetheless, the SDP relaxation can be strengthened by solving a hierarchy of moment relaxations \cite{Josz_moment-sosopf,Molzahn_Moment-BasedRelaxations,Josz_momentsumofsquares} or by a combination of lifted nonlinear cuts, valid inequalities and bound tightening methods \cite{Hijazi_PolynomialSDPcuts,Coffrin_StrengtheningSDP}, at the cost of larger SDP problems.\footnote{In fact, in many instances, moment relaxations for the OPF yield AC feasible solutions where the SDP relaxation yields inexact solutions \cite{Josz_momentsumofsquares}.} Even more recently, increased attention was given to the computationally less demanding SOCP relaxation initially proposed in \cite{Jabr_radialDNusingConicP}. The SOCP relaxation in its classical form in \cite{Jabr_radialDNusingConicP} is shown to be dominated by the SDP relaxation but recent strengthening techniques \cite{Kocuk_strongSOCP,Coffrin_Strengtheningwithboundtightening,Kocuk_Matrixminorreformulations} have shifted this paradigm.

\subsection{State-of-the-art}

There is a plethora of existing works on distributed OPF. These can be broadly classified into three categories, dual decomposition methods \cite{Kim_coarsegraineddistOPF,Kim_fastdistributedOPF,Kim_comparisonofDOPF	,Biskas_DecentralizedOPF,Magnusson_ADMMsequentialconvex,Erseghe_ADMM,Kraning_ADMM,Peng_DOPFbalancedradial,DallAnese_SDPADMM,Scott_DistributedOPFforDR}, optimality conditions decomposition (OCD) methods \cite{Nogales_DecompositionforOPF,Bakirtzis_DecentralizedDCOPF,Hug_DecentralizedOPF,Guo_Intelligentpartitioning,Minot_ParallelDCOPF} and sparse SDP decomposition methods \cite{Lam_distributedOPF,Madani_DistributedSparseSDPforOPF}. The dual decomposition techniques underlying the dual-decomposition-based distributed OPF methods in the literature can in turn be classified into two categories: region-based decompositions \cite{Kim_coarsegraineddistOPF,Kim_fastdistributedOPF,Kim_comparisonofDOPF	,Sun_ACOPFalgorithms,Biskas_DecentralizedOPF,DallAnese_SDPADMM,Magnusson_ADMMsequentialconvex,Erseghe_ADMM},\footnote{Note that the OPF problem in \cite{Magnusson_ADMMsequentialconvex} and \cite{Sun_ACOPFalgorithms} is decomposed in terms of buses, which can be thought of as the maximum number of regions in a power network.} and component-based decompositions \cite{Kraning_ADMM,Peng_DOPFbalancedradial,Scott_DistributedOPFforDR}. The focus of this study revolves around the latter decomposition techniques because they preserve privacy with respect to \emph{all} components (generators, transformers, loads, buses, lines etc.) and are flexible enough to incorporate discrete decision variables to suit a wide variety of optimization applications in power system operations such as optimal transmission switching, capacitor placement, transmission and distribution network expansion planning, optimal feeder reconfiguration, power system restoration, and vulnerability analysis, to name a few. On the downside, dual-decomposition-based AC OPF methods have no theoretical guarantee of convergence because the (primal) OPF problem is nonconvex. Nonetheless, this paper numerically shows that under the right conditions, the proposed distributed method can converge to near-optimal (possibly globally optimal) solutions. Unlike \cite{Kraning_ADMM,Peng_DOPFbalancedradial,DallAnese_SDPADMM,Lam_distributedOPF,Madani_DistributedSparseSDPforOPF,Madani_DistributedSparseSDPforOPF}, which solve a convexified version of the OPF problem, this paper tackles the nonlinear nonconvex AC OPF directly. Convex relaxations are appealing because they are computationally conducive but their main disadvantage is that they do not always yield feasible solutions. Furthermore, in contrast to \cite{Scott_DistributedOPFforDR}, the work in this paper conducts extensive numerical analysis and specifies the algorithmic parameter settings that are crucial for the convergence of the proposed component-based dual decomposition method on a vast array of test instances. On the other hand, OCD methods \cite{Nogales_DecompositionforOPF,Bakirtzis_DecentralizedDCOPF,Hug_DecentralizedOPF,Guo_Intelligentpartitioning,Minot_ParallelDCOPF} rely on matrix factorization \cite{Shyan-Lung_Parallelsolutions} to parallelize the computation of the \emph{Karush-Kuhn-Tucker} (KKT) conditions. However, as of yet, these methods are not amenable to decompositions in terms of components.

\subsection{Contributions of this work}

In contrast to most distributed AC OPF algorithms in the existing literature, the algorithm in this paper is not only tested on the classical MATPOWER \cite{MATPOWER} instances but also on the more challenging NESTA v6 \cite{NESTA} test cases, which are designed specifically to incorporate key network parameters such as line thermal limits and small angle differences, which are critical in optimization applications. To get a grasp on how difficult the problem is, the methods in \cite{Lam_distributedOPF,Erseghe_ADMM}, with the exception of \cite{Magnusson_ADMMsequentialconvex}, are only tested on MATPOWER cases with at most 118-buses. The method in \cite{Magnusson_ADMMsequentialconvex} is tested on MATPOWER's 300-bus system but does not converge after 10,000 iterations.  
%doesn't require optimal partitioning and shit.
%The The dual problem of the nonconvex AC OPF problem is unbounded therefore a modified Lagrange dual function can proposed at the expense of separability. Separability can be restored by approximating the modified Lagrange dual function (see ADMM). However, the main caveat of distributed methods that approximate the modified Lagrange dual function of such nonconvex problems (ones with unbounded original dual functions) is that they have no guarantee of converging to a globally optimal solution. On a different note, the existing literature does not have extensive convergence on distributed AC OPF on any decomposition type, let alone a component-based one.

Against this background, this paper is the first to conduct extensive numerical analysis on the application of a distributed algorithm to solve the modified Lagrange dual function of the AC OPF problem. In more detail, this paper advances the state of the art in the following ways:
\begin{itemize}
	\item Extensive numerical simulations on $72$ test cases from MATPOWER \cite{MATPOWER}, PEGASE \cite{Josz_ACdataMATPOWER} and NESTA v6 \cite{NESTA} instances show that the proposed algorithm converges to the same near-optimal (possibly globally optimal) solutions obtained from the centralized IPMs.
	\item The algorithmic parameter settings that are crucial for convergence are identified and tabulated.
	\item A deeper insight into the geometry of the modified Lagrange dual function of the OPF problem shows that this function can be nonsmooth for small values of the penalty parameters.
\end{itemize}
This type of distributed OPF analysis has not been conducted in the existing literature, let alone on a component-based dual decomposition of the OPF. Therefore the techniques developed in this paper can serve as a basis for a myriad of smart grid optimization methods that are based on AC OPF, such as security constrained unit commitment (SCUC) with contingency constraints and multiple transmission system operators (TSOs), stochastic OPF, probabilistic OPF, and multi-period OPF with demand response (DR), to name a few.

\subsection{Notation}

All vectors are column vectors unless otherwise specified, and $\boldsymbol{1}$ is an all-ones vector of length depending on the context. The inner product of two vectors $\boldsymbol{x}$, $\boldsymbol{y} \in \reals^n$ is delineated by $\left\langle \boldsymbol{x}, \boldsymbol{y} \right\rangle:=\boldsymbol{x}^{T} \boldsymbol{y}$, where $\boldsymbol{x}^{T}$ is the transpose of $\boldsymbol{x}$. The Euclidean norm of a vector $\boldsymbol{x} \in \reals^n$ is denoted by $\left\|\boldsymbol{x}\right\|:=\sqrt{\left\langle \boldsymbol{x}, \boldsymbol{x} \right\rangle}$ and the nonnegative orthant in $\reals^n$ is denoted by $\reals^n_{+}$. Also, the Hadamard product of two vectors $\boldsymbol{x}$ and $\boldsymbol{y}$ is denoted by $\boldsymbol{x} \circ \boldsymbol{y}$. Moreover, complex variables and parameters are in upper case whereas real variables and parameters are in lower case.
%All vectors are column vectors and $\left[ a;b\right] =\left[ a,b \right] $.

\subsection{Organization of the paper}

The paper starts with a formal description of the polar form OPF in general networks in Section~\ref{sec:OPFproblem}, followed by the component-based dual decomposition in Section~\ref{sec:Componentbaseddecomp}. Section~\ref{sec:modifieddual} describes the modified dual function and the proposed distributed algorithms. Section~\ref{sec:numericalevaluation} shows the numerical evaluation of the algorithms and Section~\ref{sec:conclusion} concludes the paper. Finally, Appendices~\ref{Appendix1} and~\ref{Appendix2} supplement the paper with valuable examples that provide a deeper insight on the optimality of dual-decomposition methods on nonconvex problems. 

\section{The OPF problem}\label{sec:OPFproblem}

In a power network, the OPF problem consists of finding the least-cost dispatch of power from generators to satisfy the load at all buses in a way that is governed by physical laws, such as Ohm's Law and Kirchhoff's Law, and other technical restrictions, such as transmission line thermal limit constraints. Knowing that $\Re\left\{V_{i} V_{j}^*\right\}:=v_{i}v_{j}\cos\left(\theta_{i}-\theta_{j}\right)$ and  $ \Im\left\{V_{i} V_{j}^*\right\}:=v_{i}v_{j}\sin\left(\theta_{i}-\theta_{j}\right)$, the OPF problem in \emph{polar form} can be written as
\begin{subequations}\label{eq2:opf}
	\begin{align}
		\underset {\substack{p_{i}^{g},q_{i}^{g},v_{i},\theta_{i},\\p_{ij},q_{ij},p_{ji},q_{ji}}} 
		{\mbox{ minimize}} & \sum_{(g,i)\in \mathcal{G}} f^{g}_{i}\left(p_{i}^{g}\right) & \label{eq2:objective}\\
		\text{ subject to} &  \nonumber \\
		\underline{p}_{i}^{g} & \leq p_{i}^{g} \leq \overline{p}_{i}^{g}, \qquad \qquad \qquad \ \ \ (g,i)\in \mathcal{G} & \label{eq2:Pminmax} \\
		\underline{q}_{i}^{g} & \leq q_{i}^{g} \leq \overline{q}_{i}^{g}, \qquad \qquad \qquad \ \ \ (g,i)\in \mathcal{G} & \label{eq2:Qminmax} \\
		\underline{v}_{i} & \leq v_{i} \leq \overline{v}_{i},  \ \qquad \qquad \qquad \qquad  \ \ i \in \mathcal{B} & \label{eq2:Vminmax} \\ 
		\underline{\theta}_{ij}^{\Delta} & \leq \theta_{i}-\theta_{j} \leq \overline{\theta}_{ij}^{\Delta}, \qquad \quad \quad  \ \  (i,j) \in \mathcal{L} & \label{eq2:anglediff} \\
		\sum_{(g,i)\in \mathcal{G}} p_{i}^{g}&-p_{i}^{\text{d}} =  \sum_{j\in \mathcal{B}_{i}} p_{ij} + g^{\text{sh}}_{i}v_{i}^{2},  \ \qquad  \ \ \ i\in \mathcal{B}  & \label{eq2:KCL1}  \\	
		\sum_{(g,i)\in \mathcal{G}}  q_{i}^{g}&-q_{i}^{\text{d}}=  \sum_{j\in \mathcal{B}_{i}} q_{ij} - b^{\text{sh}}_{i}v_{i}^{2},    \qquad \quad \ i\in \mathcal{B}  & \label{eq2:KCL2}  \\
		p_{ij}=& \ g^{c}_{ij} v_{i}^2-g_{ij}v_{i}v_{j}\cos\left(\theta_{i}-\theta_{j}\right)  \nonumber\\
		+& \ b_{ij}v_{i}v_{j}\sin\left(\theta_{i}-\theta_{j}\right),   \qquad \quad  (i,j) \in \mathcal{L} & \label{eq2:Pij} \\
		q_{ij}=& \ b^{c}_{ij} v_{i}^2-b_{ij} v_{i}v_{j}\cos\left(\theta_{i}-\theta_{j}\right)  \nonumber\\
		-& \ g_{ij}v_{i}v_{j}\sin\left(\theta_{i}-\theta_{j}\right),   \qquad \quad  (i,j) \in \mathcal{L} & \label{eq2:Qij} \\
		p_{ji}=& \ g^{c}_{ji} v_{j}^2-g_{ji} v_{j}v_{i}\cos\left(\theta_{j}-\theta_{i}\right) & \nonumber\\
		+& \ b_{ji}v_{j}v_{i}\sin\left(\theta_{j}-\theta_{i}\right),   \qquad \quad  (i,j) \in \mathcal{L} & \label{eq2:Pji} \\
		q_{ji}=& \ b^{c}_{ji} v_{j}^2-b_{ji} v_{j}v_{i}\cos\left(\theta_{j}-\theta_{i}\right) & \nonumber\\
		-& \ g_{ji}v_{j}v_{i}\sin\left(\theta_{j}-\theta_{i}\right),  \qquad \quad  (i,j) \in \mathcal{L} &\label{eq2:Qji} \\	
		& \sqrt{p_{ij}^2+q_{ij}^2} \leq \overline{s}_{ij}, \quad \quad \ \  (i,j) \in \mathcal{L} \cup \mathcal{L}_{t} & \label{eq2:linethermallimit} 
	\end{align}
\end{subequations}
where,  $g^{c}_{ij}:=\Re\left\{\frac{Y_{ij}^*-\mathrm{i}\frac{b^\text{ch}_{ij}}{2}}{\left|T_{ij}\right|^2}\right\}$, $b^{c}_{ij}:=\Im\left\{\frac{Y_{ij}^*-\mathrm{i}\frac{b^\text{ch}_{ij}}{2}}{\left|T_{ij}\right|^2}\right\}$, $g_{ij}:=\Re\left\{\frac{Y_{ij}^*}{T_{ij}}\right\}$, $b_{ij}:=\Im\left\{\frac{Y_{ij}^*}{T_{ij}}\right\}$, $g^{c}_{ji}:=\Re\left\{Y_{ji}^*-\mathrm{i}\frac{b^\text{ch}_{ji}}{2}\right\}$, $b^{c}_{ji}:=\Im\left\{Y_{ji}^*-\mathrm{i}\frac{b^\text{ch}_{ji}}{2}\right\}$, $g_{ji}:=\Re\left\{\frac{Y_{ji}^*}{T_{ji}^*}\right\}$ and $b_{ji}:=\Im\left\{\frac{Y_{ji}^*}{T_{ji}^*}\right\}$, and $f^{g}_{i}\left(p_{i}^{g}\right):=c2^{g}_{i}\left(p_{i}^{g}\right)^2 +c1^{g}_{i}\left(p_{i}^{g}\right)+c0^{g}_{i}$. The OPF in \eqref{eq2:opf} is a nonconvex nonlinear optimization problem that is proven to be NP-hard \cite{Lavai_0dualitygapinOPF}. The nonconvexities stem from equality constraints \cref{eq2:KCL1,eq2:KCL2,eq2:Pij,eq2:Qij,eq2:Pji,eq2:Qji}, which include nonconvex voltage bilinear terms multiplied by nonconvex sine and cosine functions of the angles, and a quadratic function of the voltage, which is also nonconvex in this equality constraint setting as it describes the boundary of the set $\left\{v^2 | v \in \left[\underline{v},\overline{v}\right]\right\}$.\footnote{The method in this paper was also applied to the OPF in \emph{rectangular form} but the results are not documented here because they were not significantly different than the polar form ones.}

\section{Component-based dual decomposition}~\label{sec:Componentbaseddecomp} 

\begin{figure}[t]
	\centering{
		\includegraphics[width=90mm] {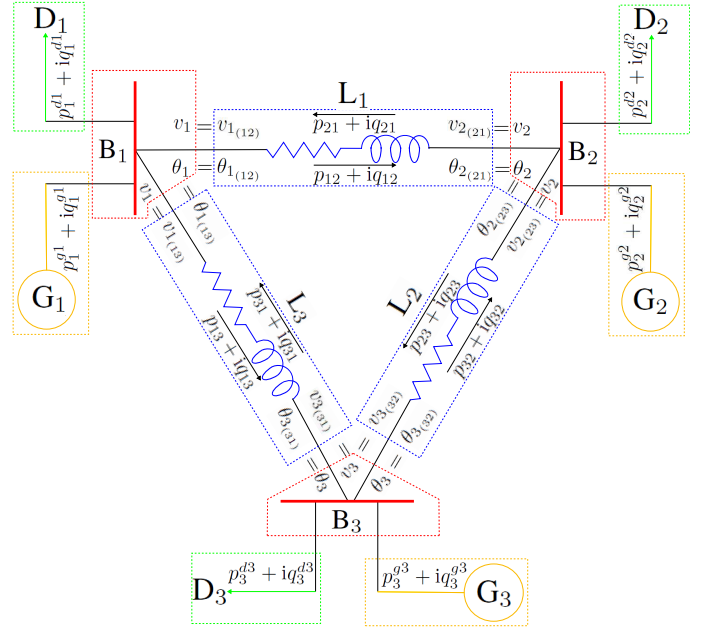}}
	\caption{A 3-bus system showing the duplication of the coupling variables and the resulting component-based decomposition.}
	\label{fig:3Bussystem}
\end{figure}
The OPF problem in its native form in \eqref{eq2:opf} is not separable in terms of components, as relaxing the coupling constraints in \eqref{eq2:KCL1} and \eqref{eq2:KCL2} is not enough to bestow a component-based separability. Towards this aim, the following variables are duplicated
\begin{align}
v_{i}=v_{i_{(ij)}},  \qquad  (i,j) \in \mathcal{L} \cup \mathcal{L}_{t}, \label{eq3:consV} \\
\theta_{i}=\theta_{i_{(ij)}},  \qquad  (i,j) \in \mathcal{L} \cup \mathcal{L}_{t}, \label{eq3:conso}
\end{align}
and the OPF problem now becomes
\begin{subequations}\label{eq3:opf}
	\begin{align}
	\underset {\substack{\boldsymbol{x}}} 
	{\mbox{ minimize}} & \sum_{(g,i)\in \mathcal{G}} f^{g}_{i}\left(p_{i}^{g}\right)   \label{eq3:objective}\\
	\text{ subject } & \text{ to \cref{eq2:Pminmax,eq2:Qminmax,eq2:KCL1,eq2:KCL2}, \cref{eq2:linethermallimit,eq3:consV,eq3:conso}} \label{eq3:same} \\
	\underline{v}_{i} & \leq v_{i_{(ij)}} \leq \overline{v}_{i},  \quad \qquad \qquad  (i,j) \in \mathcal{L} \cup \mathcal{L}_{t}  \label{eq3:Vminmax} \\ 
	\underline{\theta}_{ij}^{\Delta} & \leq \theta_{i_{(ij)}}-\theta_{j_{(ji)}} \leq \overline{\theta}_{ij}^{\Delta}, \quad \  (i,j) \in \mathcal{L}  \label{eq3:anglediff} \\
	p_{ij}= g^{c}_{ij}& v_{i_{(ij)}}^2-g_{ij}v_{i_{(ij)}}v_{j_{(ji)}}\cos\left(\theta_{i_{(ij)}}-\theta_{j_{(ji)}}\right)  \nonumber\\
	+b_{ij} & v_{i_{(ij)}}v_{j_{(ji)}}\sin\left(\theta_{i_{(ij)}}-\theta_{j_{(ji)}}\right), \  \  (i,j) \in \mathcal{L}  \label{eq3:Pij} \\
	q_{ij}= b^{c}_{ij}& v_{i_{(ij)}}^2-b_{ij} v_{i_{(ij)}}v_{j_{(ji)}}\cos\left(\theta_{i_{(ij)}}-\theta_{j_{(ji)}}\right)  \nonumber\\
	-g_{ij} & v_{i_{(ij)}}v_{j_{(ji)}}\sin\left(\theta_{i_{(ij)}}-\theta_{j_{(ji)}}\right),   \    \ (i,j) \in \mathcal{L}  \label{eq3:Qij} \\
	p_{ji}= g^{c}_{ji}& v_{j_{(ji)}}^2-g_{ji} v_{j_{(ji)}}v_{i_{(ij)}}\cos\left(\theta_{j_{(ji)}}-\theta_{i_{(ij)}}\right)  \nonumber\\
	+b_{ji} & v_{j_{(ji)}}v_{i_{(ij)}}\sin\left(\theta_{j_{(ji)}}-\theta_{i_{(ij)}}\right),   \   \ (i,j) \in \mathcal{L}  \label{eq3:Pji} \\
	q_{ji}= b^{c}_{ji}& v_{j_{(ji)}}^2-b_{ji} v_{j_{(ji)}}v_{i_{(ij)}}\cos\left(\theta_{j_{(ji)}}-\theta_{i_{(ij)}}\right)  \nonumber\\
	-g_{ji} & v_{j_{(ji)}}v_{i_{(ij)}}\sin\left(\theta_{j_{(ji)}}-\theta_{i_{(ij)}}\right),  \     \ (i,j) \in \mathcal{L} \label{eq3:Qji} 	
	\end{align}
\end{subequations}	
where $\boldsymbol{x}:=\left[  \left( \boldsymbol{x}_{i}^{g} \right)_{(g,i)\in \mathcal{G}} , \left(\boldsymbol{x}_{ij}^{\text{l}}\right)_{(i,j) \in \mathcal{L}},\left(\boldsymbol{x}_{i}^{\text{b}} \right)_{i \in \mathcal{B}} \right]$,
%\footnote{Note that $v_{i_{(ij)}}$ is equivalent to $v_{i_{(ji)}}$ and $\theta_{i_{(ij)}}$ is equivalent to $\theta_{i_{(ji)}}$.}
$\boldsymbol{x}_{i}^{g}:=[p_{i}^{g},q_{i}^{g}]$, $\boldsymbol{x}_{ij}^{\text{l}}:=[p_{ij},q_{ij},p_{ji},q_{ji},v_{i_{(ij)}},\theta_{i_{(ij)}},v_{j_{(ji)}},\theta_{j_{(ji)}}]$ and $\boldsymbol{x}_{i}^{\text{b}}:=[v_{i},\theta_{i}]$.

Let, $\boldsymbol{\lambda}_{i}:=\left[ \lambda_{p,i},\lambda_{q,i} \right] $, $\boldsymbol{\lambda}_{ij}:=\left[ \lambda_{v_{i_{(ij)}}},\lambda_{\theta_{i_{(ij)}}}\right]$, $\boldsymbol{\lambda}_{ji}:=\left[ \lambda_{v_{j_{(ji)}}},\lambda_{\theta_{j_{(ji)}}}\right]$, and $
\boldsymbol{\lambda}:=\left[\left( \boldsymbol{\lambda}_{i} \right)_{i \in \mathcal{B}}, \left(
\boldsymbol{\lambda}_{ij},\boldsymbol{\lambda}_{ji} \right)_{(i,j) \in \mathcal{L}} \right]$ 
be the vector of Lagrange multipliers associated with the coupling constraints \eqref{eq2:KCL1} and \eqref{eq2:KCL2} and the consensus constraints \eqref{eq3:consV} and \eqref{eq3:conso}. This duplication of the coupling variables along with the resulting component-based decomposition are depicted in Figure~\ref{fig:3Bussystem} for a 3-bus system. Now, by relaxing the coupling constraints \eqref{eq2:KCL1} and \eqref{eq2:KCL2} and consensus constraints \eqref{eq3:consV} and \eqref{eq3:conso}, the (partial) Lagrangian function of problem \eqref{eq3:opf} is written as
\begin{align}\label{eq:Lagrangefunction}
	L\left(\boldsymbol{x},\boldsymbol{\lambda} \right):=&\sum_{(g,i)\in \mathcal{G}} L_{i}^{g}\left(\boldsymbol{x}_{i}^{g},\boldsymbol{\lambda}_{i} \right) + \sum_{i\in \mathcal{B}} L_{i}^{\text{b}}\left(\boldsymbol{x}_{i}^{\text{b}},\boldsymbol{\lambda}_{i},\left( \boldsymbol{\lambda}_{ij}\right)_{j \in \mathcal{B}_{i}}  \right)  \nonumber \\
	+ & \sum_{(i,j) \in \mathcal{L}} 	L^{\text{l}}_{ij} \left(\boldsymbol{x}_{ij}^{\text{l}},\boldsymbol{\lambda}_{i},\boldsymbol{\lambda}_{j},\boldsymbol{\lambda}_{ij},\boldsymbol{\lambda}_{ji}\right),
\end{align}
where 
$L_{i}^{g}\left(\boldsymbol{x}_{i}^{g},\boldsymbol{\lambda}_{i} \right) :=\left( f^{g}_{i}\left(p_{i}^{g}\right)+ \left\langle \boldsymbol{\lambda}_{i} , \boldsymbol{x}_{i}^{g}\right\rangle \right)$,
\begin{align*}
	L_{i}^{\text{b}}\left(\boldsymbol{x}_{i}^{\text{b}},\boldsymbol{\lambda}_{i},\left( \boldsymbol{\lambda}_{ij}\right)_{j \in \mathcal{B}_{i}} \right):=\sum_{j\in \mathcal{B}_{i}} \left\langle \boldsymbol{\lambda}_{ij} , \boldsymbol{x}_{i}^{\text{b}}\right\rangle & + \\ 
	v_{i}^{2} \left\langle \boldsymbol{\lambda}_{i} , \left[-g_{i}^{\text{sh}},b_{i}^{\text{sh}} \right] \right\rangle &   - \left\langle \boldsymbol{\lambda}_{i} , \left[p_{i}^{\text{d}},q_{i}^{\text{d}}\right] \right\rangle,
\end{align*}
and	$L^{\text{l}}_{ij} \left(\boldsymbol{x}_{ij}^{\text{l}},\boldsymbol{\lambda}_{i},\boldsymbol{\lambda}_{j},\boldsymbol{\lambda}_{ij},\boldsymbol{\lambda}_{ji}\right):=-\left\langle \left[\boldsymbol{\lambda}_{i},\boldsymbol{\lambda}_{j},\boldsymbol{\lambda}_{ij},\boldsymbol{\lambda}_{ji} \right]  , \boldsymbol{x}_{ij}^{\text{l}}\right\rangle$. Accordingly, the Lagrange dual function is
\begin{align*}
	D\left(\boldsymbol{\lambda}\right):= & \underset {\substack{\boldsymbol{x}}} {\mbox{minimize}} 
	\  L\left(\boldsymbol{x},\boldsymbol{\lambda} \right)\nonumber \\
	& \ \text{subject to \cref{eq2:Pminmax,eq2:Qminmax,eq2:linethermallimit}, \cref{eq3:Vminmax,eq3:anglediff,eq3:Pij,eq3:Qij,eq3:Pji,eq3:Qji}}. 
\end{align*}
The Lagrange dual function can now be decomposed in terms of components as follows
\begin{align}\label{eq:LagrangeDual}
	D\left(\boldsymbol{\lambda}\right) 
	:=&\sum_{(g,i)\in \mathcal{G}} D_{i}^{g}\left(\boldsymbol{\lambda}_{i} \right)+\sum_{i \in \mathcal{B}} D^{\text{b}}_{i}\left(\boldsymbol{\lambda}_{i},\left(  \boldsymbol{\lambda}_{ij}\right)  _{j \in \mathcal{B}_{i}} \right) \nonumber \\
	  + & \sum_{(i,j) \in \mathcal{L}} D^{\text{l}}_{ij} \left(\boldsymbol{\lambda}_{i},\boldsymbol{\lambda}_{j} ,\boldsymbol{\lambda}_{ij},\boldsymbol{\lambda}_{ji}\right).
\end{align}
Finally, the dual problem is given by
\begin{align}\label{eq:dualproblem}
\underset{\boldsymbol{\lambda}}{\text{maximize}} \quad D\left(\boldsymbol{\lambda}\right).
\end{align}

The main reasons for solving the Lagrange dual function instead of the primal \eqref{eq3:opf} are that, first, the former is the pointwise infimum of a family of affine functions in $\boldsymbol{\lambda}$ and is therefore concave, even though the primal problem \eqref{eq3:opf} is nonconvex. Subsequently, first-order methods from convex optimization can be applied to solve the dual to optimality. Second, if the problem has zero duality gap, a feasible and optimal primal solution can be recovered from the dual solution. Third, the dual is separable in terms of components and can therefore be solved in a distributed fashion, thus preserving privacy. However, in this case, since the objective functions in \eqref{eq:LagrangeDual} are neither finite nor strictly convex,\footnote{Equivalently, the Lagrangian function in \eqref{eq:Lagrangefunction} is unbounded below in $\boldsymbol{x}$.} the dual function in \eqref{eq:Lagrangefunction} is unbounded. 

\section{Modified dual function and the distributed method}~\label{sec:modifieddual}

To make the Lagrangian function finite and strictly convex, it is modified as follows
\begin{align}\label{eq:Lagrangefunctionv}
	&L_{\nu}\left(\boldsymbol{x},\boldsymbol{\lambda}^{k} \right):=\sum_{(i,j) \in \mathcal{L}} 	L^{\text{l}}_{\nu,ij} \left(\boldsymbol{x}_{ij}^{\text{l}},\boldsymbol{\lambda}^{k}_{i},\boldsymbol{\lambda}^{k}_{j},\boldsymbol{\lambda}^{k}_{ij},\boldsymbol{\lambda}^{k}_{ji}\right)+ \nonumber \\
	&\sum_{(g,i)\in \mathcal{G}} L_{\nu,i}^{g}\left(\boldsymbol{x}_{i}^{g},\boldsymbol{\lambda}^{k}_{i} \right) + 
	\sum_{i\in \mathcal{B}} L_{\nu,i}^{\text{b}}\left(\boldsymbol{x}_{i}^{\text{b}},\boldsymbol{\lambda}^{k}_{i},\left( \boldsymbol{\lambda}^{k}_{ij}\right)_{j \in \mathcal{B}_{i}}  \right),
\end{align}
where 
\begin{align*}
	L_{\nu,i}^{g}\left(\boldsymbol{x}_{i}^{g},\boldsymbol{\lambda}^{k}_{i} \right) :=L_{i}^{g}\left(\boldsymbol{x}_{i}^{g},\boldsymbol{\lambda}^{k}_{i} \right)+ \frac{\nu}{2} \left\| \boldsymbol{x}_{i}^{g}-\boldsymbol{x}_{i}^{g,k} \right\|^2,
\end{align*}

\begin{align*}
L_{\nu,i}^{\text{b}}\left(\boldsymbol{x}_{i}^{\text{b}},\boldsymbol{\lambda}^{k}_{i},\left( \boldsymbol{\lambda}^{k}_{ij}\right)_{j \in \mathcal{B}_{i}} \right):=&L_{i}^{\text{b}}\left(\boldsymbol{x}_{i}^{\text{b}},\boldsymbol{\lambda}^{k}_{i},\left( \boldsymbol{\lambda}^{k}_{ij}\right)_{j \in \mathcal{B}_{i}} \right) \nonumber \\
+&\frac{\nu}{2} \left\| \boldsymbol{x}_{i}^{\text{b}}-\boldsymbol{x}_{i}^{\text{b},k} \right\|^2,
\end{align*}
and
\begin{align*}
	L^{\text{l}}_{\nu,ij} \left(\boldsymbol{x}_{ij}^{\text{l}},\boldsymbol{\lambda}^{k}_{i},\boldsymbol{\lambda}^{k}_{j},\boldsymbol{\lambda}^{k}_{ij},\boldsymbol{\lambda}^{k}_{ji}\right):=&L^{\text{l}}_{ij} \left(\boldsymbol{x}_{ij}^{\text{l}},\boldsymbol{\lambda}^{k}_{i},\boldsymbol{\lambda}^{k}_{j},\boldsymbol{\lambda}^{k}_{ij},\boldsymbol{\lambda}^{k}_{ji}\right) \nonumber \\
	+& \left\langle \boldsymbol{\nu} \circ \boldsymbol{x}^{\text{l}}_{ij},\boldsymbol{x}^{\text{l}}_{ij} \right\rangle.\footnotemark
\end{align*}
\footnotetext{\label{fn:n} $\boldsymbol{\nu}=\nu \boldsymbol{1}$.}
Consequently, the modified Lagrange dual function is
\begin{subequations}\label{eq:Dualv}
	\begin{align}
	D_{\nu}\left(\boldsymbol{\lambda}^{k}\right):= & \underset {\substack{\boldsymbol{x}}} {\mbox{minimize}} 
	\  L_{\nu}\left(\boldsymbol{x},\boldsymbol{\lambda}^{k} \right) \\
	& \ \text{subject to \cref{eq2:Pminmax,eq2:Qminmax,eq2:linethermallimit}, \cref{eq3:Vminmax,eq3:anglediff,eq3:Pij,eq3:Qij,eq3:Pji,eq3:Qji}}. 
	\end{align}
\end{subequations}
Particularly, in \eqref{eq:Dualv}, generators solve
\begin{subequations}\label{eq0:Gensubp}
	\begin{align}
	D_{\nu,i}^{g}\left(\boldsymbol{\lambda}_{i}^{k} \right):=\underset {\substack{\boldsymbol{x}_{i}^{g}}} 
	{\mbox{minimize  }} &  \vphantom{\left\| \boldsymbol{x}_{i}^{g}-\boldsymbol{x}_{i}^{g,k} \right\|^2} L_{\nu,i}^{g}\left(\boldsymbol{x}_{i}^{g},\boldsymbol{\lambda}^{k}_{i} \right)  \label{eq0:genobj}   \\
	\text{subject to } & \text{\cref{eq2:Pminmax} and \cref{eq2:Qminmax}},
	\end{align}
\end{subequations} 
whereas buses solve
\begin{align}\label{eq0:Bussubp}
	D^{\text{b}}_{\nu,i}\left(\boldsymbol{\lambda}_{i}^{k}, \left( \boldsymbol{\lambda}_{ij}^{k}\right) _{j \in \mathcal{B}_{i}}\right):=\underset {\substack{\boldsymbol{x}_{i}^{\text{b}}}} 
	{\mbox{inf  }}   L_{\nu,i}^{\text{b}}\left(\boldsymbol{x}_{i}^{\text{b}},\boldsymbol{\lambda}^{k}_{i},\left( \boldsymbol{\lambda}^{k}_{ij}\right)_{j \in \mathcal{B}_{i}} \right), 
\end{align}
and lines solve
\begin{subequations}\label{eq0:Linesubp}
	\begin{align}
	D^{\text{l}}_{\nu,ij} \left(\boldsymbol{\lambda}^{k}_{i},\boldsymbol{\lambda}^{k}_{j},\boldsymbol{\lambda}^{k}_{ij},\boldsymbol{\lambda}^{k}_{ji}\right):= \hphantom{what} & \nonumber \\
	\underset {\substack{\boldsymbol{x}_{ij}^{\text{l}}}} {\mbox{minimize  }} \hspace{2cm}  & \hspace{-2cm} L^{\text{l}}_{\nu,ij} \left(\boldsymbol{x}_{ij}^{\text{l}},\boldsymbol{\lambda}^{k}_{i},\boldsymbol{\lambda}_{j},\boldsymbol{\lambda}^{k}_{ij},\boldsymbol{\lambda}^{k}_{ji}\right)  \label{eq0:lineobj} \\
	\text{subject to} \quad \underline{v}_{i}  \leq v_{i_{(ij)}} \leq  \overline{v}_{i}, \ \underline{v}_{j} & \leq v_{j_{(ji)}} \leq  \overline{v}_{j}, \label{eq0:Vminmax} \\
	\underline{\theta}_{ij}^{\Delta}  \leq \theta_{i_{(ij)}}-\theta_{j_{(ji)}} & \leq \overline{\theta}_{ij}^{\Delta}, \label{eq0:anglediff} \\
	p_{ij}= g^{c}_{ij} v_{i_{(ij)}}^2-g_{ij}v_{i_{(ij)}}v_{j_{(ji)}}\cos&\left(\theta_{i_{(ij)}}-\theta_{j_{(ji)}}\right)  \nonumber\\
	+b_{ij}  v_{i_{(ij)}}v_{j_{(ji)}}\sin&\left(\theta_{i_{(ij)}}-\theta_{j_{(ji)}}\right), \label{eq0:Pij} \\
	q_{ij}= b^{c}_{ij} v_{i_{(ij)}}^2-b_{ij} v_{i_{(ij)}}v_{j_{(ji)}}\cos&\left(\theta_{i_{(ij)}}-\theta_{j_{(ji)}}\right)  \nonumber\\
	-g_{ij}  v_{i_{(ij)}}v_{j_{(ji)}}\sin&\left(\theta_{i_{(ij)}}-\theta_{j_{(ji)}}\right), \label{eq0:Qij} \\
	p_{ji}= g^{c}_{ji} v_{j_{(ji)}}^2-g_{ji} v_{j_{(ji)}}v_{i_{(ij)}}\cos&\left(\theta_{j_{(ji)}}-\theta_{i_{(ij)}}\right)  \nonumber\\
	+b_{ji}  v_{j_{(ji)}}v_{i_{(ij)}}\sin&\left(\theta_{j_{(ji)}}-\theta_{i_{(ij)}}\right),  \label{eq0:Pji}  \\
	q_{ji}= b^{c}_{ji} v_{j_{(ji)}}^2-b_{ji} v_{j_{(ji)}}v_{i_{(ij)}}\cos&\left(\theta_{j_{(ji)}}-\theta_{i_{(ij)}}\right)  \nonumber\\
	-g_{ji}  v_{j_{(ji)}}v_{i_{(ij)}}\sin&\left(\theta_{j_{(ji)}}-\theta_{i_{(ij)}}\right), \label{eq0:Qji} \\
	\sqrt{p_{ij}^2+q_{ij}^2} \leq \overline{s}_{ij}, \ \sqrt{p_{ji}^2+q_{ji}^2} & \leq  \overline{s}_{ij}. \label{eq0:smax}  
	\end{align}
\end{subequations}
%\footnotetext{Problem \eqref{eq0:Linesubp} can be further decomposed into 2 smaller problems, $D^{\text{l}}_{\nu,ij} \left(\boldsymbol{\lambda}^{k}_{i},\boldsymbol{\lambda}^{k}_{ij}\right)$ and $D^{\text{l}}_{\nu,ji} \left(\boldsymbol{\lambda}^{k}_{j},\boldsymbol{\lambda}^{k}_{ji}\right)$, since variables $\boldsymbol{x}_{ij}^{\text{l}}:=[p_{ij},q_{ij},v_{i_{(ij)}},\theta_{i_{(ij)}}]$ and $\boldsymbol{x}_{ji}^{\text{l}}:=[p_{ji},q_{ji},v_{j_{(ji)}},\theta_{j_{(ji)}}]$ are decoupled.}
The concave modified dual function in \eqref{eq:Dualv} can now be solved using the subgradient projection method in which, at each iteration $k$, every bus $i \in \mathcal{B}$ updates its (local) Lagrange multipliers as follows
\begin{subequations}\label{eq1:lambdaupdate}
	\begin{align}
	\boldsymbol{\lambda}_{i}^{k+1}&=\boldsymbol{\lambda}_{i}^{k}+\alpha_{i} \boldsymbol{g}_{\nu,i}^{k},  \\
	\boldsymbol{\lambda}_{ij}^{k+1}&=\boldsymbol{\lambda}_{ij}^{k}+\alpha_{ij}\boldsymbol{g}_{\nu,ij}^{k}, \ j \in \mathcal{B}_{i},
	\end{align}
\end{subequations}
where 
\begin{equation*}
\boldsymbol{g}_{\nu,i}^{k}:=	
\begin{bmatrix}
\underset{(g,i)\in \mathcal{G}}{\sum}  p_{i}^{g,k+1} -p_{i}^{\text{d}} - \underset{j\in \mathcal{B}_{i}}{\sum}  p^{k+1}_{ij} - g^{\text{sh}}_{i}\left( v^{k+1}_{i}\right) ^{2} \\
\underset{(g,i)\in \mathcal{G}}{\sum} q_{i}^{g,k+1} -q_{i}^{\text{d}} -  \underset{j\in \mathcal{B}_{i}}{\sum} q^{k+1}_{ij} + b^{\text{sh}}_{i}\left( v^{k+1}_{i}\right) ^{2}
\end{bmatrix}
\end{equation*}
and 
\begin{equation*}
\boldsymbol{g}_{\nu,ij}^{k}:=	
\begin{bmatrix}
v^{k+1}_{i}-v^{k+1}_{i_{(ij)}}\\
\theta^{k+1}_{i}-\theta^{k+1}_{i_{(ij)}}
\end{bmatrix}.
\end{equation*}
The effect of adding the proximal regularization term (with $\nu>0$) is twofold. First, it makes the local cost functions finite and strictly convex and therefore the modified dual function bounded. Second, it makes the modified dual function differentiable for large values of $\nu$. For small values of $\nu$, the concave modified dual function $D_{\nu}(\boldsymbol{\lambda}^{k})$ is typically nondifferentiable. Indeed, using \emph{Danskin}'s theorem (See Appendices~\ref{Appendix1} and~\ref{Appendix2}), the subdifferentials of $D_{\nu}(\boldsymbol{\lambda}^{k})$ are $\partial D_{\nu}(\boldsymbol{\lambda}^{k}):=	\left\{A_{c}\boldsymbol{x}:D_{\nu}(\boldsymbol{\lambda}^{k}) , \boldsymbol{x} \in \mathcal{X} \right\}$, where $\mathcal{X}$ is the feasible set defined by constraints \cref{eq2:Pminmax,eq2:Qminmax}, \cref{eq2:linethermallimit}, \cref{eq3:Pij,eq3:Qij,eq3:Pji,eq3:Qji} and $A_{c}$ is the coupling constraint matrix associated with coupling constraints \eqref{eq2:KCL1} and \eqref{eq2:KCL2} and consensus constraints \eqref{eq3:consV} and \eqref{eq3:conso}. More specifically, as the (nonconvex) transmission line subproblems in \eqref{eq0:Linesubp} can have multiple (globally) optimal solutions for a given vector $\boldsymbol{\lambda}^{k}$, the subdifferentials $\partial D_{\nu}(\boldsymbol{\lambda}^{k})$ may be not be unique and the modified dual function $D_{\nu}(\boldsymbol{\lambda}^{k})$ can be nonsmooth. 
In more detail,
\begin{equation*}
	\boldsymbol{g}_{\nu}^{k}:= \\ 	
	\begin{bmatrix}
	\begin{pmatrix}
	\boldsymbol{g}_{\nu,i}^{k}
	\end{pmatrix}_{i \in \mathcal{B}}\\
	\begin{pmatrix}
	\boldsymbol{g}_{\nu,ij}^{k}
	\end{pmatrix}_{(i,j) \in \mathcal{L} \cup \mathcal{L}_{t}}
	\end{bmatrix}\in \partial D_{\nu}\left(\boldsymbol{\lambda}^{k}\right),
\end{equation*}
which is a subgradient of $D_{\nu}(\boldsymbol{\lambda}^{k})$, may not be unique when $\nu$ is small. On the other hand, for large values of $\nu$, $\boldsymbol{g}_{\nu}^{k}$ is unique and is therefore a gradient of $D_{\nu}(\boldsymbol{\lambda}^{k})$, \text{i.e.}, $\boldsymbol{g}_{\nu}^{k}= \nabla D_{\nu}(\boldsymbol{\lambda}^{k})$.
The component-based modified dual decomposition algorithm is described in Algorithm~\ref{algorithm1}.
\begin{algorithm}[t]
	\small
	\caption{Distributed algorithm}
	\begin{algorithmic}[1]
		%		\scriptsize
		\STATE \textbf{Initialization:} $\boldsymbol{\lambda}^{1} = \boldsymbol{0}$, $\nu >> 0$, $\epsilon \leq 10^{-4}$, $\boldsymbol{x}_{i}^{\text{b},1}=\left[1,0\right]$ for all $i \in \mathcal{B}$, $\boldsymbol{x}_{i}^{g,1}=\left[\frac{\underline{p}_{i}^{g}+\overline{p}_{i}^{g}}{2}\right]$  for all $(g,i) \in \mathcal{G}$, and $\boldsymbol{x}_{i}^{\text{l},1}=\left[0,0,0,0,1,0,1,0 \right] $ for all $(i,j) \in \mathcal{L}$.
		\WHILE {$\left\| \boldsymbol{g}^{k}_{\nu}\right\| \geq \epsilon$} 
		\STATE 	\parbox[t]{\dimexpr\linewidth-0.8cm}{Generators, buses and lines solve \eqref{eq0:Gensubp}, \eqref{eq0:Bussubp} and \eqref{eq0:Linesubp} respectively in parallel, and send $\boldsymbol{x}_{i}^{g,k+1}$, $\boldsymbol{x}_{i}^{\text{b},k+1}$ and $\boldsymbol{x}_{ij}^{\text{l},k+1}$ to adjacent buses.\strut}
		\STATE 	\parbox[t]{\dimexpr\linewidth-0.8cm}{Each bus $i \in \mathcal{B}$ updates its (local) Lagrange multipliers as in \eqref{eq1:lambdaupdate} and sends $\boldsymbol{\lambda}_{i}^{k+1}$ and $\boldsymbol{\lambda}_{ij}^{k+1}$ to corresponding adjacent lines and generators.\strut}
		\STATE $k \leftarrow k + 1 $.
		\ENDWHILE
	\end{algorithmic} 
	\label{algorithm1}
\end{algorithm}
\begin{definition} \label{def:AMDgap}
	Let $P^{\dagger}_{\text{IPM}}$ be a feasible primal solution computed centrally by an IPM solver and let $ D_{\text{AMD}}(\boldsymbol{\lambda}^{\dagger})$ be a solution of the approximate modified dual function computed in a distributed fashion by Algorithms~\ref{algorithm1},~\ref{algorithm2} or~\ref{algorithm3}, initialized with the same algorithmic starting point used to find $P^{\dagger}_{\text{IPM}}$. Then the gap between the feasible primal solution $ P^{\dagger}_{\text{IPM}}$ and its associated approximate modified dual function optimal value $ D_{\text{AMD}}(\boldsymbol{\lambda}^{\dagger})$ is given by
	\begin{align}\label{eq:AMDgap}
		\text{AMDgap}:=\left( \left( P^{\dagger}_{\text{IPM}}- D_{\text{AMD}}\left(\boldsymbol{\lambda}^{\dagger}\right)\right) /P^{\dagger}_{\text{IPM}}\right) \times 100.
	\end{align}
%	\begin{align*}
%		\text{AMDgap}:=\left( \frac{P^{\dagger}_{\text{IPM}}- D_{\text{AMD}}\left(\boldsymbol{\lambda}^{\dagger}\right)}{P^{\dagger}_{\text{IPM}}}\right) \times 100.
%	\end{align*}
\end{definition}
Note that in Definition~\ref{def:AMDgap}, if $P^{\dagger}_{\text{IPM}}$ is globally optimal and $\text{AMDgap}=0$, then $ D_{\text{AMD}}(\boldsymbol{\lambda}^{\dagger})$ is an accurate approximation of the modified dual function. Also, note that unlike the classical dual function, the modified dual function $D_{\text{AMD}}(\boldsymbol{\lambda}^{\dagger})$ is \emph{not} a lower bound on the optimal solution $P^{\star}$ and thus the definition in \eqref{eq:AMDgap} instead of the classical definition 
\begin{align*}
\text{Duality gap}:=\left( \left( P^{\star}- D\left(\boldsymbol{\lambda}^{\star}\right)\right) /P^{\star}\right) \times 100.
\end{align*}

\begin{figure}[t]
	\centering{
		\psfrag{R}{\footnotesize $\left\| \boldsymbol{g}^{k}_{\nu}\right\| $ \normalsize}
		\psfrag{k}{\footnotesize Iterations ($k$) \normalsize}
		\includegraphics[width=70mm] {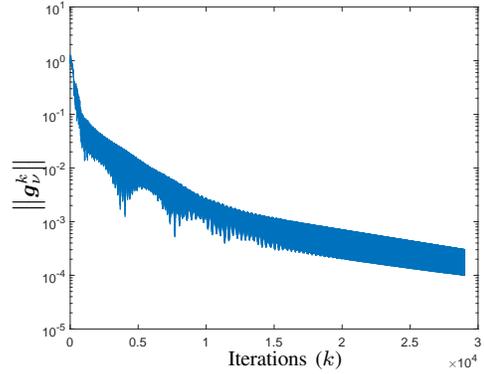}}
	\caption{Evolution of $\left\| \boldsymbol{g}^{k}_{\nu}\right\|$ when Algorithm~\ref{algorithm1} is applied to MATPOWER's case 14 with $\nu=100000$, $\alpha_{i}=100$ and $\alpha_{ij}=10000$.}
	\label{fig:case14residuals_proximal}
\end{figure}

\begin{definition} \label{def:ROgap}
	Let $P^{\dagger}_{\text{AMD}}=f\left( p_{i}^{g,\dagger}\right)$ be a feasible primal solution computed in a distributed fashion by Algorithms~\ref{algorithm1},~\ref{algorithm2} or~\ref{algorithm3}, initialized with the same algorithmic starting point used to find $P^{\dagger}_{\text{IPM}}$. Then the gap between $ P^{\dagger}_{\text{IPM}}$ and $P^{\dagger}_{\text{AMD}}$ is given by
	\begin{align*}
		\text{ROgap}:=\left( \left( P^{\dagger}_{\text{IPM}}- P^{\dagger}_{\text{AMD}}\right) /P^{\dagger}_{\text{IPM}}\right) \times 100.
	\end{align*}
%	\begin{align*}
%		\text{ROgap}:=\left( \frac{P^{\dagger}_{\text{IPM}}- P^{\dagger}_{\text{AMD}}}{P^{\dagger}_{\text{IPM}}}\right) \times 100.
%	\end{align*}
\end{definition}
However, Algorithm~\ref{algorithm1} exhibits a very slow convergence due to an oscillatory behaviour witnessed across all the considered test cases. These oscillations are illustrated in Figure~\ref{fig:case14residuals_proximal}, which shows the evolution of $\left\| \boldsymbol{g}^{k}_{\nu}\right\|$ when Algorithm~\ref{algorithm1} is applied to MATPOWER's case 14 with $\nu=100000$, $\alpha_{i}=100$ and $\alpha_{ij}=10000$. In this case, Algorithm~\ref{algorithm1} converges to a solution with an $\text{ROgap}=0.0006 \%$ and an $\text{AMDgap}=-8 \times 10 ^{-5} \%$ in $29017$ iterations.

The oscillations can be mitigated by modifying the Lagrangian function as follows
\begin{align*}
	&L_{\nu,\rho}\left(\boldsymbol{x},\boldsymbol{\lambda}^{k} \right):=\sum_{(i,j) \in \mathcal{L}} 	L^{\text{l}}_{\nu,\rho,ij} \left(\boldsymbol{x}_{ij}^{\text{l}},\boldsymbol{\lambda}^{k}_{i},\boldsymbol{\lambda}^{k}_{j},\boldsymbol{\lambda}^{k}_{ij},\boldsymbol{\lambda}^{k}_{ji}\right)+ \nonumber \\
	&\sum_{(g,i)\in \mathcal{G}} L_{\nu,i}^{g}\left(\boldsymbol{x}_{i}^{g},\boldsymbol{\lambda}^{k}_{i} \right) + 
	\sum_{i\in \mathcal{B}} L_{\rho,i}^{\text{b}}\left(\boldsymbol{x}_{i}^{\text{b}},\boldsymbol{\lambda}^{k}_{i},\left( \boldsymbol{\lambda}^{k}_{ij}\right)_{j \in \mathcal{B}_{i}}  \right),
\end{align*}
where 
\begin{align*}
&L_{\rho,i}^{\text{b}}\left(\boldsymbol{x}_{i}^{\text{b}},\boldsymbol{\lambda}^{k}_{i},\left( \boldsymbol{\lambda}^{k}_{ij}\right)_{j \in \mathcal{B}_{i}} \right):=L_{i}^{\text{b}}\left(\boldsymbol{x}_{i}^{\text{b}},\boldsymbol{\lambda}^{k}_{i},\left( \boldsymbol{\lambda}^{k}_{ij}\right)_{j \in \mathcal{B}_{i}} \right) \nonumber \\
+&\frac{\rho_{v \theta}}{2}\left( \sum_{j\in \mathcal{B}_{i}} \left(  \left( v_{i} - v^{k+1}_{i_{(ij)}} \right)^{2} +  \left( \theta_{i} - \theta^{k+1}_{i_{(ij)}} \right)^{2} \right) \right),
\end{align*}
and
\begin{align*}
&L^{\text{l}}_{\nu,\rho,ij} \left(\boldsymbol{x}_{ij}^{\text{l}},\boldsymbol{\lambda}^{k}_{i},\boldsymbol{\lambda}^{k}_{j},\boldsymbol{\lambda}^{k}_{ij},\boldsymbol{\lambda}^{k}_{ji}\right):=L^{\text{l}}_{\nu,ij} \left(\boldsymbol{x}_{ij}^{\text{l}},\boldsymbol{\lambda}^{k}_{i},\boldsymbol{\lambda}^{k}_{j},\boldsymbol{\lambda}^{k}_{ij},\boldsymbol{\lambda}^{k}_{ji}\right) \nonumber \\
&+\frac{\rho_{v \theta}}{2} \left(  \sum_{(l,m) \in \left\lbrace (i,j) \cup (j,i) \right\rbrace }  \left( v^{k}_{l} - v_{l_{(lm)}} \right)^{2} + \left( \theta^{k}_{l} - \theta_{l_{(lm)}} \right)^{2} \right).\footnotemark
\end{align*}
\footnotetext{$\boldsymbol{\nu}=\left[\nu,\nu,\nu,\nu,0,0,0,0 \right]$.}
Consequently, generators now solve \eqref{eq0:Gensubp}, transmission lines solve
\begin{subequations}\label{eq1:Linesubp}
	\begin{align}
	&D^{\text{l}}_{\nu,\rho,ij} \left(\boldsymbol{\lambda}^{k}_{i},\boldsymbol{\lambda}^{k}_{j},\boldsymbol{\lambda}^{k}_{ij},\boldsymbol{\lambda}^{k}_{ji}\right):= \nonumber \\
	&\underset {\substack{\boldsymbol{x}_{ij}^{\text{l}}}} {\mbox{minimize  }}  L^{\text{l}}_{\nu,\rho,ij} \left(\boldsymbol{x}_{ij}^{\text{l}},\boldsymbol{\lambda}^{k}_{i},\boldsymbol{\lambda}^{k}_{j},\boldsymbol{\lambda}^{k}_{ij},\boldsymbol{\lambda}^{k}_{ji}\right)   \label{eq1:Lineobj} \\
	& \hspace{2cm}\text{subject to } \text{\cref{eq0:Vminmax,eq0:anglediff,eq0:Pij,eq0:Qij,eq0:Pji,eq0:Qji,eq0:smax}},
	\end{align}
\end{subequations}
and buses solve
\begin{align}\label{eq1:Bussubp}
	&D^{\text{b}}_{\rho,i}\left(\boldsymbol{\lambda}_{i}^{k}, \left( \boldsymbol{\lambda}_{ij}^{k}\right) _{j \in \mathcal{B}_{i}}\right):=\underset {\substack{\boldsymbol{x}_{i}^{\text{b}}}} 
	{\mbox{inf  }}  L_{\rho,i}^{\text{b}}\left(\boldsymbol{x}_{i}^{\text{b}},\boldsymbol{\lambda}^{k}_{i},\left( \boldsymbol{\lambda}^{k}_{ij}\right)_{j \in \mathcal{B}_{i}} \right) .
\end{align}
The component-based modified dual decomposition algorithm with the ADMM penalty term is described in Algorithm~\ref{algorithm2}.
\begin{algorithm}[t]
	\small
	\caption{Distributed algorithm}
	\begin{algorithmic}[1]
		%		\scriptsize
		\STATE \textbf{Initialization:} $\boldsymbol{\lambda}^{1} = \boldsymbol{0}$, $\nu >> 0$, $\rho >> 0$, $\epsilon \leq 10^{-4}$, $\boldsymbol{x}_{i}^{\text{b},1}=\left[1,0\right]$ for all $i \in \mathcal{B}$, $\boldsymbol{x}_{i}^{g,1}=\left[\frac{\underline{p}_{i}^{g}+\overline{p}_{i}^{g}}{2}\right]$  for all $(g,i) \in \mathcal{G}$, and $\boldsymbol{x}_{i}^{\text{l},1}=\left[0,0,0,0,1,0,1,0 \right] $ for all $(i,j) \in \mathcal{L}$.
		\WHILE {$\left\| \boldsymbol{g}^{k}_{\nu,\rho}\right\| \geq \epsilon$} 
		\STATE 	\parbox[t]{\dimexpr\linewidth-0.8cm}{Generators and lines solve \eqref{eq0:Gensubp} and \eqref{eq1:Linesubp} respectively in parallel, and send $\boldsymbol{x}_{i}^{g,k+1}$ and $\boldsymbol{x}_{ij}^{\text{l},k+1}$ to adjacent buses.\strut}
		\STATE 	\parbox[t]{\dimexpr\linewidth-0.8cm}{Buses solve \eqref{eq1:Bussubp} in parallel and update their (local) Lagrange multipliers as in \eqref{eq1:lambdaupdate}.\strut}
		\STATE 	\parbox[t]{\dimexpr\linewidth-0.8cm}{Buses send $\boldsymbol{x}_{i}^{\text{b},k+1}$, $\boldsymbol{\lambda}_{i}^{k+1}$ and $\boldsymbol{\lambda}_{ij}^{k+1}$ to corresponding adjacent lines and generators.\strut}
		\STATE $k \leftarrow k + 1 $.
		\ENDWHILE
	\end{algorithmic} 
	\label{algorithm2}
\end{algorithm}

\begin{figure}[t]
	\centering{
		\psfrag{R}{\footnotesize $\left\| \boldsymbol{g}^{k}_{\nu,\rho}\right\| $ \normalsize}
		\psfrag{k}{\footnotesize Iterations ($k$) \normalsize}
		\includegraphics[width=70mm] {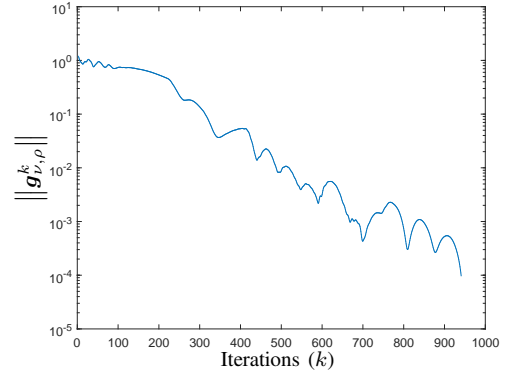}}
	\caption{Evolution of $\left\| \boldsymbol{g}^{k}_{\nu,\rho}\right\|$ when Algorithm~\ref{algorithm2} is applied to MATPOWER's case 14 with $\nu=1000$, $\rho_{v \theta}=100000$, $\alpha_{i}=100$ and $\alpha_{ij}=100000$.}
	\label{fig:case14residuals_ADMMp}
\end{figure}

The key behind the superior convergence of Algorithm~\ref{algorithm2} is the ADMM penalty term which controls the stability of the iterates. This is illustrated in Figure~\ref{fig:case14residuals_ADMMp}, which shows the evolution of $\left\| \boldsymbol{g}^{k}_{\nu,\rho}\right\|$ when Algorithm~\ref{algorithm2} is applied to MATPOWER's case 14 with $\nu=1000$, $\rho=100000$, $\alpha_{i}=100$ and $\alpha_{ij}=100000$. In this case, Algorithm~\ref{algorithm2} converges to a solution with an $\text{ROgap}=0.002 \%$ and an $\text{AMDgap}=-6.5\times 10 ^{-5} \%$ in $923$ iterations, as compared to $29017$ iterations when applying Algorithm~\ref{algorithm1} (see Figure~\ref{fig:case14residuals_proximal}).
Note that in Algorithm~\ref{algorithm2} there are two rounds of (parallel) computations at each iteration.

The oscillations in Algorithm~\ref{algorithm2} can be reduced even further by modifying \eqref{eq0:Gensubp} and \eqref{eq1:Linesubp} as follows
\begin{subequations}\label{eq3:Gensubp}
	\begin{align}
	D_{\nu,\rho,i}^{g}&\left(\boldsymbol{\lambda}_{i}^{k} \right):=\underset {\substack{\boldsymbol{x}_{i}^{g}}} 
	{\mbox{minimize  }}  \left\lbrace  \vphantom{\left\| \boldsymbol{x}_{i}^{g}-\boldsymbol{x}_{i}^{g,k} \right\|^2} L_{\nu,i}^{g}\left(\boldsymbol{x}_{i}^{g},\boldsymbol{\lambda}^{k}_{i} \right) \right. \nonumber \\
	& \left. +\rho_{pq}\left( \left(p_{i}^{g}-pc_{i}^{g,k} \right)^2 +\left(q_{i}^{g}-qc_{i}^{g,k} \right)^2 \right)  \right\rbrace  \label{eq3:genobj}   \\
	& \hspace{1cm} \text{subject to }  \text{\cref{eq2:Pminmax} and \cref{eq2:Qminmax}},
	\end{align}
\end{subequations}  
and
\begin{subequations}\label{eq3:Linesubp}
	\begin{align}
	&D^{\text{l}}_{\nu,\boldsymbol{\rho},ij} \left(\boldsymbol{\lambda}^{k}_{i},\boldsymbol{\lambda}^{k}_{j},\boldsymbol{\lambda}^{k}_{ij},\boldsymbol{\lambda}^{k}_{ji}\right):= \nonumber \\
	&\underset {\substack{\boldsymbol{x}_{ij}^{\text{l}}}} {\mbox{minimize  }} \left\lbrace \vphantom{\sum_{(l,m) \in \left\lbrace (i,j) \cup (j,i) \right\rbrace }}   L^{\text{l}}_{\nu,\rho,ij} \left(\boldsymbol{x}_{ij}^{\text{l}},\boldsymbol{\lambda}^{k}_{i},\boldsymbol{\lambda}^{k}_{j},\boldsymbol{\lambda}^{k}_{ij},\boldsymbol{\lambda}^{k}_{ji}\right) +\right. \nonumber \\
	&  \left. \frac{\rho_{pq}}{2} \left(  \sum_{(l,m) \in \left\lbrace (i,j) \cup (j,i) \right\rbrace }  \left(pc_{lm}^{k}-p_{lm}  \right)^{2} + \left(qc_{lm}^{k}-q_{lm} \right)^{2} \right) \right\rbrace \label{eq3:Lineobj} \\
	& \hspace{2cm}\text{subject to } \text{\cref{eq0:Vminmax,eq0:anglediff,eq0:Pij,eq0:Qij,eq0:Pji,eq0:Qji,eq0:smax}},
	\end{align}
\end{subequations}  
where $pc_{i}^{g,k}=-\sum_{h\in \mathcal{G}_{i} \setminus g} p_{i}^{h,k}+\sum_{j\in \mathcal{B}_{i}} p_{ij}^{k} + g^{\text{sh}}_{i}\left( v_{i}^{k}\right) ^{2}+p_{i}^{\text{d}} $, $qc_{i}^{g,k}=-\sum_{h\in \mathcal{G}_{i} \setminus g} q_{i}^{h,k}+\sum_{j\in \mathcal{B}_{i}} q_{ij}^{k} - b^{\text{sh}}_{i}\left( v_{i}^{k}\right) ^{2}+q_{i}^{\text{d}} $, $pc_{ij}^{k}=-\sum_{m\in \mathcal{B}_{i} \setminus j} p_{im}^{k}+\sum_{(g,i)\in \mathcal{G}} p_{i}^{g,k}-p_{i}^{\text{d}}-g^{\text{sh}}_{i}\left( v_{i}^{k}\right) ^{2}$, $qc_{ij}^{k}=-\sum_{m\in \mathcal{B}_{i} \setminus j} q_{im}^{k}+\sum_{(g,i)\in \mathcal{G}} q_{i}^{g,k}-q_{i}^{\text{d}}+b^{\text{sh}}_{i}\left( v_{i}^{k}\right) ^{2}$, $pc_{ji}^{k}=-\sum_{l\in \mathcal{B}_{j} \setminus i} p_{jl}^{k}+\sum_{(g,j)\in \mathcal{G}} p_{j}^{g,k}-p_{j}^{\text{d}}-g^{\text{sh}}_{j}\left( v_{j}^{k}\right) ^{2}$ and $qc_{ji}^{k}=-\sum_{l\in \mathcal{B}_{j} \setminus i} q_{lj}^{k}+\sum_{(g,j)\in \mathcal{G}} q_{j}^{g,k}-q_{j}^{\text{d}}+b^{\text{sh}}_{j}\left( v_{j}^{k}\right) ^{2}$.
Finally, the modified Lagrange dual function would be
\begin{align}\label{eq3:LagrangeDual}
	&D_{\nu,\boldsymbol{\rho}}\left(\boldsymbol{\lambda}^{k}\right) 
	:=\sum_{(i,j) \in \mathcal{L}} D^{\text{l}}_{\nu,\boldsymbol{\rho},ij} \left(\boldsymbol{\lambda}^{k}_{i},\boldsymbol{\lambda}^{k}_{j},\boldsymbol{\lambda}^{k}_{ij},\boldsymbol{\lambda}^{k}_{ji}\right) \nonumber \\
	& + \sum_{(g,i)\in \mathcal{G}} D_{\nu,\rho,i}^{g}\left(\boldsymbol{\lambda}_{i}^{k} \right) + \sum_{i \in \mathcal{B}} D^{\text{b}}_{\nu,\rho,i}\left(\boldsymbol{\lambda}_{i},\left(  \boldsymbol{\lambda}_{ij}\right)  _{j \in \mathcal{B}_{i}} \right),
\end{align}
and the associated algorithm is described in Algorithm~\ref{algorithm3}.
\begin{algorithm}[t]
	\small
	\caption{Distributed algorithm}
	\begin{algorithmic}[1]
		%		\scriptsize
		\STATE \textbf{Initialization:} Same as in Algorithm~\ref{algorithm2}.
		\WHILE {$\left\| \boldsymbol{g}^{k}_{\nu,\boldsymbol{\rho}}\right\| \geq \epsilon$} 
		\STATE 	\parbox[t]{\dimexpr\linewidth-0.8cm}{Generators and lines solve \eqref{eq3:Gensubp} and \eqref{eq3:Linesubp} respectively in parallel, and send $\boldsymbol{x}_{i}^{g,k+1}$ and $\boldsymbol{x}_{ij}^{\text{l},k+1}$ to adjacent buses.\strut}
		\STATE 	\parbox[t]{\dimexpr\linewidth-0.8cm}{Buses solve \eqref{eq1:Bussubp} in parallel and update their (local) Lagrange multipliers as in \eqref{eq1:lambdaupdate}.\strut}
		\STATE 	\parbox[t]{\dimexpr\linewidth-0.8cm}{Buses send $\boldsymbol{x}_{i}^{\text{b},k+1}$, $\boldsymbol{\lambda}_{i}^{k+1}$, $\boldsymbol{\lambda}_{ij}^{k+1}$, $\boldsymbol{xc}^{k}_{ij}$ and $\boldsymbol{xc}^{g,k}_{i}$ to corresponding adjacent lines and generators.\footnotemark\strut}
		\STATE $k \leftarrow k + 1 $.
		\ENDWHILE
	\end{algorithmic} 
	\label{algorithm3}
\end{algorithm}
\footnotetext{$\boldsymbol{xc}^{g,k}_{i}=\left[ pc_{i}^{g,k},qc_{i}^{g,k}\right] $ and $\boldsymbol{xc}^{k}_{ij}=\left[ pc_{ij}^{k},qc_{ij}^{k}\right] $.}
\begin{figure}[t]
	\centering{
		\psfrag{R}{\footnotesize $\left\| \boldsymbol{g}^{k}_{\nu,\boldsymbol{\rho}}\right\| $ \normalsize}
		\psfrag{k}{\footnotesize Iterations ($k$) \normalsize}
		\includegraphics[width=70mm] {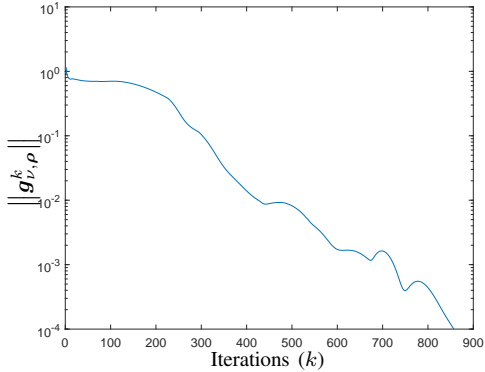}}
	\caption{Evolution of $\left\| \boldsymbol{g}^{k}_{\nu,\boldsymbol{\rho}}\right\|$ when Algorithm~\ref{algorithm3} is applied to MATPOWER's case 14 with $\nu=1000$, $\rho_{v \theta}=100000$, $\rho_{pq}=1000$, $\alpha_{i}=100$ and $\alpha_{ij}=100000$.}
	\label{fig:case14residuals_ADMMp2}
\end{figure}
The convergence of Algorithm~\ref{algorithm3} on MATPOWER's case 14 is illustrated in Figure~\ref{fig:case14residuals_ADMMp2}, which shows the evolution of $\left\| \boldsymbol{g}^{k}_{\nu,\rho}\right\|$ with $\nu=1000$, $\rho_{v \theta}=100000$, $\rho_{pq}=1000$, $\alpha_{i}=100$ and $\alpha_{ij}=100000$. In this case, Algorithm~\ref{algorithm3} converges to a solution with an $\text{ROgap}=0.0008 \%$ and an $\text{AMDgap}=-7\times 10 ^{-5} \%$ in $857$ iterations as compared to $923$ iterations when applying Algorithm~\ref{algorithm2} (see Figure~\ref{fig:case14residuals_ADMMp}).

The evolution from Algorithm~\ref{algorithm1} to Algorithm~\ref{algorithm3} results in a faster convergence but comes at the expense of more message exchanges. In fact, on many test instances, Algorithm~\ref{algorithm2} can strike a good trade-off between communication and convergence speed and applying Algorithm~\ref{algorithm3} only marginally improves the convergence speed, which does not warrant the extra communication requirements associated with it. Due to a lack of space, the next section only numerically evaluates Algorithm~\ref{algorithm3} and shows that, with the right parameter settings, it converges on all the 72 considered test cases.\footnote{The method can easily be extended to incorporate discrete variables such as transformer taps and shunt capacitor banks.} 

\section{Numerical evaluation}\label{sec:numericalevaluation}

Algorithm~\ref{algorithm3} is implemented in MATLAB and the interfacing between AMPL and MATLAB is made possible by AMPL's application programming interface. The simulations are all conducted on a computing platform with 10 Intel Xeon E5-2687W v3 CPUs at 3.10GHz, 64-bit operating system, and 128GB RAM. In all simulations, AMPL \cite{AMPL} is used as a frontend modeling language for the optimization problems along with KNITRO 10.2 \cite{KNITRO} as a backend solver for the nonconvex original OPF problem in \eqref{eq2:opf} and the nonconvex line and bus subproblems in \eqref{eq1:Linesubp} and \eqref{eq1:Bussubp} respectively. Generator subproblems are convex and admit closed-form solutions (see \cite{Peng_DOPFbalancedradial}).
\begin{table}[!t]
	%\normalsize
	\centering
	\renewcommand{\arraystretch}{1.3}
	\caption{Convergence of Algorithm~\ref{algorithm3} on MATPOWER instances.}
	\resizebox{\linewidth}{!}{%
		\begin{tabular}{| c | c | c | c | c | c | c | c |}
			\cline{2-6}
			\multicolumn{1}{c|}{} & \multicolumn{3}{c|}{Objective (\$)} & \multicolumn{2}{c|}{Gap (\%)} &  \multicolumn{1}{c}{} & \multicolumn{1}{c}{} \\\cline{1-8}
			Case   & $P^{\dagger}_{\text{IPM}}$  & $P^{\dagger}_{\text{AMD}}$  &  $D_{\nu,\boldsymbol{\rho}}\left(\boldsymbol{\lambda}^{\dagger}\right)$  & AMD  & RO   & $I$ & set \\\hline
			5	&	17551.89	&	17551.16	&	17552.02	&	-7.35E-04	&	4.14E-03	&	3911	&	A	\\\hline
			6ww	&	3143.97	&	3143.82	&	3143.97	&	3.36E-06	&	4.94E-03	&	918	&	B	\\\hline
			9	&	5296.69	&	5296.30	&	5296.69	&	-8.52E-06	&	7.25E-03	&	630	&	B	\\\hline
			14	&	8081.52	&	8082.16	&	8081.53	&	-7.12E-05	&	-7.88E-03	&	857	&	C	\\\hline
			24	&	63352.20	&	63352.33	&	63352.21	&	-6.47E-06	&	-2.05E-04	&	924	&	B	\\\hline
			30	&	576.89	&	576.97	&	576.89	&	-1.10E-04	&	-1.34E-02	&	2763	&	B	\\\hline
			30	&	8906.14	&	8906.79	&	8906.14	&	-1.11E-06	&	-7.25E-03	&	1017	&	B	\\\hline
			39	&	41864.18	&	41864.22	&	41864.23	&	-1.16E-04	&	-9.82E-05	&	7468	&	B	\\\hline
			57	&	41737.79	&	41736.02	&	41737.79	&	-1.51E-06	&	4.24E-03	&	1305	&	B	\\\hline
			89	&	5819.81	&	5819.90	&	5819.91	&	-1.84E-03	&	-1.59E-03	&	10927	&	D	\\\hline
			118	&	129660.69	&	129660.22	&	129660.75	&	-4.27E-05	&	3.66E-04	&	1168	&	B	\\\hline
			300	&	719725.10	&	719724.47	&	719725.38	&	-3.88E-05	&	8.72E-05	&	11755	&	E	\\\hline	
	\end{tabular}}
	\label{tab:ADMMmatpower}
\end{table}
\begin{table}[!t]
	%\normalsize
	\centering
	\renewcommand{\arraystretch}{1.3}
	\caption{Convergence of Algorithm~\ref{algorithm3} on NESTA v5 instances.}
	\resizebox{\linewidth}{!}{%
		\begin{tabular}{| c | c | c | c | c | c | c | c |}
			\cline{2-6}
			\multicolumn{1}{c|}{} & \multicolumn{3}{c|}{Objective (\$)} & \multicolumn{2}{c|}{Gap (\%)} &  \multicolumn{1}{c}{} & \multicolumn{1}{c}{} \\\cline{1-8}
			Case   & $P^{\dagger}_{\text{IPM}}$  & $P^{\dagger}_{\text{AMD}}$  &  $D_{\nu,\boldsymbol{\rho}}\left(\boldsymbol{\lambda}^{\dagger}\right)$  & AMD  & RO  & $I$ & set \\\hline															
			\multicolumn{8}{|c|}{Normal Operating Conditions} \\\hline															
			3	&	5812.64	&	5812.11	&	5812.64	&	-3.92E-06	&	9.09E-03	&	855	&	B	\\\hline
			4	&	156.43	&	156.49	&	156.43	&	-8.81E-04	&	-3.87E-02	&	708	&	B	\\\hline
			5	&	17551.89	&	17551.66	&	17551.93	&	-2.16E-04	&	1.31E-03	&	3351	&	F	\\\hline
			6\_c	&	23.21	&	23.21	&	23.21	&	-3.84E-03	&	-1.96E-02	&	928	&	G	\\\hline
			6\_ww	&	3143.97	&	3143.82	&	3143.97	&	3.36E-06	&	4.94E-03	&	918	&	B	\\\hline
			9	&	5296.69	&	5296.30	&	5296.69	&	-8.39E-06	&	7.24E-03	&	630	&	B	\\\hline
			14	&	244.05	&	244.03	&	244.06	&	-1.85E-03	&	8.22E-03	&	2544	&	B	\\\hline
			24	&	63352.20	&	63352.33	&	63352.21	&	-6.49E-06	&	-2.00E-04	&	924	&	B	\\\hline
			29	&	29895.49	&	29895.62	&	29897.32	&	-6.10E-03	&	-4.35E-04	&	45660	&	L	\\\hline
			30\_as	&	803.13	&	803.05	&	803.13	&	-4.54E-04	&	9.82E-03	&	1512	&	B	\\\hline
			30\_fsr	&	575.77	&	575.82	&	575.78	&	-1.10E-03	&	-9.26E-03	&	1566	&	B	\\\hline
			30	&	204.97	&	204.92	&	204.97	&	-8.29E-04	&	2.20E-02	&	3725	&	B	\\\hline
			39	&	96505.52	&	96505.50	&	96505.53	&	-1.24E-05	&	1.57E-05	&	5915	&	B	\\\hline
			57	&	1143.27	&	1143.25	&	1143.27	&	-3.33E-05	&	2.36E-03	&	6204	&	B	\\\hline
			73	&	189764.08	&	189764.44	&	189764.08	&	-2.70E-06	&	-1.92E-04	&	1034	&	B	\\\hline
			89	&	5819.81	&	5819.94	&	5819.92	&	-2.01E-03	&	-2.22E-03	&	10973	&	I	\\\hline
			118	&	3718.64	&	3718.77	&	3718.68	&	-1.22E-03	&	-3.66E-03	&	7423	&	B	\\\hline
			162	&	4230.23	&	4230.07	&	4230.23	&	-1.50E-04	&	3.58E-03	&	22387	&	E	\\\hline
			189	&	849.29	&	849.32	&	849.31	&	-1.50E-03	&	-3.41E-03	&	26116	&	E	\\\hline
			300	&	16891.28	&	16885.77	&	16891.55	&	-1.62E-03	&	3.26E-02	&	97225	&	J	\\\hline
			\multicolumn{8}{|c|}{Congested Operating Conditions (API)} \\\hline															
			3	&	367.74	&	367.83	&	367.74	&	-3.57E-04	&	-2.53E-02	&	2443	&	K	\\\hline
			4	&	767.27	&	767.38	&	767.26	&	1.35E-03	&	-1.42E-02	&	4835	&	B	\\\hline
			5	&	2998.54	&	2998.75	&	2998.75	&	-6.97E-03	&	-6.90E-03	&	20800	&	K	\\\hline
			6\_c	&	814.40	&	814.53	&	814.43	&	-3.26E-03	&	-1.59E-02	&	6642	&	K	\\\hline
			6\_ww	&	273.76	&	273.57	&	273.77	&	-3.48E-03	&	7.13E-02	&	1280	&	K	\\\hline
			9	&	656.60	&	656.61	&	656.61	&	-1.09E-03	&	-1.06E-03	&	8939	&	K	\\\hline
			14	&	325.56	&	325.87	&	325.78	&	-6.88E-02	&	-9.50E-02	&	12461	&	K	\\\hline
			24	&	6421.37	&	6423.98	&	6423.25	&	-2.93E-02	&	-4.06E-02	&	18371	&	K	\\\hline
			29	&	295782.68	&	295772.77	&	295781.62	&	3.58E-04	&	3.35E-03	&	33102	&	L	\\\hline
			30\_as	&	571.13	&	570.70	&	571.12	&	9.48E-04	&	7.55E-02	&	30485	&	B	\\\hline
			30\_fsr	&	372.14	&	369.32	&	372.28	&	-3.92E-02	&	7.56E-01	&	29885	&	M	\\\hline
			30	&	415.53	&	415.80	&	415.59	&	-1.48E-02	&	-6.70E-02	&	27173	&	K	\\\hline
			39	&	7466.25	&	7466.42	&	7466.27	&	-2.65E-04	&	-2.22E-03	&	40216	&	B	\\\hline
			57	&	1430.65	&	1430.85	&	1430.80	&	-1.06E-02	&	-1.35E-02	&	12057	&	K	\\\hline
			73	&	20123.98	&	20121.49	&	20125.17	&	-5.93E-03	&	1.23E-02	&	15427	&	E	\\\hline
			89	&	4288.02	&	4290.80	&	4289.81	&	-4.17E-02	&	-6.47E-02	&	83052	&	F	\\\hline
			118	&	10325.27	&	10330.53	&	10326.14	&	-8.40E-03	&	-5.09E-02	&	20723	&	N	\\\hline
			162	&	6111.68	&	6111.93	&	6111.68	&	-3.42E-05	&	-4.20E-03	&	15734	&	B	\\\hline
			189	&	1982.82	&	1984.41	&	1983.21	&	-1.94E-02	&	-8.01E-02	&	39572	&	O	\\\hline
			300	&	22866.01	&	22865.65	&	22867.11	&	-4.80E-03	&	1.58E-03	&	130433	&	T	\\\hline
			\multicolumn{8}{|c|}{Small Angle Difference Conditions (SAD)} \\\hline															
			3	&	5992.72	&	5993.25	&	5992.72	&	-9.57E-06	&	-8.82E-03	&	386	&	P	\\\hline
			4	&	324.02	&	324.04	&	324.05	&	-9.78E-03	&	-4.80E-03	&	953	&	P	\\\hline
			5	&	26423.32	&	26421.77	&	26428.28	&	-1.88E-02	&	5.86E-03	&	3569	&	P	\\\hline
			6\_c	&	24.43	&	24.43	&	24.44	&	-3.29E-02	&	-2.02E-02	&	1390	&	P	\\\hline
			6\_ww	&	3149.51	&	3149.70	&	3149.51	&	-1.36E-04	&	-6.28E-03	&	814	&	P	\\\hline
			9	&	5590.09	&	5590.09	&	5590.09	&	-3.78E-05	&	6.43E-06	&	807	&	P	\\\hline
			14	&	244.15	&	244.11	&	244.15	&	-1.26E-03	&	1.43E-02	&	4668	&	Q	\\\hline
			24	&	79804.96	&	79778.89	&	79805.13	&	-2.05E-04	&	3.27E-02	&	23248	&	R	\\\hline
			29	&	46933.26	&	46926.47	&	46940.55	&	-1.55E-02	&	1.45E-02	&	29251	&	L	\\\hline
			30\_as	&	914.44	&	914.84	&	914.51	&	-7.63E-03	&	-4.38E-02	&	5408	&	P	\\\hline
			30\_fsr	&	577.73	&	577.84	&	577.94	&	-3.68E-02	&	-1.95E-02	&	3121	&	P	\\\hline
			30	&	205.11	&	205.17	&	205.15	&	-1.95E-02	&	-2.58E-02	&	6707	&	P	\\\hline
			39	&	97219.04	&	97219.70	&	97219.00	&	3.97E-05	&	-6.73E-04	&	30125	&	P	\\\hline
			57	&	1143.88	&	1144.48	&	1144.49	&	-5.31E-02	&	-5.18E-02	&	16366	&	P	\\\hline
			73	&	235241.70	&	235253.53	&	235241.72	&	-7.46E-06	&	-5.03E-03	&	7623	&	P	\\\hline
			89	&	5827.01	&	5827.31	&	5827.32	&	-5.23E-03	&	-5.09E-03	&	43516	&	B	\\\hline
			118	&	4324.17	&	4324.60	&	4325.25	&	-2.51E-02	&	-9.90E-03	&	8197	&	B	\\\hline
			162	&	4369.19	&	4369.41	&	4369.66	&	-1.07E-02	&	-5.09E-03	&	26386	&	P	\\\hline
			189	&	914.61	&	914.65	&	914.66	&	-5.71E-03	&	-4.10E-03	&	24138	&	S	\\\hline
			300	&	16910.23	&	16905.29	&	16910.69	&	-2.72E-03	&	2.92E-02	&	107723	&	P	\\\hline																		
	\end{tabular}}
	\label{tab:ADMMNESTA}
\end{table}

The centralized IPM solutions, shown under $P^{\dagger}_{\text{IPM}}$ in Tables~\ref{tab:ADMMmatpower} and~\ref{tab:ADMMNESTA}, are initialized as in Algorithm~\ref{algorithm2}. This same initialization is also used as a starting point for the IPM solver at each iteration $k$ in Algorithm~\ref{algorithm3}. The parameter settings of Algorithm~\ref{algorithm3} are summarized in Table~\ref{tab:paramsetting} and the results are shown in Tables~\ref{tab:ADMMmatpower} and~\ref{tab:ADMMNESTA} for MATPOWER \cite{MATPOWER}, PEGASE \cite{Josz_ACdataMATPOWER} and NESTA v6 \cite{NESTA} instances respectively. The NESTA test cases are designed specifically to incorporate key network parameters such as line thermal limits and small angle differences, which are critical in optimization applications.

Tables~\ref{tab:ADMMmatpower} and~\ref{tab:ADMMNESTA} show that for $\epsilon=10^{-4}$, and after careful individualized tuning of parameters (see Table~\ref{tab:paramsetting}), Algorithm~\ref{algorithm3} converges to feasible solutions with negligible $\text{AMDgap}$ and $\text{ROgap}$ on all the 72 test cases.\footnote{Note that the stopping criterion in Algorithms~\ref{algorithm1},~\ref{algorithm2} and~\ref{algorithm3} requires a central authority to compute the norm of the subgradient; nonetheless, if a central authority is unavailable, the stopping criterion can be defined as in \cite{Magnusson_ADMMsequentialconvex} or \cite{Kraning_ADMM}.}

The glimmerings of a principled way of setting the parameters of Algorithm~\ref{algorithm3} are not apparent in Tables~\ref{tab:ADMMmatpower} and~\ref{tab:ADMMNESTA}. However, extensive simulations show that they can be clustered in a summarizing table (Table~\ref{tab:paramsetting}) of plausible parameter settings. Some parameter settings, like B, K and P for example, seem to work on the most number of test cases. This stands in contrast to settings A, C, D, G, I, J, M, N, O, Q, R, S and T which are tailored specifically to their respective test cases in Table~\ref{tab:ADMMNESTA}. 
\begin{table}[!t]
	\scriptsize
	\centering
	\renewcommand{\arraystretch}{1.3}
	\caption{Summarized parameter settings of Algorithm~\ref{algorithm3}.}
	%	\resizebox{\linewidth}{!}{%
	\begin{tabular}{| c | c | c | c | c | c |}
		\cline{1-6}
		Setting   & $\nu$  & $\rho_{pq}$ & $\rho_{v \theta}$  & $\alpha_{i}$ & $\alpha_{ij}$  \\\hline
		A	&	 3,000 	&	 30 	&	 300,000 	&	 300 	 & 	 300,000 	\\\hline
		B	&	 1,000 	&	 100 	&	 10,000 	&	 100 	 & 	 10,000 	\\\hline
		C	&	 1,000 	&	 1,000 	&	 100,000 	&	 100 	 & 	 100,000 	\\\hline
		D	&	 100 	&	 1 	&	 10,000 	&	 10 	 & 	 10,000 	\\\hline
		E	&	 5,000 	&	 500 	&	 50,000 	&	 500 	 & 	 50,000 	\\\hline
		F	&	 3,000 	&	 300 	&	 300,000 	&	 300 	 & 	 300,000 	\\\hline
		G	&	 100 	&	 10 	&	 1,000 	&	 10 	 & 	 1,000 	\\\hline
		H	&	 5,000 	&	 500 	&	 500,000 	&	 500 	 & 	 500,000 	\\\hline
		I	&	 100 	&	 10 	&	 10,000 	&	 10 	 & 	 10,000 	\\\hline
		J	&	 10,000 	&	 100 	&	 100,000 	&	 1,000 	 & 	 100,000 	\\\hline
		K	&	 10,000 	&	 1,000 	&	 10,000 	&	 1,000 	 & 	 10,000 	\\\hline
		L	&	 1,000 	&	 100 	&	 100,000 	&	 100 	 & 	 100,000 	\\\hline
		M	&	 8,000 	&	 800 	&	 800,000 	&	 800 	 & 	 800,000 	\\\hline
		N	&	 5,000 	&	 500 	&	 100,000 	&	 500 	 & 	 100,000 	\\\hline
		O	&	 80,000 	&	 8,000 	&	 100,000 	&	 8,000 	 & 	 100,000 	\\\hline
		P	&	 10,000 	&	 1,000 	&	 100,000 	&	 1,000 	 & 	 100,000 	\\\hline
		Q	&	 10,000 	&	 1,000 	&	 100,000 	&	 1,000 	 & 	 100,000 	\\\hline
		R	&	 1,000 	&	 10 	&	 10,000 	&	 100 	 & 	 10,000 	\\\hline
		S	&	 50,000 	&	 5,000 	&	 500,000 	&	 5,000 	 & 	 500,000 	\\\hline
		T	&	8,000	&	800	&	100,000	&	800	&	100,000	\\\hline
	\end{tabular}
	\label{tab:paramsetting}
\end{table}
There are three key contributors behind the convergence of Algorithm~\ref{algorithm3} on all the $72$ cases. First, parameters $\nu$, $\rho_{pq}$ and $\rho_{v \theta}$ are set to high values, typically in the ranges $\left[ 100, 80000\right] $, $\left[ 1, 8000\right] $ and $\left[ 1000, 800000\right] $, respectively. Second, most test cases require setting $\rho_{v \theta}$ to at least 2 orders of magnitude larger than $\nu$ and 3 orders of magnitude larger than $\rho_{pq}$. Third, this disproportion in setting $\nu$, $\rho_{pq}$ and $\rho_{v \theta}$ is reflected exactly in setting the values of the step size in the multiplier update \eqref{eq1:lambdaupdate}. More specifically, the step size $\alpha_{ij}$ is also 2 orders of magnitude larger than $\alpha_{i}$ in these test cases. In fact, $\alpha_{i}$ is set to $0.1 \nu$ and $\alpha_{ij}$ is set equal to $\rho_{v \theta}$. To see the significance of this, all the test instances with this specific parameter tuning would otherwise either diverge or require more than $10^6$ iterations to converge. Some notoriously difficult cases are MATPOWER's case 5, NESTA's cases 30$\_$fsr$\_$API and 189$\_$API for which very few other parameter settings (which are not shown here due space limitations), besides the corresponding ones in Table~\ref{tab:ADMMmatpower}, seem to make Algorithm~\ref{algorithm3} converge. Furthermore, even after exhaustive parameter tuning, the convergence on some test instances (like NESTA's 300 bus systems) is substantially slower than others. Nonetheless, this still suggests that Algorithm~\ref{algorithm3} converges even on these \emph{difficult} test instances.

Algorithm~\ref{algorithm3} is (theoretically) not guaranteed to converge to feasible solutions, let alone to globally optimal ones. However, as shown in Tables~\ref{tab:ADMMmatpower} and~\ref{tab:ADMMNESTA} and in Appendices~\ref{Appendix1} and~\ref{Appendix2}, the right starting point combined with the right parameter settings can result in a convergence to feasible near-optimal (possibly globally optimal) solutions (corroborated by tight convex relaxations \cite{Kocuk_strongSOCP,Coffrin_Strengtheningwithboundtightening,Hijazi_PolynomialSDPcuts,Coffrin_StrengtheningSDP,Kocuk_Matrixminorreformulations}), despite having no guarantees that the subproblems in \eqref{eq1:Linesubp} are solved to global optimality. Moreover, as shown in Appendix~\ref{Appendix2}, case-specific parameter settings can lead to globally optimal solutions irrespective of the choice of algorithmic starting point.
%Diminishing step size algorithms do not work here because it would break the full decentralized property. Also, diminishing step size means slower convergence because the dual function becomes flatter when $\rho$ is large. So the step size has to be large too. 
%OPF is ill-conditioned \cite{Madani_DistributedSparseSDPforOPF}.

\section{Conclusion}\label{sec:conclusion}

The founding premise of this work is that, given the right algorithmic parameter settings, the method is numerically demonstrated to converge to feasible near-optimal (possibly globally optimal) solutions to the nonconvex AC OPF problem; corroborated by tight convex relaxations, on all the $72$ considered test cases.
%The method is numerically demonstrated to converge to the same solutions obtained from the centralized IPMs, on all the $72$ considered test cases.
Despite the absence of a principled way to set up the parameters of the algorithm, this work demonstrates that, first, the proximal and the ADMM penalty parameters should be set to at least 100. Second, the ADMM penalty parameter for the voltage and angle terms is set to at least 1 order of magnitude larger than the proximal penalty parameter in order to ensure differentiability of the modified dual function. Third, most test cases require setting the ADMM penalty parameter for the voltage and angle terms to at least 3 orders of magnitude larger than the ADMM penalty parameter for the active and reactive power terms to witness convergence. These three results not only affect the speed of convergence, but can mean the difference between convergence and divergence. Future work will consist of investigating different accelerated subgradient methods to speed-up the convergence of the method.

\appendices
\section{Modified dual - example 1}\label{Appendix1}

\begin{figure}
	\centering
	\begin{subfigure}[t]{0.24\textwidth}
		\psfrag{x}{\footnotesize $x_{1}$ \normalsize}
		\psfrag{y}{\footnotesize $x_{2}$ \normalsize}
		\includegraphics[width=48mm]{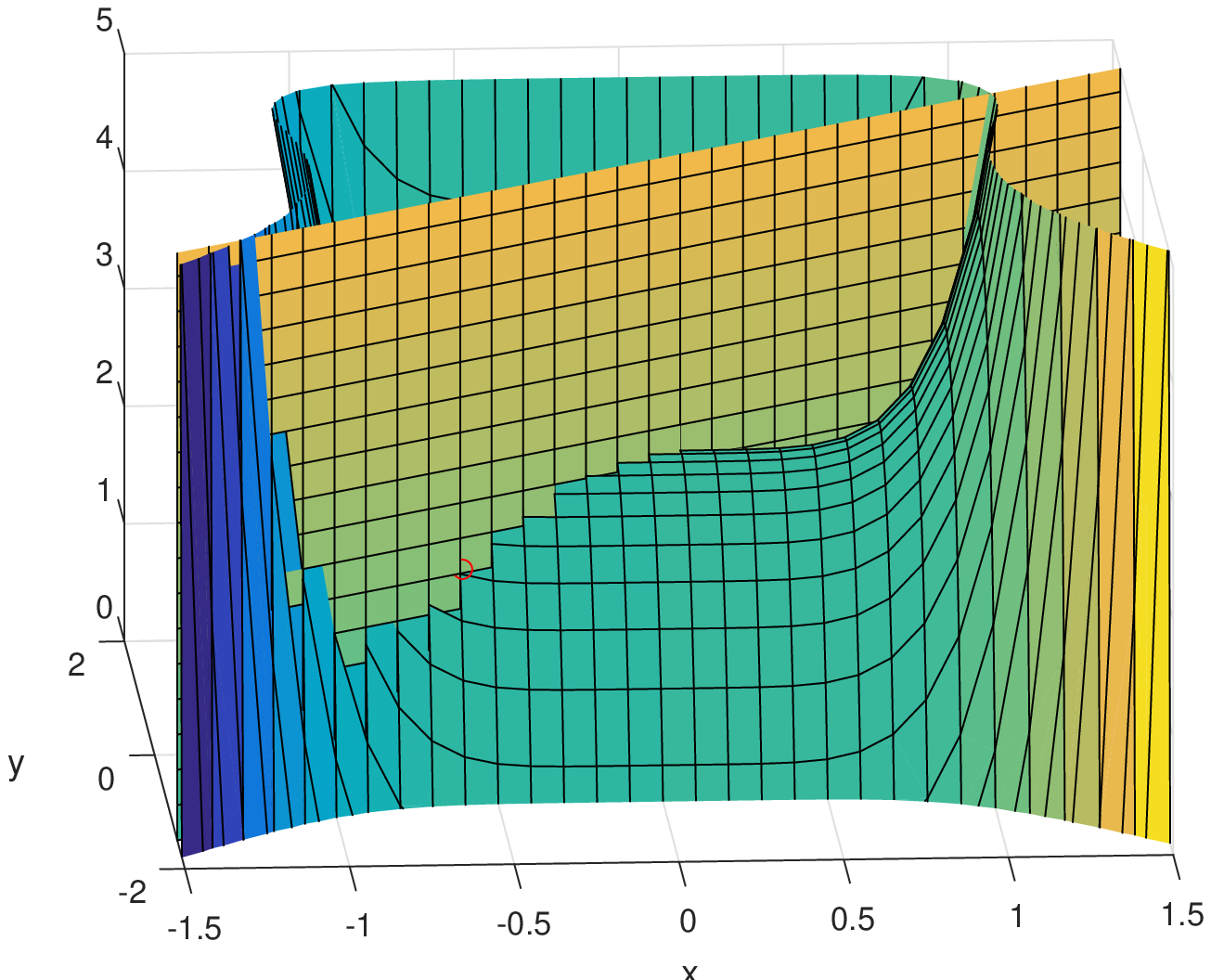}
		\caption{}
		\label{fig:example1primal3d}   
	\end{subfigure}             
	\begin{subfigure}[t]{0.24\textwidth}
		\psfrag{x}{\footnotesize $x_{1}$ \normalsize}
		\includegraphics[width=48mm]{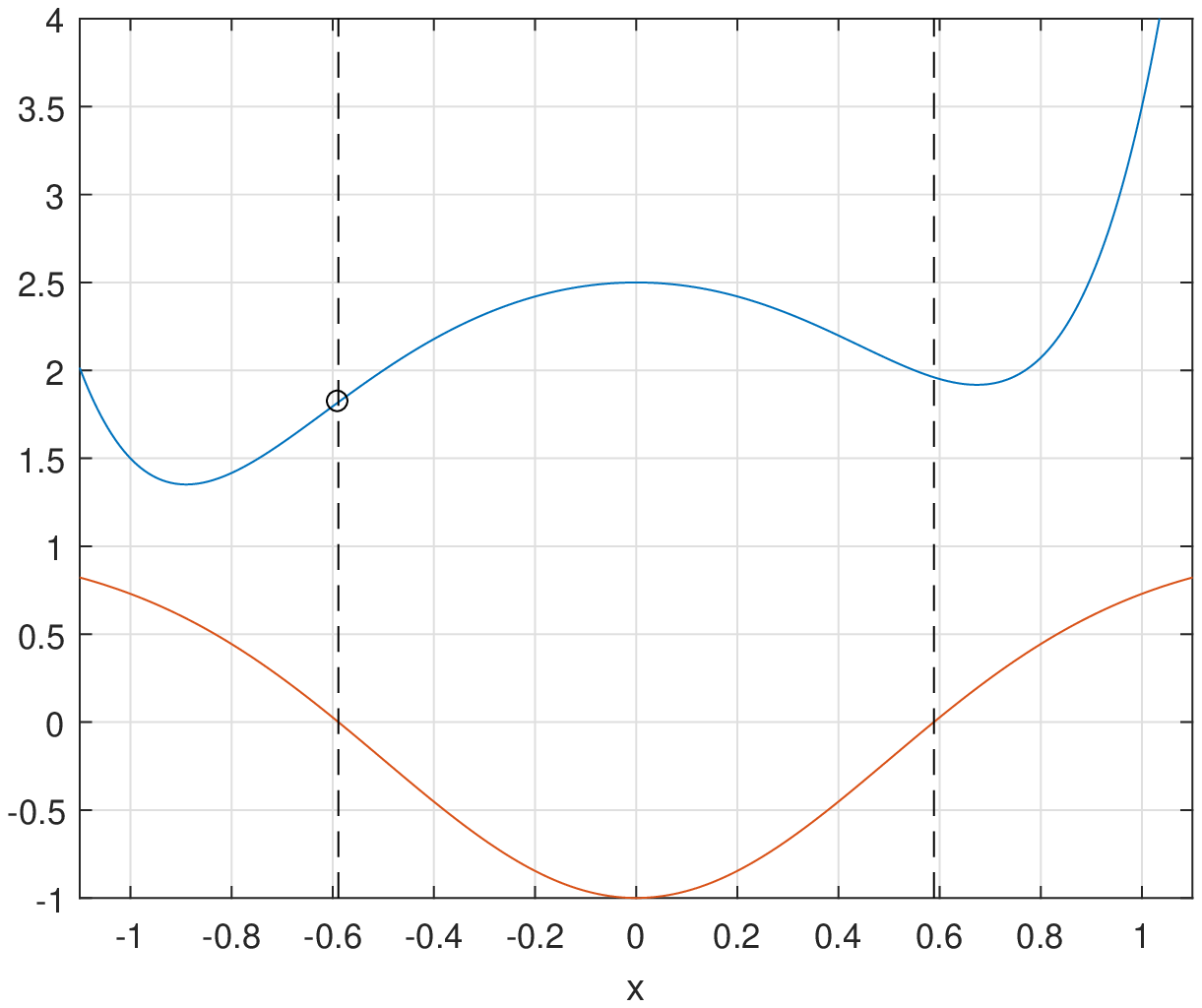}
		\caption{}
		\label{fig:example1primal2d}
	\end{subfigure}             
	\caption{Problem \eqref{example1primal} (a) and its 2D equivalent in (b). The optimal point and value are $\boldsymbol{x}^{\star}=[-0.5887,-0.5887]$, $p^{\star}=1.8194$ (shown as a circle). The suboptimal point and value are $\boldsymbol{x}^{\dagger}=[0.5887,0.5887]$, $p^{\dagger}=1.9608$.}
	\label{fig:example1primal}
\end{figure}

\begin{figure}
	\centering
	\begin{subfigure}[t]{0.24\textwidth}
		\psfrag{l}{\scriptsize $\lambda$ \normalsize}
		\psfrag{Dual}{\scriptsize $D\left( \lambda\right) $ \normalsize}
		\includegraphics[width=48mm]{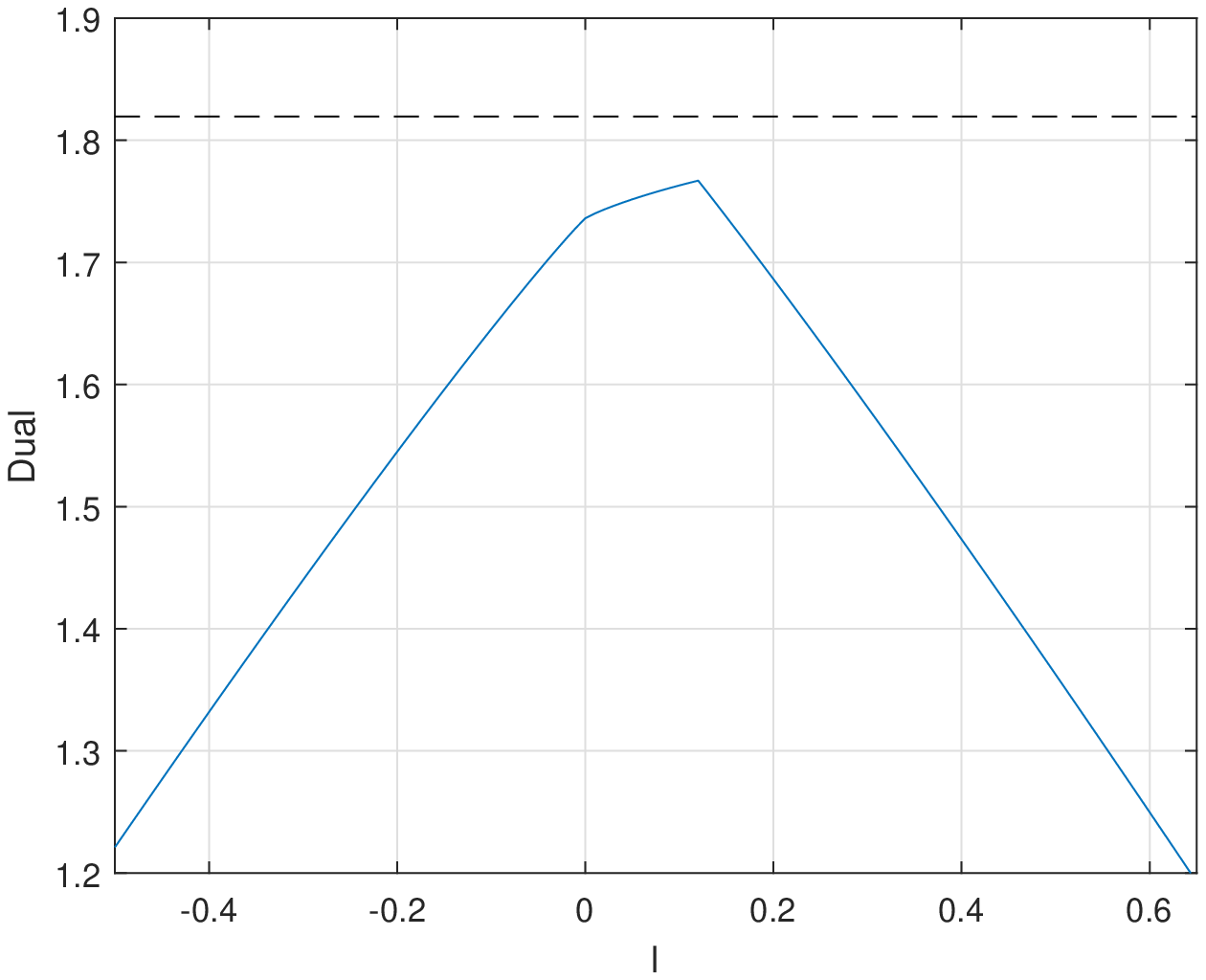}
		\caption{}
		\label{fig:example1dual1}   
	\end{subfigure}             
	\begin{subfigure}[t]{0.24\textwidth}
		\psfrag{l}{\scriptsize $\lambda$ \normalsize}
		\includegraphics[width=48mm]{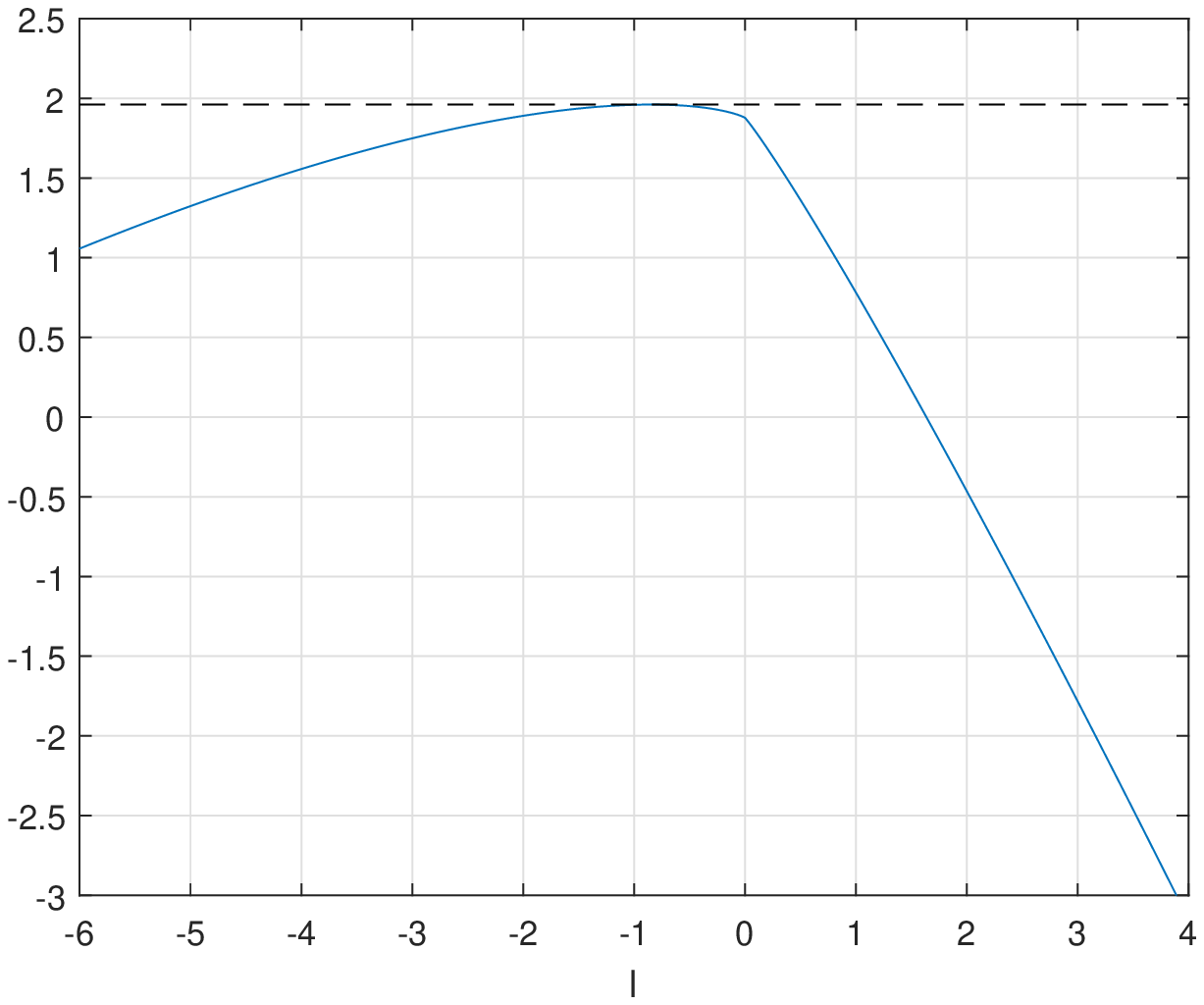}
		\caption{}
		\label{fig:example1dual2}
	\end{subfigure}             
	\caption{The Lagrange dual function of problem \eqref{example1primal}. The dashed lines in (a) and (b) show $p^{\star}$ and $p^{\dagger}$ respectively.}
	\label{fig:example1dual}
\end{figure}

\begin{figure}
	\centering
	\begin{subfigure}[t]{0.24\textwidth}
		\psfrag{R}{\scriptsize Residuals \normalsize}
		\psfrag{I}{\scriptsize Iterations ($k$) \normalsize}
		\psfrag{P}{\tiny Primal residuals \normalsize}
		\psfrag{D}{\tiny Dual residuals \normalsize}
		\includegraphics[width=48mm]{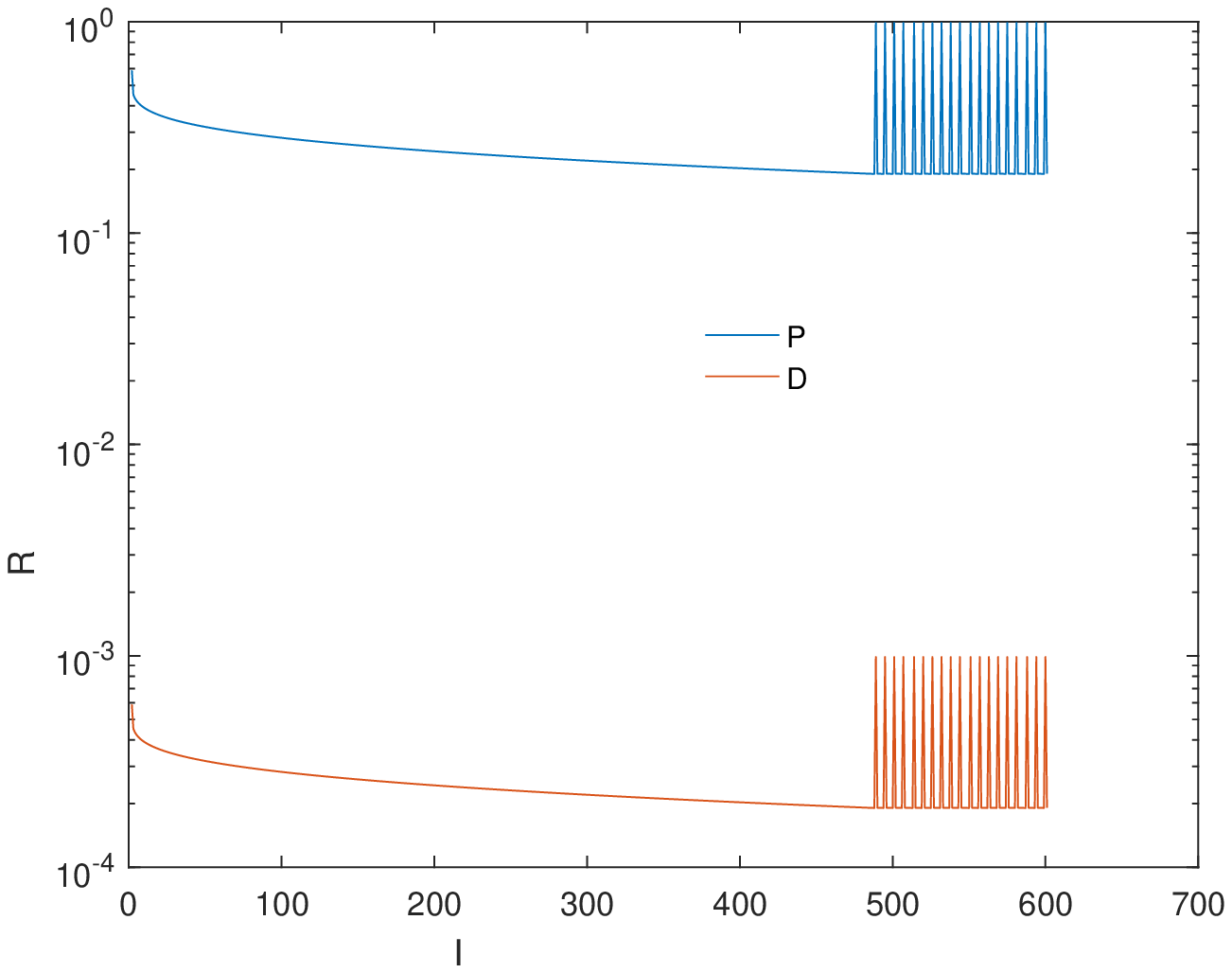}
		\caption{Step size=$0.001$.}
		\label{fig:example1dualsol1}   
	\end{subfigure}             
	\begin{subfigure}[t]{0.24\textwidth}
		\psfrag{R}{\scriptsize Residuals \normalsize}
		\psfrag{I}{\scriptsize Iterations ($k$) \normalsize}
		\psfrag{P}{\tiny Primal residuals \normalsize}
		\psfrag{D}{\tiny Dual residuals \normalsize}
		\includegraphics[width=48mm]{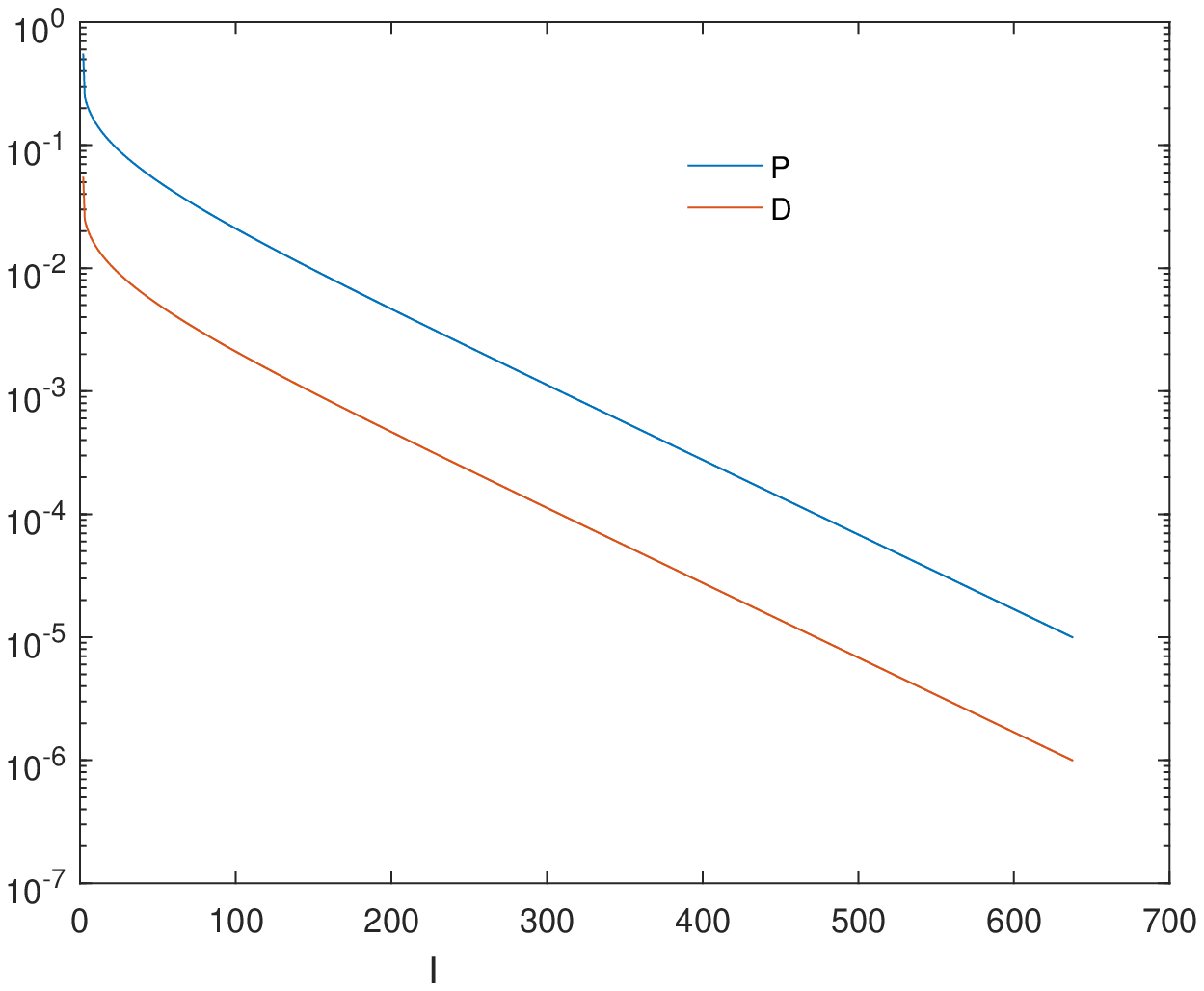}
		\caption{Step size=$0.1$.}
		\label{fig:example1dualsol2}
	\end{subfigure}             
	\caption{Primal and dual residuals of the subgradient projection method with $\lambda^{1}=0$, applied to solving \eqref{example1dualp}, and when \eqref{example1dualf} is solved to optimality (a) and suboptimality (b).}
	\label{fig:example1pdresiduals}
\end{figure}

Consider the nonconvex problem
\begin{subequations}\label{example1primal}
	\begin{align}
	\underset {\substack{x_{1},x_{2}}} 
	{\mbox{minimize  }}   2x_{1}^6+x_{2}^5 & -2x_{2}^2+2.5  \\
	\text{subject to } \ \ \qquad x_{1}&=x_{2}, \\
	-2\exp^{(-2x_{2}^2)}+1& \leq 0,
	\end{align}
\end{subequations}
shown in Figure~\ref{fig:example1primal}. 
Let $\boldsymbol{x}:=\left[ x_{1},x_{2}\right]  \in \mathcal{X}$, where $\mathcal{X}:=\mathcal{X}_{1} \times \mathcal{X}_{2}$, $\mathcal{X}_{1}=\reals$ and $\mathcal{X}_{2}:=\left\lbrace x_{2} \in \reals | -2\exp^{(-2x_{2}^2)}+1 \leq 0 \right\rbrace $. Also, let $f(\boldsymbol{x})=2x_{1}^6+x_{2}^5 -2x_{2}^2+2.5$ $\left( f:\reals^{2} \mapsto \reals\right) $. The (partial) Lagrangian function of problem \eqref{example1primal} is defined as $L\left( \boldsymbol{x}, \lambda\right):= f(\boldsymbol{x}) +\lambda\left( x_{1}-x_{2} \right)$, and the Lagrange dual function of problem \eqref{example1primal} is defined as
\begin{align}\label{example1dualf}
D\left( \lambda\right):=\underset {\substack{\boldsymbol{x} \in \mathcal{X}}} 
{\mbox{inf  }}  L\left( \boldsymbol{x}, \lambda\right).
\end{align}
Consequently, the Lagrange dual problem is
\begin{align}\label{example1dualp}
\underset {\substack{\lambda}} 
{\mbox{maximize  }} D\left( \lambda\right).
\end{align}
The Lagrange dual function in \eqref{example1dualf} is concave, as it is the pointwise infimum of a family of affine functions of $\lambda$, despite the nonconvexity of problem \eqref{example1primal}. The concave Lagrange dual function \eqref{example1dualf} is shown in Figure~\ref{fig:example1dual}.  
The dual function in Figure~\ref{fig:example1dual1}, which is obtained by solving \eqref{example1dualf} to (global) optimality, shows that problem \eqref{example1primal} has a nonzero duality gap. More interestingly, the approximated dual function in Figure~\ref{fig:example1dual2}, which is obtained from solving \eqref{example1dualf} to suboptimality by listing all the KKT points of problem \eqref{example1dualf} and selecting the first suboptimal point, has an optimal value of $d^{\dagger}=p^{\dagger}=1.9608$. This can be interpreted as an inaccurate approximation of the dual function and is a result of not solving the \eqref{example1dualf} to global optimality. Formally, any method for solving the Lagrange dual function of a nonconvex problem can converge to a suboptimal point if problem \eqref{example1dualf} is consistently (at each iteration) solved to suboptimality. And indeed, as shown in Figure~\ref{fig:example1dualsol2}, the subgradient projection method applied to the dual function converges to the suboptimal point $\boldsymbol{x}^{\dagger}$ and value $d^{\dagger}=p^{\dagger}=1.9608$, when the algorithm is started with $x_{2}^{1}=1$ and when selecting the first suboptimal point from the list of KKT points of problem \eqref{example1dualf}. 
Moreover, the same convergence behaviour is obtained by solving \eqref{example1dualf} using an IPM solver\footnote{IPM solvers only guarantee local optimality.} (IPOPT 3.12.5 \cite{IPOPT}, KNITRO 10.2 \cite{KNITRO}) with a starting point $\boldsymbol{x}^{0}_{\text{IPM}}=[0.5,0.5]$ at each iteration.

Furthermore, the Lagrange dual function in Figure~\ref{fig:example1dual1} is nonsmooth. The dual function in Figure~\ref{fig:example1dual1} is nondifferentiable at $\lambda=0$ and $\lambda=0.1203$. This is because \eqref{example1dualf} can have multiple (globally) optimal solutions for a given $\lambda$, and as a consequence, the subdifferentials $\partial D\left(\boldsymbol{\lambda}\right)$ may be not be unique. Indeed, using Danskin's theorem \cite{Danskin,onatheoremofDanskin,nonlinearpogramming}, the subdifferentials of $D\left(\boldsymbol{\lambda}\right)$ are $\partial D\left(\boldsymbol{\lambda}\right):=\left\{A_{c}\boldsymbol{x}:D\left(\boldsymbol{\lambda}\right) , \boldsymbol{x} \in \mathcal{X} \right\}$ ($A_{c}=[1,-1]$). The effect of nondifferentiability on the convergence of the subgradient projection method is apparent in Figure~\ref{fig:example1dualsol1} which shows the oscillations of the dual (and primal) residuals when $D\left( \lambda^{k}\right)$ approaches its maximum value of $d^{\star}=1.7670$ at $\lambda=0.1203$.

Both issues of nonzero duality gap and nondifferentiability can be addressed by modifying the Lagrange function as follows
\begin{align*}
L_{\rho}\left( \boldsymbol{x}, \lambda\right):=L\left( \boldsymbol{x}, \lambda\right)+\frac{\rho}{2}\left\|  x_{1}-x_{2} \right\| ^{2},
\end{align*}
which is also known as the augmented Lagrange function, and the augmented Lagrange dual function would be
\begin{align}\label{eq:example1dualfA}
D_{\rho}\left( \lambda\right):=\underset {\substack{\boldsymbol{x} \in \mathcal{X}}} 
{\mbox{inf  }} L_{\rho}\left( \boldsymbol{x}, \lambda\right).
\end{align}
\begin{figure}
	\centering
	\begin{subfigure}[t]{0.24\textwidth}
		\psfrag{l}{\scriptsize $\lambda$ \normalsize}
		\psfrag{Dual}{\scriptsize $D_{\rho}\left( \lambda\right) $ \normalsize}
		\includegraphics[width=48mm]{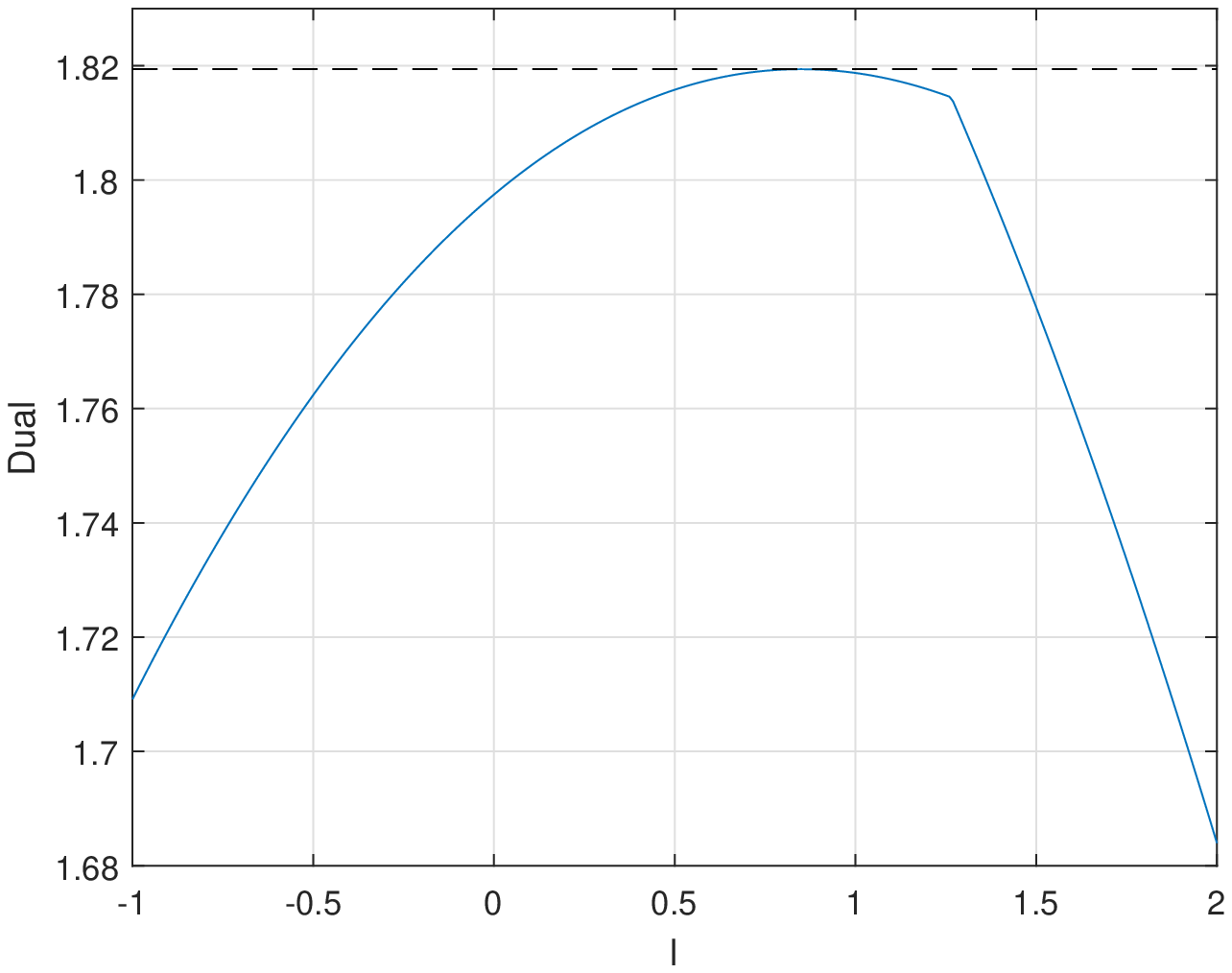}
		\caption{}
		\label{fig:example1ALRdual1}   
	\end{subfigure}             
	\begin{subfigure}[t]{0.24\textwidth}
		\psfrag{l}{\scriptsize $\lambda$ \normalsize}
		\includegraphics[width=48mm]{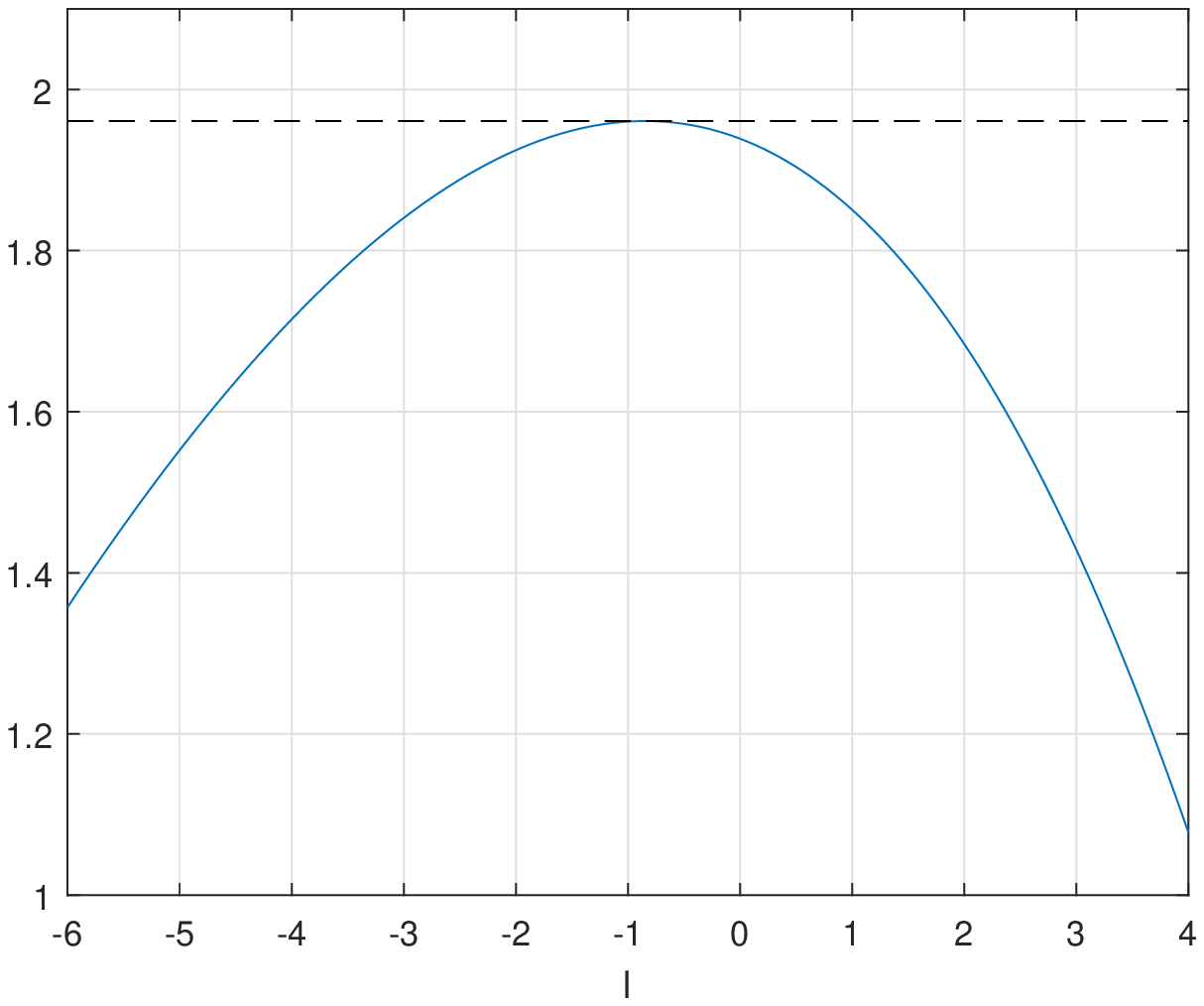}
		\caption{}
		\label{fig:example1ALRdual2}
	\end{subfigure}             
	\caption{The augmented Lagrange dual function of problem \eqref{example1primal}, with $\rho=10$. The dashed lines in (a) and (b) show $p^{\star}$ and $p^{\dagger}$ respectively.}
	\label{fig:example1ALRdual}
\end{figure}
As shown in Figure~\ref{fig:example1ALRdual1}, the problem now has a zero modified duality gap, which is the gap between the optimal primal value $p^{\star}$ and the optimal modified dual value $d_{M}^{\star}:=D_{\rho}\left( \lambda^{\star}\right)$.\footnote{The gap between the optimal primal value $p^{\star}$ and the optimal modified dual value $d_{M}^{\star}$ is called modified duality gap to distinguish it from the classical definition of duality gap, which is the gap between the optimal primal value $p^{\star}$ and the optimal dual value $d^{\star}:=D\left( \lambda^{\star}\right)$.} This should not be surprising as for very large values of $\rho$, the augmented Lagrangian regularization term would be equivalent to a barrier function \cite{Bertsekas_convexoptalg2015}. Also, for $\rho \geq 2$, the augmented Lagrange dual function is smooth over the set $[0,\lambda^{\star}]$. 

However, the augmented Lagrange dual function in \eqref{eq:example1dualfA} is not separable in terms of sets of variables ($\mathcal{X}_{1}$ and $\mathcal{X}_{2}$). Nevertheless, ADMM can be used to decouple these sets of variables ($\mathcal{X}_{1}$ and $\mathcal{X}_{2}$), by using alternate minimizations over these sets. In particular, given the current iterates $\left(x_{1}^{k},x_{2}^{k},\lambda^{k} \right) $, ADMM generates a new iterate $\left(x_{1}^{k+1},x_{2}^{k+1},\lambda^{k+1} \right) $ as follows
\begin{align}
x_{1}^{k+1} \in & \ \underset {\substack{x_{1} \in \mathcal{X}_{1}}} {\argmin } \ L_{\rho}\left(x_{1},x_{ 2}^{k},\lambda^{k} \right), \label{eq:ADMMx1} \\
x_{2}^{k+1} \in & \ \underset {\substack{x_{2} \in \mathcal{X}_{2}}} {\argmin } \ L_{\rho}\left(x_{1}^{k+1},x_{2},\lambda^{k} \right), \label{eq:ADMMx2} \\
\lambda^{k+1}=& \ \lambda^{k}+\rho \left(x_{1}^{k+1}-x_{2}^{k+1} \right). \label{eq:lupdate}
\end{align}
\begin{figure}
	\centering
	\begin{subfigure}[t]{0.24\textwidth}
		\psfrag{R}{\scriptsize Residuals \normalsize}
		\psfrag{I}{\scriptsize Iterations ($k$) \normalsize}
		\psfrag{P}{\tiny Primal residuals \normalsize}
		\psfrag{D}{\tiny Dual residuals \normalsize}
		\includegraphics[width=48mm]{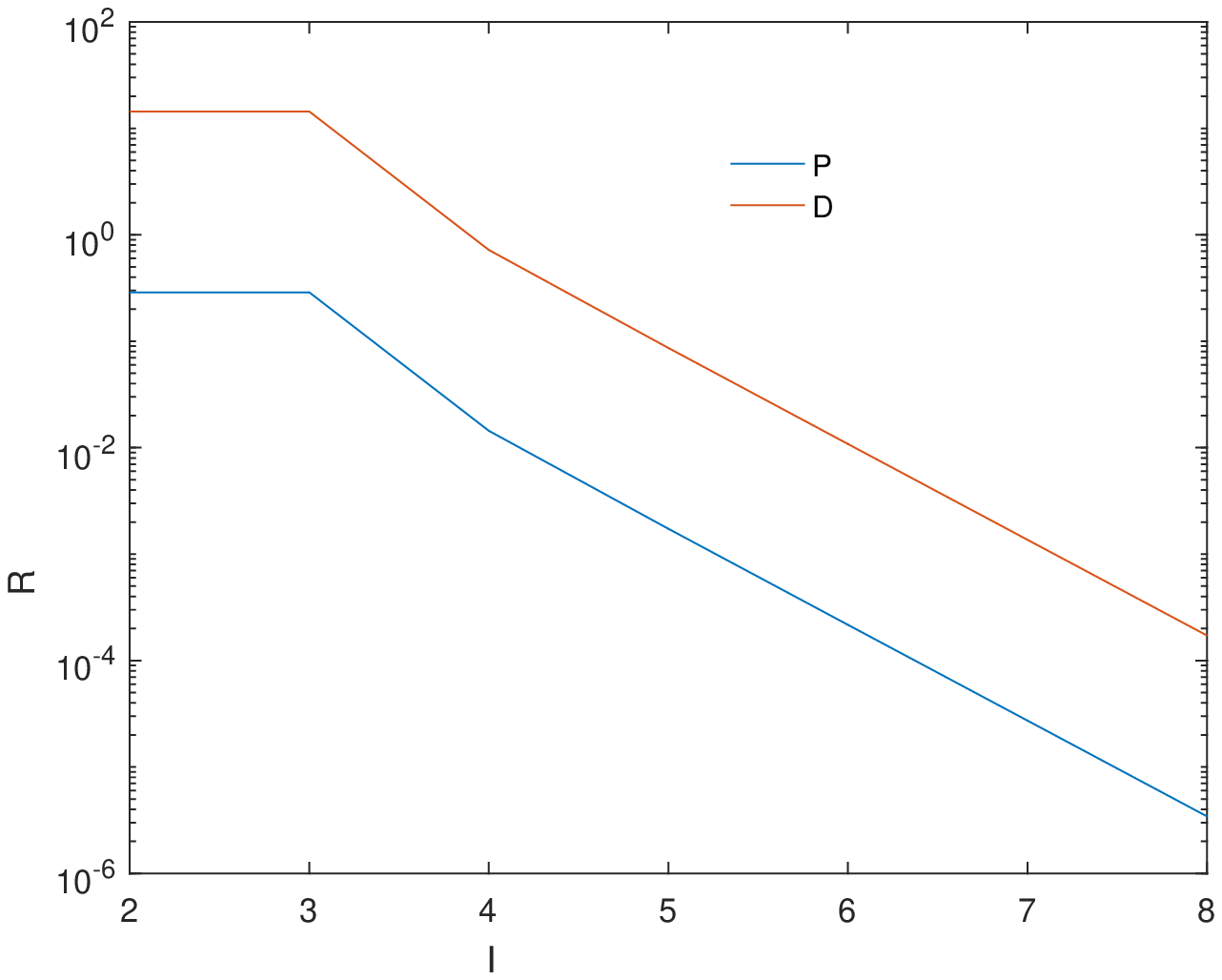}
		\caption{$x_{2}^{1}=-1$}
		\label{fig:example1pdresidualsADMM1}   
	\end{subfigure}             
	\begin{subfigure}[t]{0.24\textwidth}
		\psfrag{R}{\scriptsize Residuals \normalsize}
		\psfrag{I}{\scriptsize Iterations ($k$) \normalsize}
		\psfrag{P}{\tiny Primal residuals \normalsize}
		\psfrag{D}{\tiny Dual residuals \normalsize}
		\includegraphics[width=48mm]{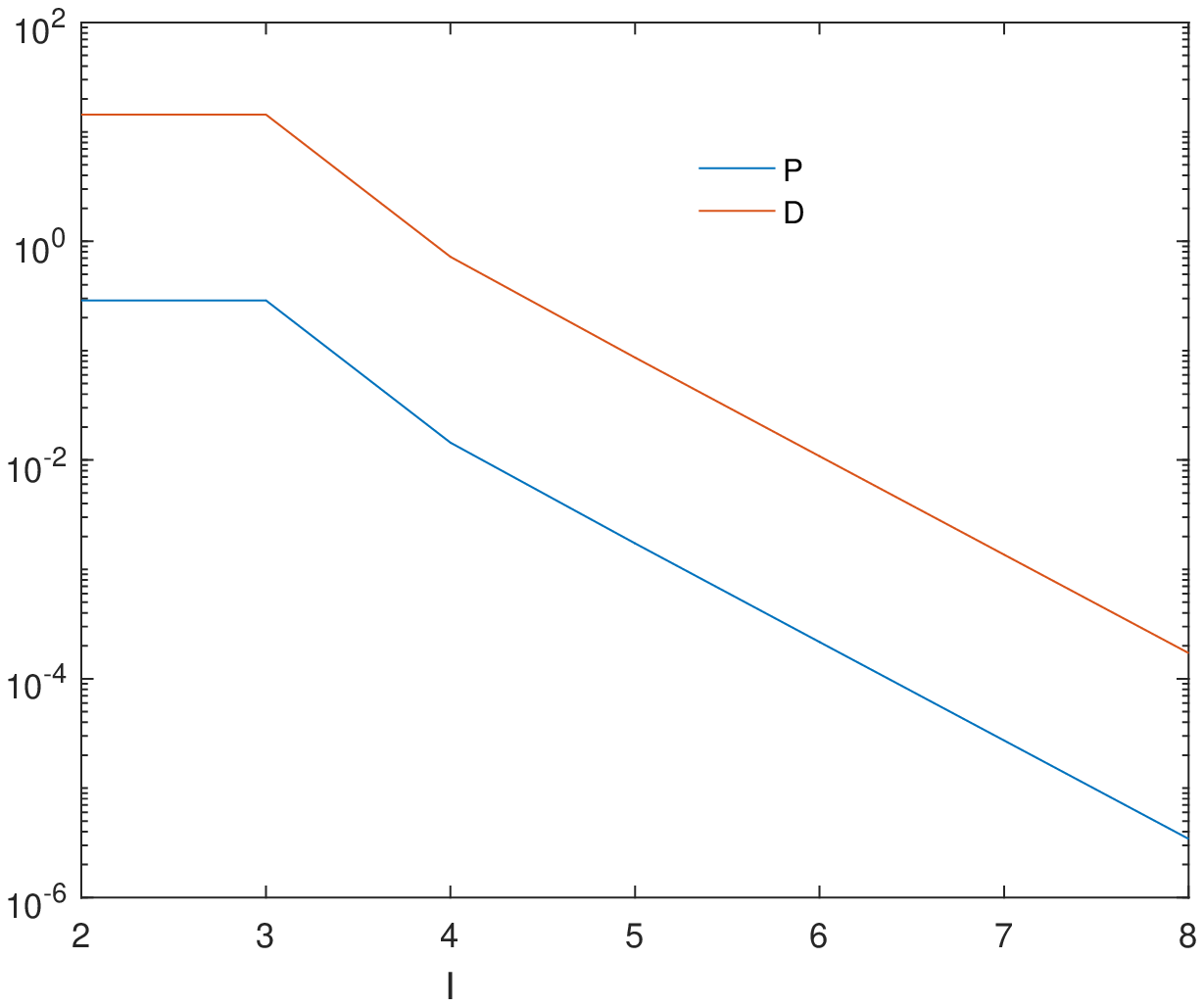}
		\caption{$x_{2}^{1}=1$}
		\label{fig:example1pdresidualsADMM2}
	\end{subfigure}             
	\caption{Primal and dual residuals of ADMM ($\rho=50$), for $x_{2}^{1}=-1$ (a) and $x_{2}^{1}=1$ (b), and when \eqref{eq:ADMMx1} and \eqref{eq:ADMMx2} are solved to optimality.}
	\label{fig:example1pdresidualsADMM}
\end{figure}
Figure~\ref{fig:example1pdresidualsADMM} shows the primal and dual residuals of ADMM with $\rho=50$ and for two different starting points. The starting point $x_{2}^{1}=-1$ (Figure~\ref{fig:example1pdresidualsADMM1}) leads to a convergence to $d_{M}^{\star}=p^{\star}$, whereas $x_{2}^{1}=1$ (Figure~\ref{fig:example1pdresidualsADMM2}) leads to a convergence to $d_{M}^{\dagger}=p^{\dagger}$. Furthermore, for $\rho=2$, ADMM with $x_{2}^{1}=-1$ converges to $d_{M}^{\star}=p^{\star}$ in $39$ iterations as compared to $14$ for $\rho=10$ and $8$ for $\rho=50$. 

\begin{figure}[t]
	\centering{
		\psfrag{R}{\scriptsize Residuals \normalsize}
		\psfrag{I}{\scriptsize Iterations ($k$) \normalsize}
		\psfrag{P}{\tiny Primal residuals \normalsize}
		\psfrag{D}{\tiny Dual residuals \normalsize}
		\includegraphics[width=60mm]{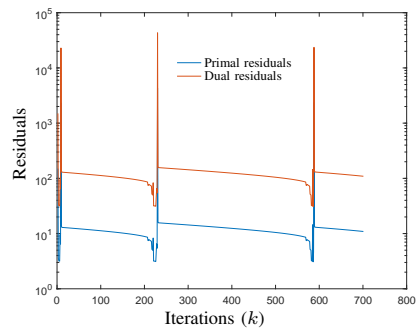}}
	\caption{Primal and dual residuals of ADMM ($\rho=10$), for $x_{2}^{1}=-1$ and $\boldsymbol{x}^{0}_{\text{IPM}}=[1,1]$.}
	\label{fig:example1ADMMsub}
\end{figure}

To underscore the effect of solving \eqref{eq:ADMMx1} and \eqref{eq:ADMMx2} to suboptimality (as might be the case when using an IPM solver), an IPM solver is used to solve \eqref{eq:example1dualfA} for $\rho=10$. The IPM solver is initialized with $\boldsymbol{x}^{0}_{\text{IPM}}=[1,1]$ at each iteration $k$ and the algorithm is initialized with $x_{2}^{1}=-1$. The result, shown in Figure~\ref{fig:example1ADMMsub}, is an oscillatory behaviour which is a due to the iterates alternating between the modified dual function (Figure~\ref{fig:example1ALRdual1}), obtained by solving \eqref{eq:example1dualfA} to optimality, and the approximate modified dual function (Figure~\ref{fig:example1ALRdual2}), obtained by solving \eqref{eq:example1dualfA} to suboptimality. In this specific case, the oscillations recur and the subgradient projection method does not converge. In other cases with different $\boldsymbol{x}^{0}_{\text{IPM}}$, the algorithm eventually converges but very slowly.\footnote{Note that in this example we actually know which IPM solver starting point leads to convergence but in many other practical problems one does not have this information.} The main reason why this is important is that in many cases, like the OPF problem, solving \eqref{eq:ADMMx1} and \eqref{eq:ADMMx2} to optimality can be time consuming (not ideal for real-time applications) and therefore IPM solvers are used instead of GNLP solvers. In these cases it is best to initialize both the algorithm and the IPM solvers (at each iteration) with the same starting point.\footnote{Note that this is not always obvious as most IPM solvers, when not given a starting point, select a trivial one [0,0], which might not be an ideal starting point for the problem at hand.} Indeed, in this example, initializing both the algorithm and the IPM solver at each iteration with the same starting point results in a convergence to the same solutions obtained when the subproblems are solved to global optimality.

\begin{figure}
	\centering
	\begin{subfigure}[t]{0.24\textwidth}
		\psfrag{R}{\scriptsize Residuals \normalsize}
		\psfrag{I}{\scriptsize Iterations ($k$) \normalsize}
		\psfrag{P}{\tiny Primal residuals \normalsize}
		\psfrag{D}{\tiny Dual residuals \normalsize}
		\includegraphics[width=48mm]{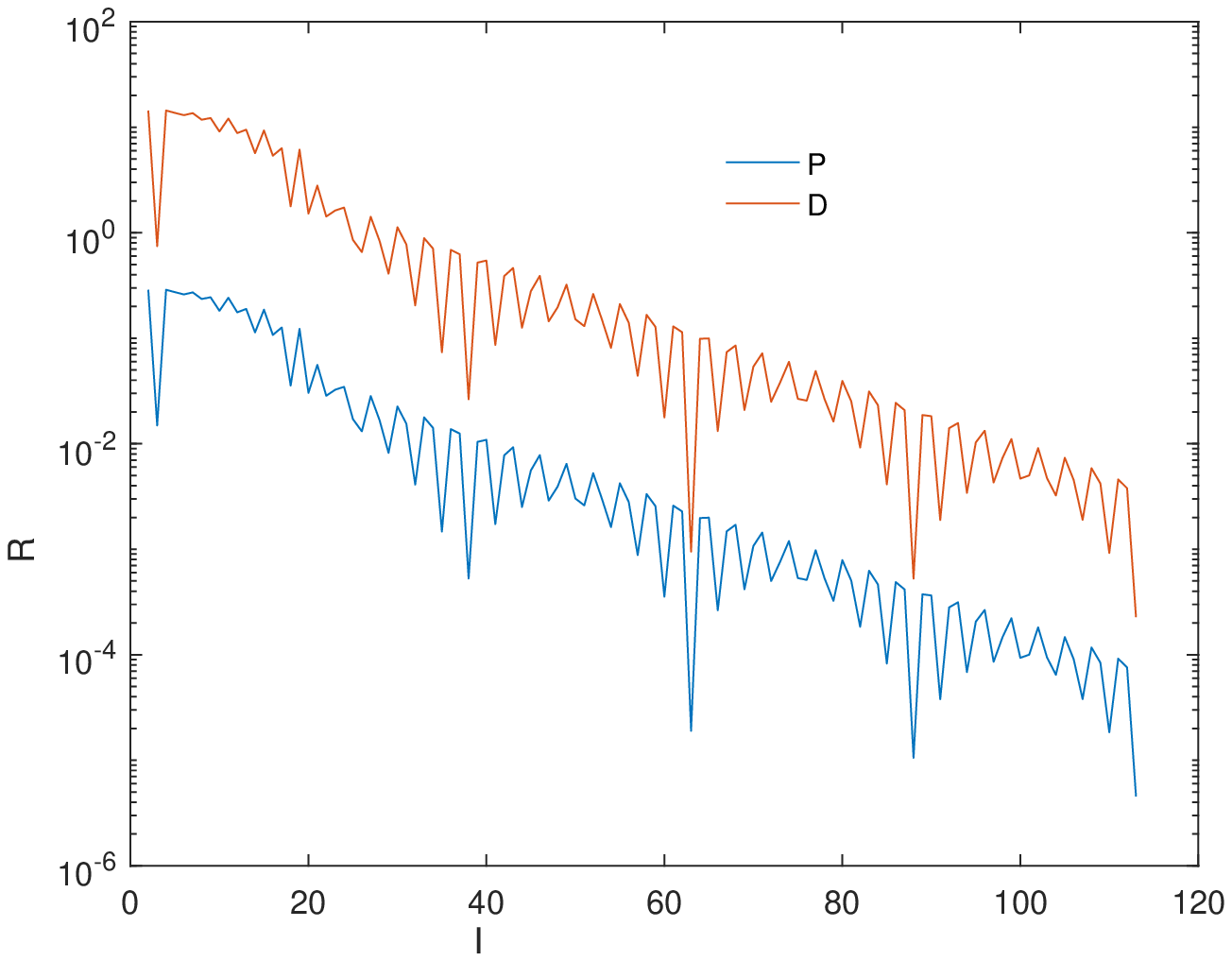}
		\caption{$\boldsymbol{x}^{1}=[-1,-1]$}
		\label{fig:example1proximal1}   
	\end{subfigure}             
	\begin{subfigure}[t]{0.24\textwidth}
		\psfrag{R}{\scriptsize Residuals \normalsize}
		\psfrag{I}{\scriptsize Iterations ($k$) \normalsize}
		\psfrag{P}{\tiny Primal residuals \normalsize}
		\psfrag{D}{\tiny Dual residuals \normalsize}
		\includegraphics[width=48mm]{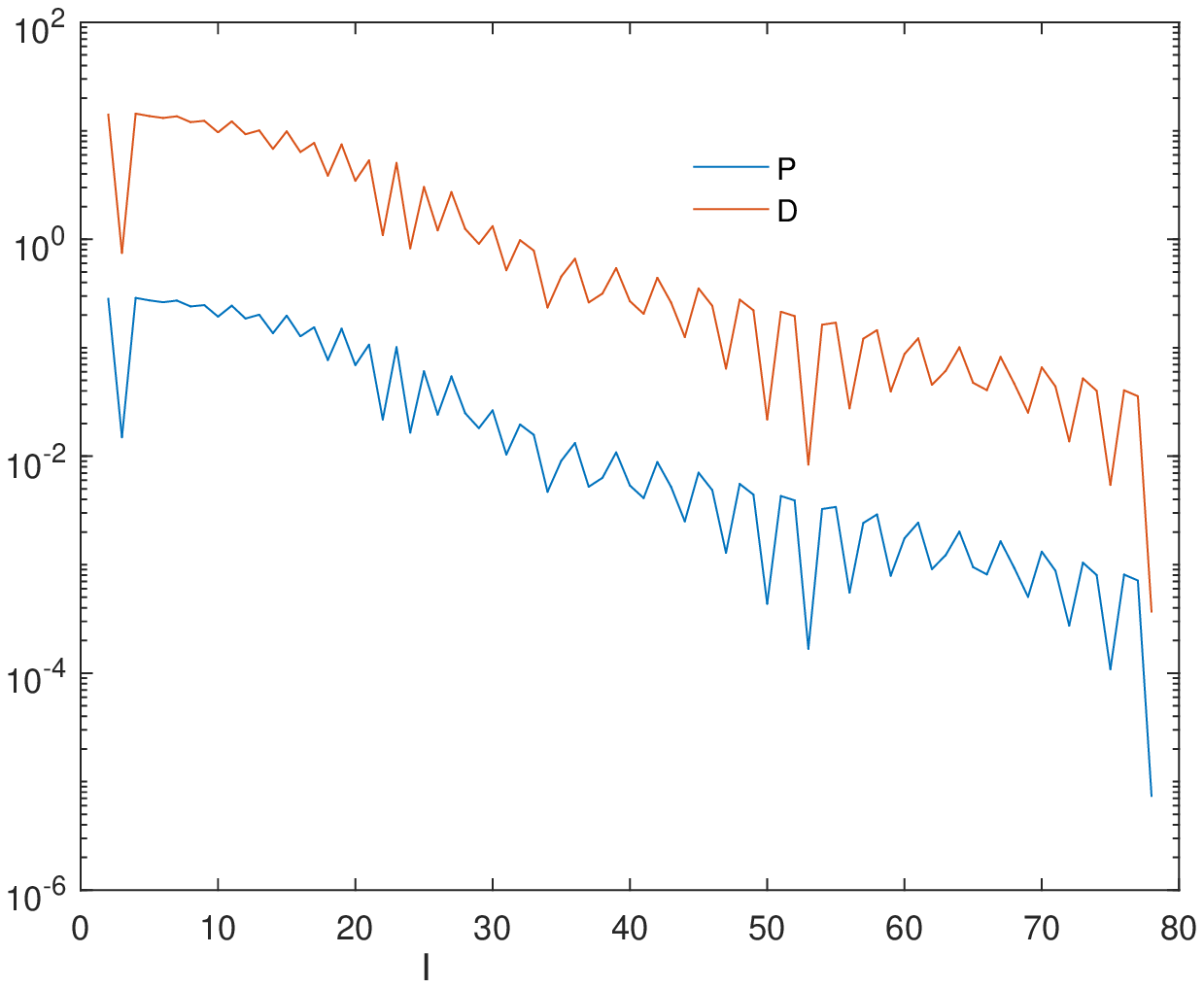}
		\caption{$\boldsymbol{x}^{1}=[1,1]$}
		\label{fig:example1proximal2}
	\end{subfigure}             
	\caption{Primal and dual residuals of the proximal method with $\nu=50$, for $\boldsymbol{x}^{1}=[-1,-1]$ (a) and $\boldsymbol{x}^{1}=[1,1]$ (b).}
	\label{fig:example1proximal}
\end{figure}

Finally, Figure~\ref{fig:example1proximal} shows the convergence of the proximal method with $\nu=50$, for $\boldsymbol{x}^{1}=[-1,-1]$ (Figure~\ref{fig:example1proximal1}) and $\boldsymbol{x}^{1}=[1,1]$ (Figure~\ref{fig:example1proximal2}).
In particular, given the current iterates $\left(x_{1}^{k},x_{2}^{k},\lambda^{k} \right) $, the proximal method generates a new iterate $\left(x_{1}^{k+1},x_{2}^{k+1},\lambda^{k+1} \right) $ as follows
\begin{align}
x_{1}^{k+1} \in & \ \underset {\substack{x_{1} \in \mathcal{X}_{1}}} {\argmin } \ \left\lbrace L\left( x_{1}, x_{2}^{k}, \lambda\right)+\frac{\nu}{2}\left\|  x_{1}-x_{1}^{k} \right\| ^{2} \right\rbrace , \label{eq:proxx1} \\
x_{2}^{k+1} \in & \ \underset {\substack{x_{2} \in \mathcal{X}_{2}}} {\argmin } \ \left\lbrace L\left( x_{1}^{k}, x_{2}, \lambda\right)+\frac{\nu}{2}\left\|  x_{2}-x_{2}^{k} \right\| ^{2} \right\rbrace , \label{eq:proxx2} \\
\lambda^{k+1}=& \ \lambda^{k}+\nu \left(x_{1}^{k+1}-x_{2}^{k+1} \right). \label{eq:lupdateprox}
\end{align}
Just like ADMM, the proximal method is yet another method for approximating the modified dual function, and therefore, the same observations seen when applying ADMM above are witnessed when applying the proximal method. The only difference is that the proximal method takes longer than ADMM to converge due the oscillatory behaviour seen in Figure~\ref{fig:example1proximal}. However, the superior convergence of ADMM comes at the expense of more message exchanges.

\section{Modified dual - example 2}\label{Appendix2}

\begin{figure}
	\centering
	\begin{subfigure}[t]{0.24\textwidth}
		\psfrag{x}{\footnotesize $x_{1}$ \normalsize}
		\psfrag{y}{\footnotesize $x_{2}$ \normalsize}
		\includegraphics[width=48mm]{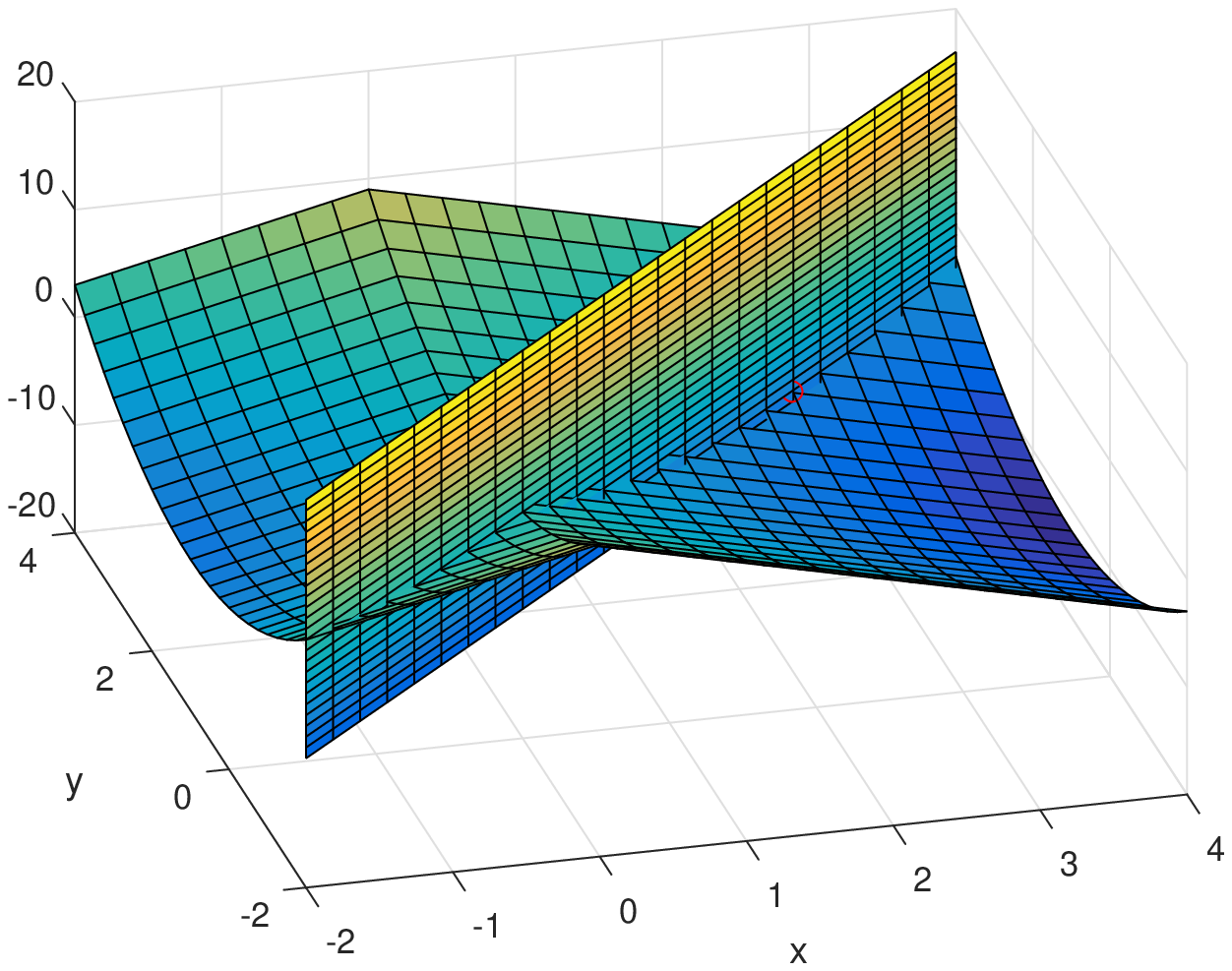}
		\caption{}
		\label{fig:example2primal3d}   
	\end{subfigure}             
	\begin{subfigure}[t]{0.24\textwidth}
		\psfrag{x}{\footnotesize $x_{1}$ \normalsize}
		\includegraphics[width=48mm]{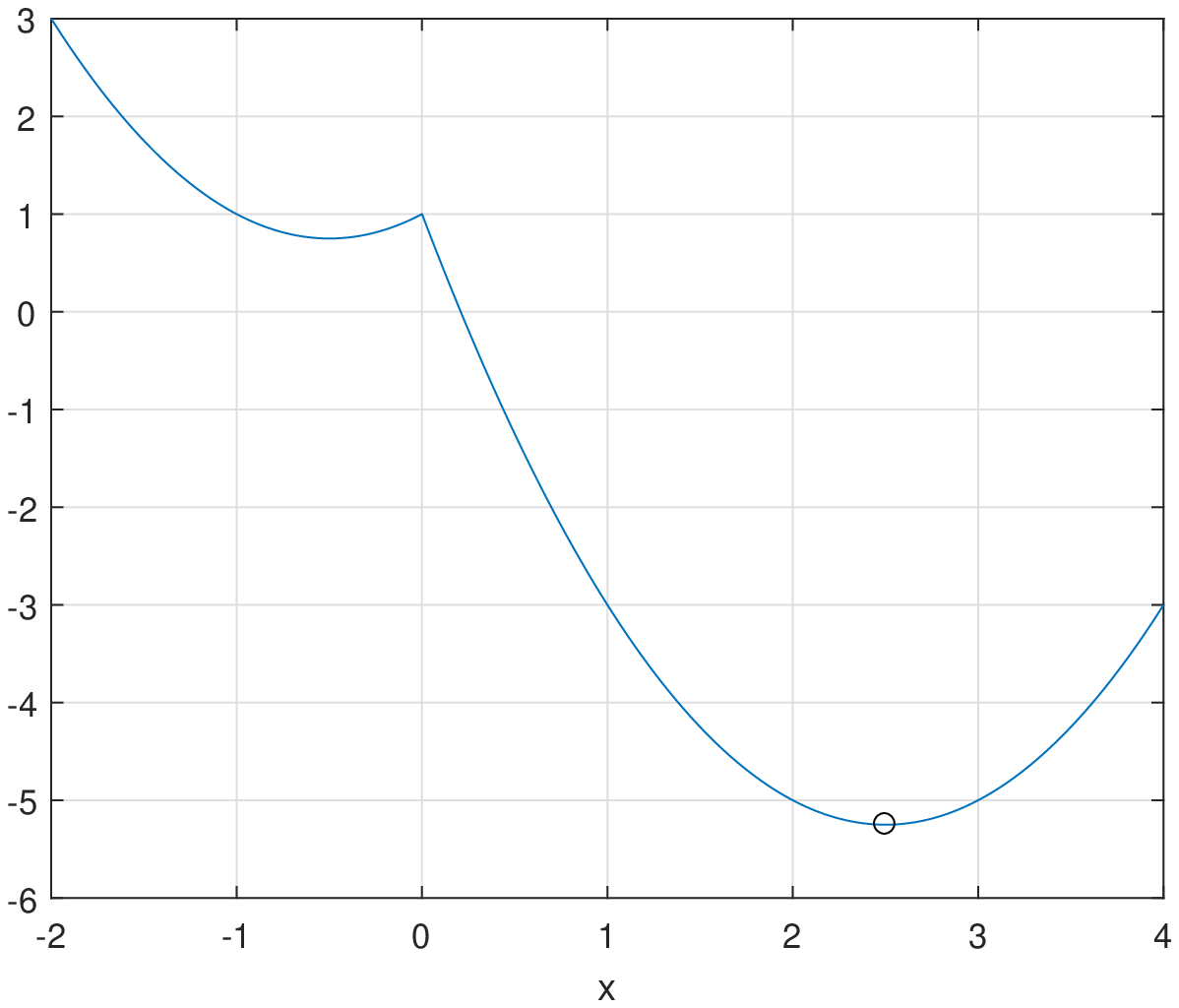}
		\caption{}
		\label{fig:example2primal2d}
	\end{subfigure}             
	\caption{Problem \eqref{example2primal} (a) and its 2D equivalent in (b). The optimal point and value are $\boldsymbol{x}^{\star}=[2.5,2.5]$, $p^{\star}=-5.25$ (shown as a circle). The suboptimal point and value are $\boldsymbol{x}^{\dagger}=[-0.5,-0.5]$, $p^{\dagger}=0.75$.}
	\label{fig:example2primal}
\end{figure}

\begin{figure}
	\centering
	\begin{subfigure}[t]{0.24\textwidth}
		\psfrag{l}{\scriptsize $\lambda$ \normalsize}
		\psfrag{Dual}{\scriptsize $D_{\rho}\left( \lambda\right) $ \normalsize}
		\includegraphics[width=48mm]{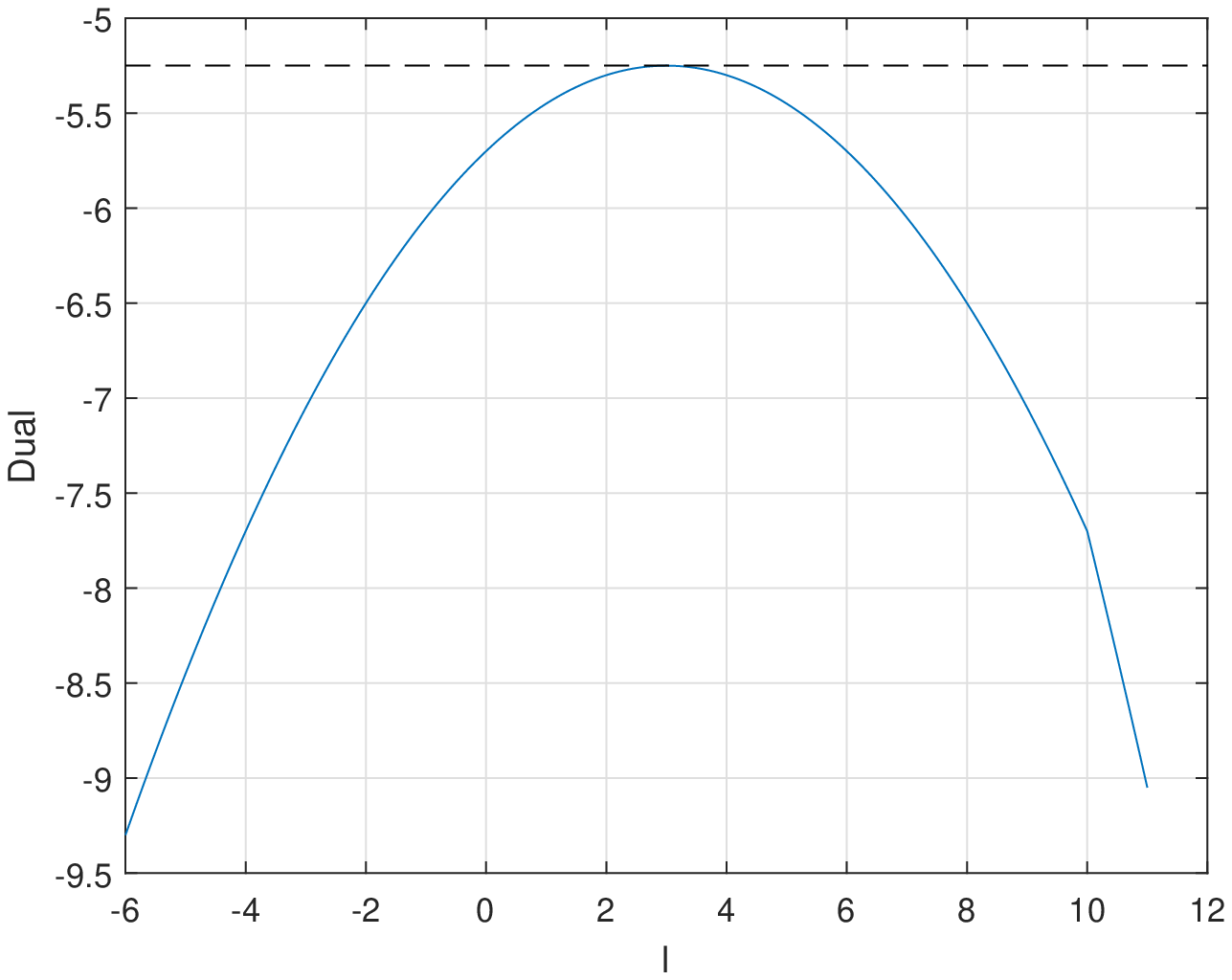}
		\caption{}
		\label{fig:example2ALRdual1}   
	\end{subfigure}             
	\begin{subfigure}[t]{0.24\textwidth}
		\psfrag{l}{\scriptsize $\lambda$ \normalsize}
		\includegraphics[width=48mm]{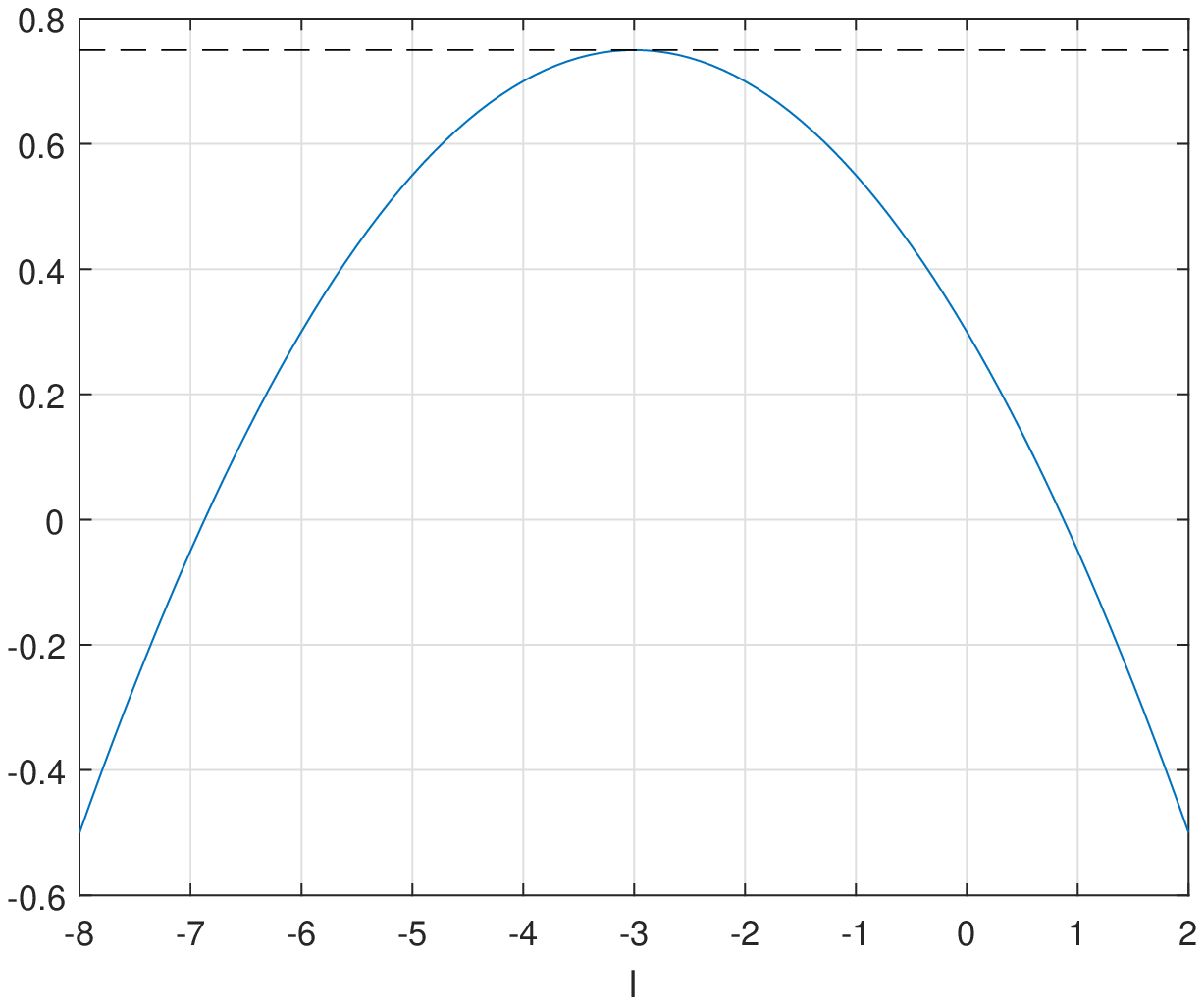}
		\caption{}
		\label{fig:example2ALRdual2}
	\end{subfigure}             
	\caption{The augmented Lagrange dual function of problem \eqref{example2primal}, with $\rho=10$. The dashed lines in (a) and (b) show $p^{\star}$ and $p^{\dagger}$ respectively.}
	\label{fig:example2ALRdual}
\end{figure}

Consider the nonconvex problem
\begin{subequations}\label{example2primal}
	\begin{align}
	\underset {\substack{x_{1},x_{2}}} 
	{\mbox{minimize  }}   -3\left| x_{1}\right|  & + \left(x_{2}-1 \right)^{2}   \\
	\text{subject to } \ \ \qquad x_{1}&=x_{2},
	\end{align}
\end{subequations}
shown in Figure~\ref{fig:example2primal}. 
Let $\boldsymbol{x}:=\left[ x_{1},x_{2}\right]  \in \mathcal{X}$, where $\mathcal{X}:=\mathcal{X}_{1} \times \mathcal{X}_{2}=\reals^2$, $\mathcal{X}_{1}=\reals$ and $\mathcal{X}_{2}=\reals$. Also, let $f(\boldsymbol{x})=-3\left| x_{1}\right| + \left(x_{2}-1 \right)^{2}$ $\left( f:\reals^{2} \mapsto \reals\right) $. The (partial) Lagrange dual function of problem \eqref{example2primal} is written as in \eqref{example1dualf}. The Lagrangian of \eqref{example2primal} is unbounded below in $\boldsymbol{x}$ and the dual function therefore takes the value $-\infty$. However, by modifying the Lagrange dual function as in \eqref{eq:example1dualfA}, the problem now has zero modified duality gap and the augmented Lagrange dual function is smooth over the set $[0,\lambda^{\star}]$, as shown in Figure~\ref{fig:example2ALRdual}. 

\begin{figure}[t]
	\centering
	\begin{subfigure}[t]{0.24\textwidth}
		\psfrag{R}{\scriptsize Residuals \normalsize}
		\psfrag{I}{\scriptsize Iterations ($k$) \normalsize}
		\psfrag{P}{\tiny Primal residuals \normalsize}
		\psfrag{D}{\tiny Dual residuals \normalsize}
		\includegraphics[width=48mm]{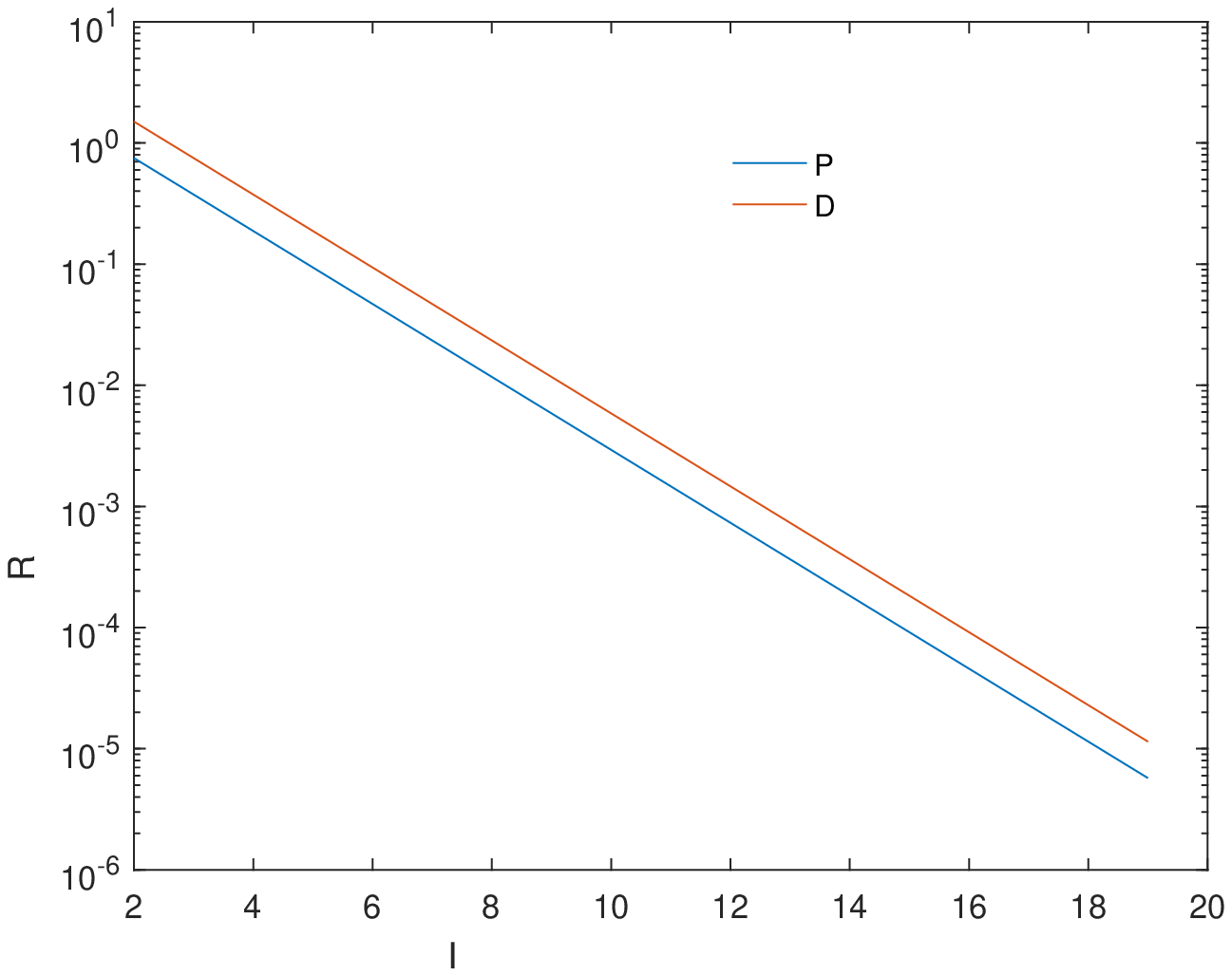}
		\caption{$x_{2}^{1}=1$}
		\label{fig:example2pdresidualsADMM1}   
	\end{subfigure}             
	\begin{subfigure}[t]{0.24\textwidth}
		\psfrag{R}{\scriptsize Residuals \normalsize}
		\psfrag{I}{\scriptsize Iterations ($k$) \normalsize}
		\psfrag{P}{\tiny Primal residuals \normalsize}
		\psfrag{D}{\tiny Dual residuals \normalsize}
		\includegraphics[width=48mm]{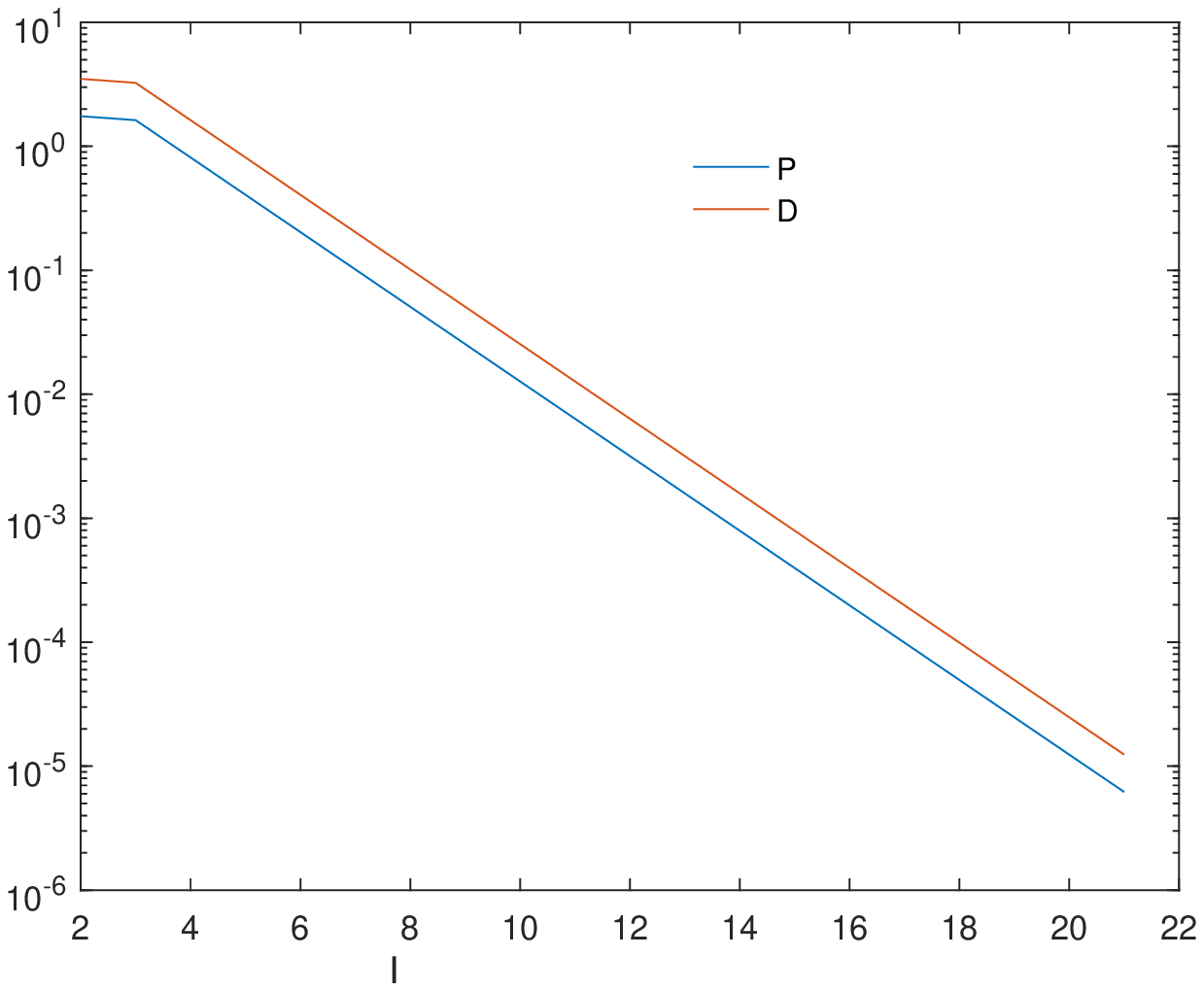}
		\caption{$x_{2}^{1}=-1$}
		\label{fig:example2pdresidualsADMM2}
	\end{subfigure}             
	\caption{Primal and dual residuals of ADMM ($\rho=2$), for $x_{2}^{1}=1$ (a) and $x_{2}^{1}=-1$ (b), and when \eqref{eq:ADMMx1} and \eqref{eq:ADMMx2} are solved to optimality. In both (a) and (b) ADMM converges to $d_{M}^{\star}=p^{\star}$.}
	\label{fig:example2pdresidualsADMM}
\end{figure}

An interesting observation is that, for $\rho=2$ and when \eqref{eq:ADMMx1} and \eqref{eq:ADMMx2} are solved to global optimality, the convergence of ADMM in this example is insensitive to the choice of starting point. In these settings, ADMM always converges to $d_{M}^{\star}=p^{\star}$ and the choice of starting point only affects the speed of convergence, as shown in Figure~\ref{fig:example2pdresidualsADMM}.   
However, for $\rho=10$ and higher, ADMM again becomes sensitive to the choice of starting point. A starting point $x_{2}^{1}=-1$ leads to a convergence to $d_{M}^{\dagger}=p^{\dagger}$ in $42$ iterations, whereas $x_{2}^{1}=1$ leads to a convergence to $d_{M}^{\star}=p^{\star}$ in $49$ iterations. Furthermore, for $\rho=2$, ADMM with $x_{2}^{1}=1$ converges in $19$ iterations as compared to $49$ for $\rho=10$ and $120$ for $\rho=50$. This highlights the fact that increasing $\rho$ does not always translate to a faster convergence. In fact, the effect of $\rho$ on the convergence of ADMM is problem-specific in practice. This should not be surprising as $\rho$ is also considered as the step size in the multiplier update in \eqref{eq:lupdate}. The step size can certainly be adjusted separately but this will be at the expense of more parameter tuning, which results in the loss of generality and simplicity of the method. 

Finally, to underscore the effect of solving \eqref{eq:ADMMx1} and \eqref{eq:ADMMx2} to suboptimality (as might be the case when using an IPM solver), an IPM solver is used in the following two cases for $\rho=2$. In case 1, where the IPM solver is initialized with $\boldsymbol{x}^{0}_{\text{IPM}}=[-1,-1]$ at each iteration $k$ and the algorithm with $x_{2}^{1}=1$, ADMM converges to the suboptimal point $d_{M}^{\dagger}=p^{\dagger}$ in $19$ iterations, suggesting that \eqref{eq:ADMMx1} and \eqref{eq:ADMMx2} are consistently solved to suboptimality. In case 2, where the IPM solver is initialized with $\boldsymbol{x}^{0}_{\text{IPM}}=[1,1]$ at each iteration $k$ and the algorithm with $x_{2}^{1}=1$, ADMM converges to the optimal point $d_{M}^{\star}=p^{\star}$, similar to the convergence in Figure~\ref{fig:example2pdresidualsADMM1}. 

The observations drawn from the examples above can be summarized as follows:
\begin{itemize}
	\item The subproblems in the augmented Lagrangian relaxation have to be solved to global optimality in order to witness a zero modified duality gap.
	\item A distributed method that approximates the modified dual function is (theoretically) not guaranteed to converge to a global optimum if the primal problem is nonconvex \cite{Boyd_ADMM}. Nonetheless, it is possible to witness a zero modified duality gap in such methods if the following conditions hold:
	\begin{itemize}
		\item The subproblems are solved to global optimality at each iteration.
		\item The algorithm is initialized with the same starting point that leads the centralized IPM starting point to a globally optimal solution of the primal.
		\item The penalty parameters and the step sizes in the multiplier update are tailored specifically to the problem at hand, keeping in mind that some parameter settings can make the method insensitive to the choice of algorithmic starting point.
	\end{itemize}
\end{itemize}

\section*{Acknowledgment}

This research is funded by ARENA. The authors would like to thank Dr. Paul Scott for the valuable discussions on the topic.

\bibliographystyle{IEEEtran}
{\footnotesize
\bibliography{DistributedOPF}}

\begin{thebibliography}{10}
\providecommand{\url}[1]{#1}
\csname url@samestyle\endcsname
\providecommand{\newblock}{\relax}
\providecommand{\bibinfo}[2]{#2}
\providecommand{\BIBentrySTDinterwordspacing}{\spaceskip=0pt\relax}
\providecommand{\BIBentryALTinterwordstretchfactor}{4}
\providecommand{\BIBentryALTinterwordspacing}{\spaceskip=\fontdimen2\font plus
\BIBentryALTinterwordstretchfactor\fontdimen3\font minus
  \fontdimen4\font\relax}
\providecommand{\BIBforeignlanguage}[2]{{%
\expandafter\ifx\csname l@#1\endcsname\relax
\typeout{** WARNING: IEEEtran.bst: No hyphenation pattern has been}%
\typeout{** loaded for the language `#1'. Using the pattern for}%
\typeout{** the default language instead.}%
\else
\language=\csname l@#1\endcsname
\fi
#2}}
\providecommand{\BIBdecl}{\relax}
\BIBdecl

\bibitem{IPMforOPF}
R.~Jabr, A.~Coonick, and B.~Cory, ``A primal-dual interior point method for
  optimal power flow dispatching,'' \emph{Power Systems, IEEE Transactions on},
  vol.~17, no.~3, pp. 654--662, Aug 2002.

\bibitem{MATPOWER}
R.~Zimmerman, C.~Murillo-S\'{a}nchez, and R.~Thomas, ``{MATPOWER}: Steady-state
  operations, planning, and analysis tools for power systems research and
  education,'' \emph{Power Systems, IEEE Transactions on}, vol.~26, no.~1, pp.
  12--19, Feb 2011.

\bibitem{Kocuk_strongSOCP}
B.~Kocuk, S.~S. Dey, and X.~A. Sun, ``Strong {SOCP} relaxations for the optimal
  power flow problem,'' \emph{Operations Research}, 2016.

\bibitem{Coffrin_Strengtheningwithboundtightening}
C.~Coffrin, H.~Hijazi, and P.~Van~Hentenryck, ``Strengthening convex
  relaxations with bound tightening for power network optimization,'' in
  \emph{Principles and Practice of Constraint Programming}, ser. Lecture Notes
  in Computer Science, G.~Pesant, Ed.\hskip 1em plus 0.5em minus 0.4em\relax
  Springer International Publishing, 2015, vol. 9255, pp. 39--57.

\bibitem{Hijazi_PolynomialSDPcuts}
H.~Hijazi, C.~Coffrin, and P.~V. Hentenryck, ``Polynomial {SDP} cuts for
  optimal power flow,'' in \emph{2016 Power Systems Computation Conference
  (PSCC)}, June 2016, pp. 1--7.

\bibitem{Coffrin_StrengtheningSDP}
C.~Coffrin, H.~Hijazi, and P.~V. Hentenryck, ``Strengthening the {SDP}
  relaxation of {AC} power flows with convex envelopes, bound tightening, and
  valid inequalities,'' \emph{IEEE Transactions on Power Systems}, vol.~PP,
  no.~99, pp. 1--1, 2016.

\bibitem{Kocuk_Matrixminorreformulations}
B.~Kocuk, S.~S. Dey, and X.~A. Sun, ``Matrix minor reformulation and
  {SOCP}-based spatial branch-and-cut method for the {AC} optimal power flow
  problem,'' \emph{arXiv preprint arXiv:1703.03050}, 2017.

\bibitem{Lavai_0dualitygapinOPF}
J.~Lavaei and S.~Low, ``Zero duality gap in optimal power flow problem,''
  \emph{Power Systems, IEEE Transactions on}, vol.~27, no.~1, pp. 92--107, Feb
  2012.

\bibitem{Molzahn_investigationofnon0gap}
D.~K. Molzahn, B.~C. Lesieutre, and C.~L. DeMarco, ``Investigation of non-zero
  duality gap solutions to a semidefinite relaxation of the optimal power flow
  problem,'' in \emph{System Sciences (HICSS), 2014 47th Hawaii International
  Conference on}.\hskip 1em plus 0.5em minus 0.4em\relax IEEE, 2014, pp.
  2325--2334.

\bibitem{Kocuk_InexactnessofSDP}
B.~Kocuk, S.~S. Dey, and X.~A. Sun, ``Inexactness of {SDP} relaxation and valid
  inequalities for optimal power flow,'' \emph{IEEE Transactions on Power
  Systems}, vol.~31, no.~1, pp. 642--651, Jan 2016.

\bibitem{Josz_moment-sosopf}
C.~Josz, J.~Maeght, P.~Panciatici, and J.~Gilbert, ``Application of the
  moment-{SOS} approach to global optimization of the {OPF} problem,''
  \emph{Power Systems, IEEE Transactions on}, vol.~30, no.~1, pp. 463--470, Jan
  2015.

\bibitem{Molzahn_Moment-BasedRelaxations}
D.~Molzahn and I.~Hiskens, ``Sparsity-exploiting moment-based relaxations of
  the optimal power flow problem,'' \emph{Power Systems, IEEE Transactions on},
  vol.~30, no.~6, pp. 3168--3180, Nov 2015.

\bibitem{Josz_momentsumofsquares}
\BIBentryALTinterwordspacing
C.~Josz and D.~K. Molzahn, ``{Moment/Sum-of-Squares Hierarchy for Complex
  Polynomial Optimization},'' aug 2015. [Online]. Available:
  \url{http://arxiv.org/abs/1508.02068}
\BIBentrySTDinterwordspacing

\bibitem{Jabr_radialDNusingConicP}
R.~Jabr, ``Radial distribution load flow using conic programming,'' \emph{Power
  Systems, IEEE Transactions on}, vol.~21, no.~3, pp. 1458--1459, Aug 2006.

\bibitem{Kim_coarsegraineddistOPF}
B.~H. Kim and R.~Baldick, ``Coarse-grained distributed optimal power flow,''
  \emph{IEEE Transactions on Power Systems}, vol.~12, no.~2, pp. 932--939, May
  1997.

\bibitem{Kim_fastdistributedOPF}
R.~Baldick, B.~H. Kim, C.~Chase, and Y.~Luo, ``A fast distributed
  implementation of optimal power flow,'' \emph{IEEE Transactions on Power
  Systems}, vol.~14, no.~3, pp. 858--864, Aug 1999.

\bibitem{Kim_comparisonofDOPF}
B.~H. Kim and R.~Baldick, ``A comparison of distributed optimal power flow
  algorithms,'' \emph{IEEE Transactions on Power Systems}, vol.~15, no.~2, pp.
  599--604, May 2000.

\bibitem{Biskas_DecentralizedOPF}
P.~N. Biskas and A.~G. Bakirtzis, ``Decentralised {OPF} of large multiarea
  power systems,'' \emph{IEE Proceedings - Generation, Transmission and
  Distribution}, vol. 153, no.~1, pp. 99--105, Jan 2006.

\bibitem{Magnusson_ADMMsequentialconvex}
S.~Magn\'{u}sson, P.~C. Weeraddana, and C.~Fischione, ``A distributed approach
  for the optimal power flow problem based on {ADMM} and sequential convex
  approximations,'' \emph{IEEE Transactions on Control of Network Systems},
  vol.~2, no.~3, pp. 238--253, Sept 2015.

\bibitem{Erseghe_ADMM}
T.~Erseghe, ``Distributed optimal power flow using {ADMM},'' \emph{IEEE
  Transactions on Power Systems}, vol.~29, no.~5, pp. 2370--2380, Sept 2014.

\bibitem{Kraning_ADMM}
M.~Kraning, E.~Chu, J.~Lavaei, and S.~Boyd, ``Dynamic network energy management
  via proximal message passing,'' \emph{Found. Trends Optim.}, vol.~1, no.~2,
  pp. 73--126, Jan. 2014.

\bibitem{Peng_DOPFbalancedradial}
Q.~Peng and S.~H. Low, ``Distributed optimal power flow algorithm for radial
  networks, i: Balanced single phase case,'' \emph{IEEE Transactions on Smart
  Grid}, pp. 1--11, to be published.

\bibitem{DallAnese_SDPADMM}
E.~Dall'Anese, H.~Zhu, and G.~B. Giannakis, ``Distributed optimal power flow
  for smart microgrids,'' \emph{IEEE Transactions on Smart Grid}, vol.~4,
  no.~3, pp. 1464--1475, Sept 2013.

\bibitem{Scott_DistributedOPFforDR}
P.~Scott and S.~Thi\'{e}baux, ``Distributed multi-period optimal power flow for
  demand response in microgrids,'' in \emph{Proceedings of the 2015 ACM Sixth
  International Conference on Future Energy Systems}, ser. e-Energy '15.\hskip
  1em plus 0.5em minus 0.4em\relax New York, NY, USA: ACM, 2015, pp. 17--26.

\bibitem{Nogales_DecompositionforOPF}
F.~J. Nogales, F.~J. Prieto, and A.~J. Conejo, ``A decomposition methodology
  applied to the multi-area optimal power flow problem,'' \emph{Annals of
  Operations Research}, vol. 120, no.~1, pp. 99--116, 2003.

\bibitem{Bakirtzis_DecentralizedDCOPF}
A.~G. Bakirtzis and P.~N. Biskas, ``A decentralized solution to the {DC}-{OPF}
  of interconnected power systems,'' \emph{IEEE Transactions on Power Systems},
  vol.~18, no.~3, pp. 1007--1013, Aug 2003.

\bibitem{Hug_DecentralizedOPF}
G.~Hug-Glanzmann and G.~Andersson, ``Decentralized optimal power flow control
  for overlapping areas in power systems,'' \emph{IEEE Transactions on Power
  Systems}, vol.~24, no.~1, pp. 327--336, Feb 2009.

\bibitem{Guo_Intelligentpartitioning}
J.~Guo, G.~Hug, and O.~K. Tonguz, ``Intelligent partitioning in distributed
  optimization of electric power systems,'' \emph{IEEE Transactions on Smart
  Grid}, vol.~7, no.~3, pp. 1249--1258, May 2016.

\bibitem{Minot_ParallelDCOPF}
A.~Minot, Y.~M. Lu, and N.~Li, ``A parallel primal-dual interior-point method
  for dc optimal power flow,'' in \emph{2016 Power Systems Computation
  Conference (PSCC)}, June 2016, pp. 1--7.

\bibitem{Lam_distributedOPF}
A.~Y.~S. Lam, B.~Zhang, and D.~N. Tse, ``Distributed algorithms for optimal
  power flow problem,'' in \emph{2012 IEEE 51st IEEE Conference on Decision and
  Control (CDC)}, Dec 2012, pp. 430--437.

\bibitem{Madani_DistributedSparseSDPforOPF}
R.~Madani, A.~Kalbat, and J.~Lavaei, ``{Distributed Computation for Sparse
  Semidefinite Programming with Applications to Power Optimization Problems}.''

\bibitem{Sun_ACOPFalgorithms}
A.~X. Sun, D.~T. Phan, and S.~Ghosh, ``Fully decentralized {AC} optimal power
  flow algorithms,'' in \emph{2013 IEEE Power Energy Society General Meeting},
  July 2013, pp. 1--5.

\bibitem{Shyan-Lung_Parallelsolutions}
S.-L. Lin and J.~E.~V. Ness, ``Parallel solution of sparse algebraic
  equations,'' in \emph{Conference Proceedings Power Industry Computer
  Application Conference}, May 1993, pp. 380--386.

\bibitem{NESTA}
C.~Coffrin, D.~Gordon, and P.~Scott, ``{NESTA}, the {NICTA} energy system test
  case archive,'' \emph{CoRR}, vol. abs/1411.0359, 2014.

\bibitem{Josz_ACdataMATPOWER}
C.~Josz, S.~Fliscounakis, J.~Maeght, and P.~Panciatici, ``{AC} power flow data
  in {MATPOWER} and {QCQP} format: i{T}esla, {RTE} snapshots, and {PEGASE},''
  \emph{arXiv preprint arXiv:1603.01533}, 2016.

\bibitem{AMPL}
R.~Fourer, D.~M. Gay, and B.~Kernighan, \emph{Algorithms and Model Formulations
  in Mathematical Programming}, S.~W. Wallace, Ed.\hskip 1em plus 0.5em minus
  0.4em\relax New York, NY, USA: Springer-Verlag New York, Inc., 1989.

\bibitem{KNITRO}
R.~H. Byrd, J.~Nocedal, and R.~A. Waltz, ``Knitro: An integrated package for
  nonlinear optimization,'' in \emph{Large Scale Nonlinear Optimization,
  35–59, 2006}.\hskip 1em plus 0.5em minus 0.4em\relax Springer Verlag, 2006,
  pp. 35--59.

\bibitem{IPOPT}
A.~W{\"a}chter and L.~T. Biegler, ``On the implementation of an interior-point
  filter line-search algorithm for large-scale nonlinear programming,''
  \emph{Mathematical Programming}, vol. 106, no.~1, pp. 25--57, 2006.

\bibitem{Danskin}
J.~M. Danskin, \emph{{The Theory of Max-Min and Its Applications to Weapons
  Allocation Problems}}.\hskip 1em plus 0.5em minus 0.4em\relax New York:
  Springer-Verlag, 1967.

\bibitem{onatheoremofDanskin}
P.~Bernhard and A.~Rapaport, ``On a theorem of {D}anskin with an application to
  a theorem of {V}on {N}eumann-{S}ion,'' \emph{Nonlinear Analysis: Theory,
  Methods \& Applications}, vol.~24, no.~8, pp. 1163 -- 1181, 1995.

\bibitem{nonlinearpogramming}
D.~P. Bertsekas, \emph{Nonlinear programming}.\hskip 1em plus 0.5em minus
  0.4em\relax Athena Scientific, 1999.

\bibitem{Bertsekas_convexoptalg2015}
D.~Bertsekas, \emph{Convex optimization algorithms}.\hskip 1em plus 0.5em minus
  0.4em\relax Athena Scientific Belmont, 2015.

\bibitem{Boyd_ADMM}
S.~Boyd, N.~Parikh, E.~Chu, B.~Peleato, and J.~Eckstein, ``Distributed
  optimization and statistical learning via the alternating direction method of
  multipliers,'' \emph{Foundations and Trends{\textregistered} in Machine
  Learning}, vol.~3, no.~1, pp. 1--122, 2011.

\end{thebibliography}

\end{document}